\documentclass[11pt,a4paper]{article}
\pdfoutput=1
\usepackage{jheppub}

\usepackage{hyperref}

\usepackage[english]{babel}

\usepackage{amsmath}
\usepackage{amssymb}

\usepackage{epstopdf}

\usepackage{verbatim}

\usepackage{xcolor}
\usepackage{physics}
\usepackage{graphicx} 

\usepackage{tabularx}
\usepackage{multirow}
\usepackage{multicol}
\usepackage{pbox}

\usepackage[normalem]{ulem} 
\usepackage{comment}

\def\apjl{ApJ}%
\def\apjs{ApJS}%
\def\ao{Appl.\ Opt.}%
\def\aap{A\&A}%

\makeatother

\usepackage{babel}
\begin{document}

\author[a]{Alessia Platania} 

\affiliation[a]{Perimeter Institute for Theoretical Physics, 31 Caroline St. N., Waterloo, ON N2L 2Y5, Canada}
  
\emailAdd{aplatania@perimeterinstitute.ca}

\title{Causality, unitarity and stability in quantum gravity: \\a non-perturbative perspective}

\begin{abstract}
{Resumming  quantum fluctuations at the level of the gravitational path integral is expected to result in non-local effective actions and thus in a non-trivial momentum dependence of the propagator. Which properties the (dressed) graviton propagator has to satisfy and whether they can all be met are  key open questions. In this work we present criteria and conditions for the momentum dependence of a graviton propagator which is consistent with unitarity, causality, and stability in a non-perturbative setting. To this end, we revisit several aspects of these conditions, highlighting some caveats and subtleties that got lost in recent discussions, and spelling out others that to our best knowledge have not been studied in detail. We discuss the consequences of these concepts for the properties of the graviton propagator. Finally, we provide examples of  propagators satisfying unitarity and causality, while avoiding tachyonic and vacuum instabilities, and allowing for an analytic Wick rotation.}
\end{abstract}

\maketitle

\section{Introduction}


The conundrum of quantum gravity (QG) originally arose as a fundamental incompatibility between renormalizability and unitarity in the perturbative quantization of Einstein-Hilbert~\cite{tHooft:1974toh,Goroff:1985sz,Goroff:1985th,vandeVen:1991gw} and quadratic gravity~\cite{Stelle:1977ry}. Subsequent attempts to reconcile unitarity and renormalizability resulted in a proliferation of different approaches to the problem of QG. 

The formulation of the Standard Model of particle physics was crucially driven by the interplay between theoretical ideas and  experimental tests: inputs from the theory drove the experiments, and the experiments constrained the models and tested their assumptions. QG lives in a different universe: quantum gravitational effects may be too tiny to be directly detected in experiments. Discriminating between different proposals based on observations appears currently out of reach. 
Lacking experiments, theoretical investigations of the quantum aspects of gravity grope in the dark. Consistency, which is far from being trivial in the realm of QG, becomes a key requirement to guide theoretical studies. 

First of all, a fundamental quantum theory of gravity ought to be renormalizable: it must be predictive, delivering (finite) observables parametrized by a few free parameters only.
In the modern, Wilsonian understanding of renormalization~\cite{Wilson:1973jj}, the renormalization group (RG) trajectory of a renormalizable quantum field theory (QFT) ends up in a (free or interacting) fixed point in the ultraviolet (UV), and the number of free parameters of the theory is dictated by the codimension of its critical hypersurface. Power counting suggests that Einstein gravity cannot be asymptotically free, due to the negative mass dimension of the Newton coupling. Nonetheless, the existence of an ``asymptotically safe'' (i.e., interacting) fixed point for gravity~\cite{1976W}---whether fundamental or stemming from string theory~\cite{deAlwis:2019aud,Basile:2021euh}---remains a compelling possibility~\cite{Eichhorn:2017egq,Donoghue:2019clr,Pawlowski:2020qer,Bonanno:2020bil}, which can be investigated using, e.g., lattice techniques~\cite{Laiho:2016nlp,Loll:2019rdj} or the Functional Renormalization Group (FRG)~\cite{Dupuis:2020fhh}\footnote{In an RG setup, the bare action is not guessed, rather, it is derived as an UV fixed point of the RG flow (modulo reconstruction problem~\cite{Manrique:2008zw,Morris:2015oca,Fraaije:2022uhg})}.
The FRG has proven useful 
in several approaches to QG, including  asymptotically safe gravity~\cite{Souma:1999at,Lauscher:2002sq,Litim:2003vp,Codello:2006in,Machado:2007ea,Benedetti:2009rx,Manrique:2011jc,Dietz:2012ic,Ohta:2013uca,Dona:2013qba,Codello:2013fpa,Christiansen:2014raa,Falls:2014tra,Becker:2014qya,Christiansen:2015rva,Meibohm:2015twa,Oda:2015sma,Biemans:2016rvp,Eichhorn:2016esv,  Falls:2016msz,Gies:2016con,Biemans:2017zca,Christiansen:2017cxa,Hamada:2017rvn,Platania:2017djo,Falls:2017lst,Eichhorn:2018nda,Eichhorn:2019yzm,Knorr:2021slg,Baldazzi:2021orb,Bonanno:2021squ,Fehre:2021eob,Knorr:2022ilz}, unimodular quantum gravity~\cite{Eichhorn:2013xr,deBrito:2019umw,deBrito:2020rwu,deBrito:2020xhy}, Ho\v{r}ava–Lifshitz gravity~\cite{Contillo:2013fua,DOdorico:2015pil}, Lorentz-symmetry-violating models~\cite{Knorr:2018fdu,Eichhorn:2019ybe}, matrix and tensor models~\cite{Eichhorn:2013isa,Eichhorn:2017xhy,Eichhorn:2018phj,Eichhorn:2019hsa}, group field theory~\cite{Benedetti:2015yaa,Geloun:2016qyb,BenGeloun:2018ekd,Lahoche:2018hou,Pithis:2020kio} and string theory~\cite{deAlwis:2019aud,Basile:2021euh,Basile:2021krk,Basile:2021krr,Gao:2022ojh,Ferrero:2022dpk}. Crucially, the FRG can be used as an alternative to the path integral to compute the \emph{quantum effective action}~\cite{Codello:2015oqa,Knorr:2018kog,Knorr:2019atm,Ohta:2020bsc,Bonanno:2021squ,Knorr:2021niv}. In turn, the quantum effective action can be exploited to connect first-principle computations in QG with experiments and observations, e.g., to explore quantum corrections to cosmological and black-hole spacetimes (see~\cite{Bonanno:2006eu,Falls:2012nd,torres15,Koch:2015nva,Bonanno:2015fga,Bonanno:2016rpx,Kofinas:2016lcz,Falls:2016wsa,Bonanno:2016dyv,Bonanno:2017gji,Bonanno:2017kta,Bonanno:2017zen,Bonanno:2018gck,Liu:2018hno,Majhi:2018uao,Anagnostopoulos:2018jdq,Adeifeoba:2018ydh,Pawlowski:2018swz,Gubitosi:2018gsl,Platania:2019qvo,Platania:2019kyx,Bonanno:2019ilz,Held:2019xde,Bosma:2019aiu,Knorr:2022kqp} for some simple models and~\cite{Bonanno:2017pkg,Platania:2020lqb} for reviews), and to compute scattering amplitudes~\cite{Draper:2020bop,Draper:2020knh,Knorr:2020bjm,Ferrero:2021lhd,Knorr:2022lzn}. Finally, the knowledge quantum effective action is paramount to establish whether the theory satisfies all known consistency criteria.

Indeed, beyond renormalizability, a consistent theory of QG ought to preserve at least some of the properties of (local) QFTs, such as causality, unitarity and stability. In the context of effective field theories (EFT), these conditions have been translated in strong and very precise bounds on the Wilson coefficients~\cite{Adams:2006sv,Cheung:2016yqr,Bellazzini:2016xrt,deRham:2017zjm, deRham:2017xox, deRham:2018qqo, DeRham:2018bgz, Alberte:2019lnd, Alberte:2019xfh, Alberte:2019zhd, Alberte:2020bdz, Alberte:2020jsk, deRham:2021fpu,Herrero-Valea:2022lfd, deRham:2022hpx}. Yet, their non-perturbative realization beyond an EFT setup requires accounting for the momentum dependence of couplings (aka, the form factors~\cite{Knorr:2019atm}) in the quantum effective action. Logarithmic form factors  stemming from perturbative computations at one-loop order are the simplest realization of this momentum dependence, and their presence already brings important consequences for unitarity, causality and stability~\cite{Donoghue:2018lmc,Donoghue:2019ecz,Donoghue:2019fcb,Donoghue:2021meq}. Going beyond one-loop order, the form factors can be much more complicated and ultimately ought to be derived by integrating out all quantum gravitational fluctuations, e.g., at the level of the path integral or using the FRG. In particular assessing unitarity, as well as causality and stability, requires investigating the effective action beyond a polynomial expansion in momenta~\cite{Wetterich:2019qzx,Draper:2020bop,Platania:2020knd,Knorr:2021niv,Bonanno:2021squ,Fehre:2021eob}, 
since such truncations can potentially generate fictitious poles~\cite{Kuntz:2019qcf,Platania:2020knd}.

The goal of the present work is to determine  conditions for the non-perturbative realization of causality, unitarity and stability, in particular at the level of the dressed graviton propagator, and to introduce models satisfying all of them while allowing for an analytic Wick rotation. To this end, we first revisit various aspects of (non-perturbative) unitarity, causality and stability, we examine some of their caveats and subtle details, and discuss their applicability to QG. Specifically, in Sect.~\ref{sect:Non-perturbative-unitarity} we argue why (non-perturbative) unitarity is generally best studied at the level of the effective action. Approximations to the effective action based on truncated derivative expansions naturally lead to the appearance of fictitious poles in the dressed propagator. The resulting fake ghosts decouple dynamically when a sufficiently high number of operators is considered, in that the modulus of their initially-negative residue decreases and vanishes in the limit where no approximation is employed~\cite{Platania:2020knd}. We discuss this residue decoupling mechanism in Sect.~\ref{sect:polology}, and we provide
further numerical evidence of its validity in the  case of effective actions constructed with various entire and non-entire form factors. Next, in Sect.~\ref{sect:causality} we collect and analyze various notions and definitions of causality that have appeared in the literature, and attempt to clarify their relation. We also discuss tachyonic and vacuum (in)stabilities, differentiating between the way they arise classically and at the quantum level. In this course, we also highlight that the tachyonic case comes with several ambiguities, and that tachyonic instabilities are not necessarily problematic, since in some cases they can be cured by the interaction terms in the quantum effective action. We show this by providing some arguments and also by constructing an explicit example for a scalar model. In the context of causality, we pick the definition of microscopic causality discussed in~\cite{Donoghue:2019fcb}, which is based on the structure of the Fourier modes of the propagator, and we generalize the analysis from the case of unstable ghosts studied in~\cite{Donoghue:2019fcb} to the case of a general propagator.
The types of poles and their implications for unitarity, causality and stability are summarized in Tab.~\ref{tab:polesconditions}. Our analysis indicates that to avoid acausalities and (tachyonic and vacuum) instabilities, the dressed propagator should be free from complex-conjugate poles and poles with negative width. In addition, to allow for an analytic Wick rotation no essential singularities should occur. In Sect.~\ref{sect:complexpoles}, we highlight that logarithmic quantum corrections that naturally arise at the level of the gravitational effective action can only alleviate some problems, but are not enough to retain unitarity, causality and stability. This motivates our study in Sect.~\ref{sect:goodpropy}, where we construct models where causality, unitarity, stability and analytic continuation can all be preserved. Interestingly, the common feature of these models is the presence of  branch cuts. We summarize our findings in Sect.~\ref{sect:conclusions}.

\section{Non-perturbative unitarity, effective actions, and the RG}\label{sect:Non-perturbative-unitarity}

In this section we review some basic concepts intertwined with unitarity, that will be useful throughout the manuscript, including the notion of unstable particle and how it is related to quantum effects, the optical theorem and its non-perturbative character, and the role of quantum effective actions.

\subsection{Non-perturbative character of the optical theorem and spectral density}

In the context of QFT the condition that the S-matrix is unitary, $S^{\dagger}S=\mathbb{I}$, implies the optical theorem: for any initial state $|i\rangle$ and final state $|f\rangle$, the transfer matrix $T$ -- defined by $S=\mathbb{I}+i T$ -- satisfies the relation
\begin{equation}\label{opti}
i\langle f|T^\dagger- T|i \rangle=\langle f|T^\dagger T|i\rangle \;.
\end{equation}
For a unitary theory the right-hand side of this equation has to be positive for any initial and final state.

While one \textit{could} use perturbation theory to expand the left- and right-hand sides of the optical theorem (provided that this perturbative expansion does not break down at any scale) and verify unitarity order by order, this is not necessary if the quantum effective action $\Gamma_0$ is known. In fact, any scattering amplitude can be computed from the functional derivatives of the effective action according to 
\begin{equation}
\langle f|S|i\rangle\propto\langle\Omega|T\left\{ \phi(x_{1})\dots\phi(x_{n})\right\} |\Omega\rangle_{(c)}=\left[\frac{\delta^{n}\Gamma_0[\phi]}{\delta\phi(x_{1})\dots\delta\phi(x_{n})}\right]_{\phi=0} \;,
\end{equation}
where $\phi(x_{i})$ are interacting quantum fields and $|\Omega\rangle$
is the vacuum of the fully-interacting theory. 
Therefore, the optical theorem does not need perturbation theory, rather it can be seen as a set of non-perturbative relations between scattering amplitudes and cross sections. In particular, if the effective action is known, one could in principle use it to evaluate $\langle f|T^\dagger T|i\rangle$ and verify that it is positive for any initial and final states. 

On the right-hand side of Eq.~\eqref{opti}, one can insert an identity operator to express $\langle f|T^\dagger T|i\rangle$ as a sum over all possible intermediate states
\begin{equation}\label{eq:optth}
	T_{if}-T_{fi}^{\dagger}=i\,\sum_{n}T_{fn}^{\dagger}T_{in}\;\;.
\end{equation}
The unitarity condition thus holds if there are no negative-norm states (ghosts) in the set of all possible asymptotic states. Indeed, if ghosts exist in the spectrum of the theory, the space of asymptotic states is no longer complete, i.e., it is no longer a Fock space. In this case the ``identity'' would carry some minus signs that would also enter the sum in Eq.~\eqref{eq:optth}. Unitarity thus has to do with the field content of the theory. In the context of QFT, this information is encoded in (the pole structure of) the dressed propagator $\mathbf{\Delta}(q^2)=\sum_i\Delta_i(q^2)\mathbf{t}_i$, where $\mathbf{t}_i$ denotes a tensorial structure and~$\Delta_i(q^2)\equiv i D_i(q^2)$ is the correspondent scalar part. For a unitary theory the corresponding spectral density,
\begin{equation}\label{spectralpropyrelation}
\rho(q^2)=-\lim_{\epsilon\to0}\,[\pi^{-1}\mathrm{Im}\left(D(q^2+i\epsilon)\right)]\,,
\end{equation}
has to be real and positive definite (at least for asymptotic states, see the discussion in  Sect.~\ref{subsect:assumpunitarity}). Indeed, for $|i\rangle=|f\rangle$ the optical theorem yields relations of the form $|\mathcal{M}|^2=2\mathrm{Im}\mathcal{M}$, where $\mathcal{M}$ is the transition amplitude $|i\rangle \to |f\rangle$. Specializing to the case where $|i\rangle=|f\rangle$ is a one-particle state, one finds that (\textit{i}) the optical theorem entails a non-trivial relation between the spectral density $\rho$ and the vertices of the theory, and that (\textit{ii}) the unitarity condition implies the positivity of $\rho\propto |\mathcal{M}_{1\to1}|^2$.

\begin{figure}
\centering\includegraphics[scale=0.5]{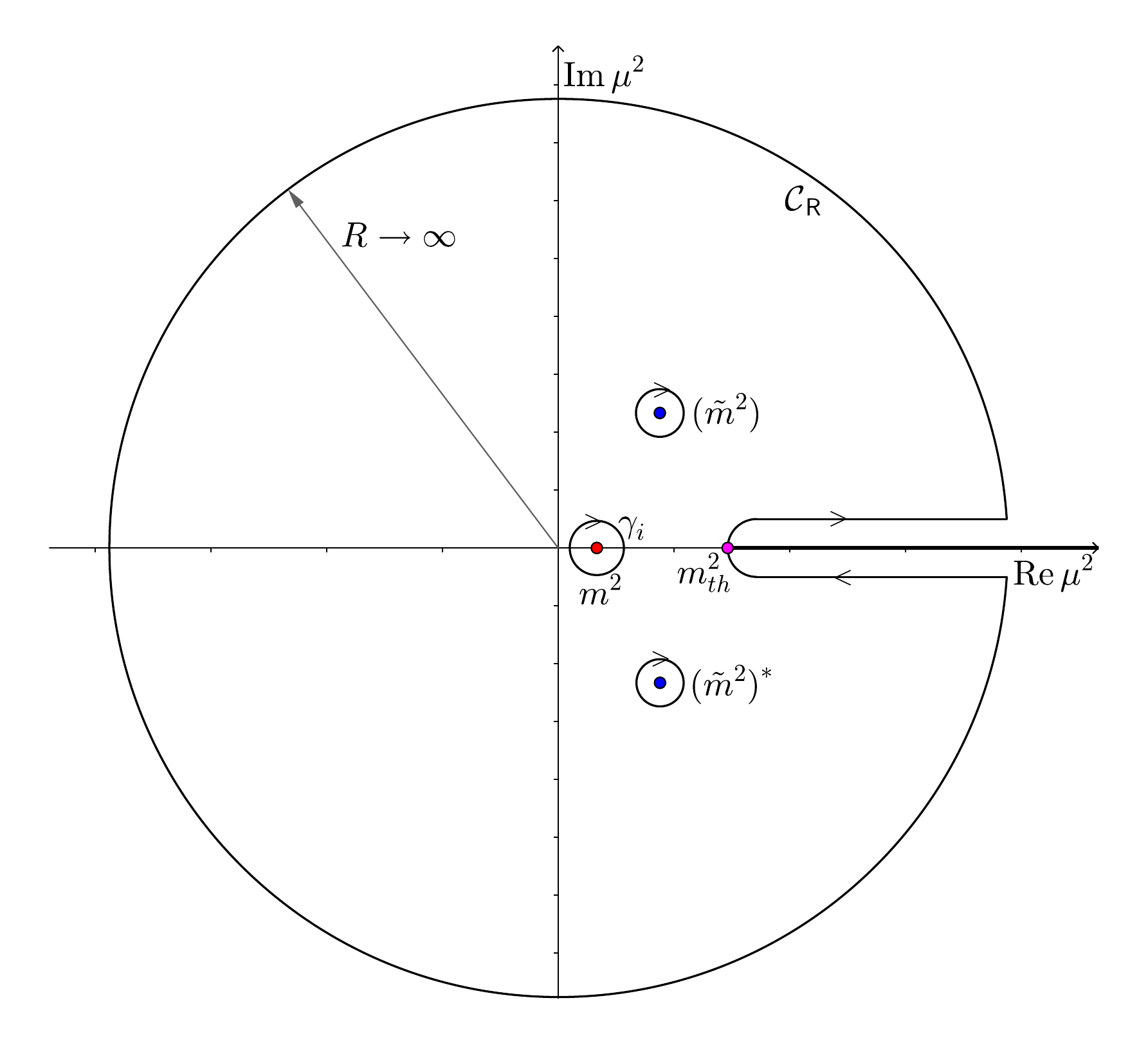}
\caption{Integration contour for Eq.~\eqref{eq:fullydressed}.\label{cauchypoles}}
\end{figure}

A dressed propagator can in principle feature multiple poles, each one corresponding to a degree of freedom of the theory. Assuming that each of these poles has multiplicity one, the contribution of each pole and/or branch singularity to the propagator can be isolated with the aid of the Cauchy integral formula. The propagator can thus be written as a sum of single-pole propagators and, possibly, a continuum part. In what follows we will work in Lorentzian signature, using the mostly negative convention (+ - - -), and we will assume that $D(q^2)$ has no essential singularities invalidating the use of the Cauchy formula. For a scalar propagator with a single branch cut on the real axis\footnote{Let us remark that, in principle, there could be multiple branch cuts, and they do not generally lie on the real axis. An example of this situation will be presented in Sect.~\ref{sect:goodpropy}.}
\begin{align} \label{eq:fullydressed}
\Delta(q^{2})&=i D(q^{2})=i\left\{ \frac{1}{2\pi i}\oint_{\Gamma}\frac{D(\mu^{2})}{\mu^{2}-q^{2}}d\mu^{2}\right\} \nonumber \\
&=i\left\{ \sum_{n}\frac{R_{n}}{q^{2}-m_{n}^{2}}+\sum_{n}\left(\frac{\tilde{R}_{n}}{q^{2}-(\tilde{m}_{n}^{2})}+\frac{\tilde{R}_{n}^{*}}{q^{2}-(\tilde{m}_{n}^{2})^{*}}\right)+\int_{m_{th}^{2}}^{\infty}\frac{\sigma(\mu^{2})}{q^{2}-\mu^{2}}d\mu^{2}\right\} \,,
\end{align}
where $m_n$ are renormalized masses, and $\sigma(q^{2})=\mathrm{Im}\{i\Delta(q^{2})\}$ for $q^{2}>m_{th}^{2}$, with $m_{th}$ being a threshold mass for the production of a resonance or a multi-particle state. The integration contour $\Gamma=\mathcal{C}_{R}\,\cup\,\gamma_{i}$ is shown in Fig.~\ref{cauchypoles}.
Eq.~\eqref{eq:fullydressed} is {the most general
form of a dressed propagator}. There can be contributions from stable particles (ghosts) and possibly bound states, characterized by isolated poles with positive (negative) residues. Pairs of complex-conjugate poles would appear in the form of isolated pairs of poles coming with complex-conjugate masses and residues. The residues of the complex-conjugate poles must be complex conjugates themselves to preserve the reflection-positivity condition of the propagator, $D^{\ast}(z)=D(z^{\ast})$. Finally, unstable particles or multi-particle states correspond to branch cut singularities above or below a threshold mass $m_{th}$. 

The total spectral density reads
\begin{equation}
\rho(q^{2})=-\lim_{\epsilon\to0}\,[\pi^{-1}\mathrm{Im}\left(D(q^2+i\epsilon)\right)]=\sum_{n}R_{n}\delta(q^{2}-m_{n}^{2})+\sigma(q^{2})\theta(q^{2}-m_{th}^{2}) \,.
\end{equation}
The contribution of complex-conjugate poles to the spectral density cancels out, implying that they do not affect unitarity. Stable particles/ghosts contribute with Dirac deltas in the expression of the spectral density. Unstable particles/ghosts contribute instead with a continuum part $\sigma(q^2)$. Unitarity thus holds if
\begin{equation}
R_{n}\geq0\;\;\forall n\qquad \mathrm{and}\qquad\sigma(q^{2})\geq0\,,
\end{equation}
namely, if there are no stable ghosts, and if unstable ghosts come in the form of Merlin modes~\cite{Donoghue:2019ecz,Donoghue:2019fcb}.
These conditions guarantee that the spectral density is positive. Moreover, the spectral density must be real and the normalization condition
\begin{equation}
\int_{0}^{\infty}\rho(\mu^{2})d\mu^{2}=\sum_{n}R_{n}+\int_{0}^{\infty}\sigma(\mu^{2})d\mu^{2}=1
\end{equation}
is to be imposed.

The K\"{a}llen-Lehmann spectral representation of the (dressed) propagator~\cite{Kallen:1952zz,Lehmann:1954xi}, from which Eq.~\eqref{spectralpropyrelation} follows, is very general and does not rely on any perturbative expansion. It encodes all-order, non-perturbative information on the dressed propagator, and thereby on the degrees of freedom of the full theory. The spectral density can be derived from the dressed propagator, which in turn can be computed from the effective action $\Gamma_0$. This makes the quantum effective action $\Gamma_0$ crucial to establish unitarity in non-perturbative QFTs, and a useful tool in the case of QFTs which are perturbative at all scales.

\subsection{Effective actions and the renormalization group}\label{subsect:effactfrg}

The effective action $\Gamma_0$ is obtained by integrating out all quantum fluctuations at the level of the functional integral. One possible way to obtain $\Gamma_0$ is via the FRG equation \cite{Wetterich:1992yh,Morris:1993qb,Reuter:1993kw,Reuter:1996cp}
\begin{equation}\label{frgeq}
k\partial_k \Gamma_k= \frac{1}{2}\mathrm{STr}\left\{(\Gamma_k^{(2)}+\mathcal{R}_k)^{-1}k\partial_k \mathcal{R}_k\right\}\,.
\end{equation}
Here $\Gamma_k$ is the so-called effective average action. At the level of the path integral, it can be thought of as the effective action obtained by the integration of fluctuations with momenta~$p^2\gtrsim k^2$. Its variation with respect to the RG scale $k$ is induced by the variation of the infrared regulator $\mathcal{R}_k$ and by the corresponding RG-scale dependent inverse propagator. The supertrace Str stands for both a sum over discrete indices and an integral over continuum variables, i.e., either momenta or spacetime coordinates. The RG flow of $\Gamma_k$ connects smoothly the UV regime, where $\Gamma_k$ approaches a bare (fixed point) action $S$, and the infrared limit $k\to0$, where all fluctuations are integrated out. In the latter limit, $\Gamma_k$ reduces to the quantum effective action $\Gamma_0$.

The effective average action $\Gamma_k$ contains all invariants according to the underlying symmetries. In the infrared limit $k\to0$, as a result of the integration of fluctuations along all momentum scales, the quantum effective action is expected to be non-local. This is the case even at the one-loop level, where the integration of fluctuating modes give rise to logarithmic corrections to the classical action.

One of the challenges of the FRG approach to QG is the impossibility to solve Eq.~\eqref{frgeq} exactly. In order to obtain an approximate solution to this equation and investigate the UV completion of the theory, one has to resort to ``truncations'' of the theory space: an ansatz for $\Gamma_k$ is chosen, that includes a manageable number of operators. The effective average action $\Gamma_k$ is expanded using, e.g., a derivative expansion, and then this expansion is truncated to a finite truncation order $N$ to make the computation of the beta functions feasible. On the one hand, this method is useful to determine some key features of the RG flow of $\Gamma_k$, such as its fixed points and universality properties. On the other hand, this method does not allow to infer the unitarity of the theory. In fact, if a derivative expansion of $\Gamma_k$ is truncated at a finite order, the corresponding propagator derived in the limit $k\to0$  displays a finite number of unphysical poles~\cite{Platania:2020knd}. As we will explain in detail in the next section, reinforcing the arguments in \cite{Platania:2020knd}, establishing unitarity in QFT requires the knowledge of the full effective action, as truncations of its derivative expansion would lead to the appearance of unphysical poles.

\subsection{Absorptive part of the dressed propagator and unstable particles}

We have argued that the full spectral density is to be computed using the dressed propagator, which in turn can be derived from the effective action~$\Gamma_0$.

Beyond being convenient, in some cases using fully-dressed quantities (aka, resumming all quantum effects) is crucial to count the degrees of freedom of a theory correctly~\cite{Fradkin:1981vx}. Quantum effects can for instance make a particle appearing in the bare theory unstable~\cite{Fradkin:1981vx,Donoghue:2019fcb}, thus removing it from the spectrum of asymptotic states~\cite{Veltman:1963th}. In what follows we review this mechanism using a simple example.

Starting from the bare theory of a single scalar particle with bare mass $m_0$, resumming quantum effects would generally lead to a dressed propagator of the form
\begin{equation}
\Delta(q^{2})=\frac{i}{q^{2}-m_0^{2}+\Sigma(q^{2})}\,.
\end{equation}
Here $\Sigma(q^{2})$ is the self-energy generated by vacuum polarization. The (bare) Feynman propagator has a pole at $q^{2}=m_0^{2}$. Turning interactions on, if $\Sigma(k^{2})$ does not lead to extra
poles other than the one present in the bare theory, quantum effects can renormalize the mass $m_0$ and can also make the particle unstable. This happens when the self-energy $\Sigma(q^2)$ has an imaginary part, as it gives the particle a non-zero decay width. The renormalized mass is defined by the condition $m^{2}-m_0^{2}+\mathrm{Re}\{\Sigma(m^{2})\}=0$, while the imaginary part of~$\Sigma(q^2)$ is related to the total decay rate of the particle. Let us assume that $\Sigma(q^{2})$ does not add extra
poles. In this case, the sum over all cuts on the right-hand side of the optical theorem reads
\begin{equation}
\begin{aligned}\label{eq:absprop1}
 & 2\mathrm{Im}\left[i\Delta(q^{2}+i\epsilon)\theta(q_{0})\right]= \\
 & = \frac{i\theta(q_{0})}{q^{2}-m_0^{2}+\mathrm{Re}\{\Sigma\}+i\mathrm{Im}\{\Sigma\}+i\epsilon}-\frac{i\theta(q_{0})}{q^{2}-m_0^{2}+\mathrm{Re}\{\Sigma\}-i\mathrm{Im}\{\Sigma\}-i\epsilon} \\
 & = \frac{(+i)\left(2\epsilon+2\mathrm{Im}\{\Sigma\}\right)\theta(q_{0})(-i)}{\left(q^{2}-m_0^{2}+\mathrm{Re}\{\Sigma\}+i\mathrm{Im}\{\Sigma\}+i\epsilon\right)\left(q^{2}-m_0^{2}+\mathrm{Re}\{\Sigma\}-i\mathrm{Im}\{\Sigma\}-i\epsilon\right)} \\
 & =\frac{2\epsilon\,\theta(q_{0})}{\left(q^{2}-m_0^{2}+\mathrm{Re}\{\Sigma\}\right)^{2}+\left(\mathrm{Im}\{\Sigma\}+\epsilon\right)^{2}}+ 2\left\{ \Delta(q^{2}+i\epsilon)\,\mathrm{Im}\{\Sigma\}\,\Delta^{*}(q^{2}+i\epsilon)\right\} \theta(q_{0})\,,
\end{aligned}
\end{equation}
where we have omitted the dependence of $\Sigma(q^2)$ on $q^2$ for shortness. The sum above defines the absorptive part of the propagator and, for a theory with one single pole, the result depends on whether $\mathrm{Im}\{\Sigma(q^2)\}$ vanishes or not. In particular
\begin{align}\label{eq:absprop2}
\lim_{\epsilon\to0}2\mathrm{Im}\left[i\Delta(q^{2}+i\epsilon)\theta(q_{0})\right] & =\begin{cases}
2\pi R\,\delta\left(q^{2}-m^{2}\right)\theta(q_{0}) & \mathrm{if}\,\,\,\mathrm{Im}\{\Sigma(q^2)\}=0\\
2\left\{ \Delta(q^{2})\,\mathrm{Im}\{\Sigma(q^{2})\}\,\Delta^{*}(q^{2})\right\} \theta(q_{0}) & \mathrm{if}\,\,\,\mathrm{Im}\{\Sigma(q^2)\}\neq0
\end{cases}\,,
\end{align}
$R$ being the residue at the corresponding pole. In the case $\mathrm{Im}\{\Sigma(q^2)\}=0$, provided that $m^{2}>0$, the absorptive part of the propagator corresponds to an intermediate (stable) one-particle state with mass $m$. 
In the case $\mathrm{Im}\{\Sigma(q^2)\}\neq0$, quantum effects have made the particle of mass $m$ unstable, so that the first term in the absorptive part of the propagator
vanishes. The spectral density, which is proportional to the absorptive part of the propagator,  thus depends crucially on the form of the self-energy $\Sigma(q^2)$.

Since unstable particles are not part of the space of asymptotic states~\cite{Veltman:1963th}, they do not contribute to the sum in Eq.~\eqref{eq:optth}. Consequently, determining which degrees of freedom enter the sum over intermediate states in the optical theorem requires resumming quantum fluctuations first. Note that perturbative expansions break down in the branch cut regions, so that describing unstable particles requires all-order, non-perturbative effects. This motivates further the use of dressed quantities to assess the unitarity of QFTs.

\subsection{Discussion and assumptions}\label{subsect:assumpunitarity}

Let us end this section with some words of caution.

The arguments reviewed in this section are well understood in the context of QFTs of matter, but generally get more involved in the context of gauge theories. In particular, it is not yet clear whether and under which conditions the spectral density of a gauge field has to be positive. For instance, QCD is expected to be unitary, but the spectral density arising from the gluon propagator is not positive definite~\cite{Hayashi:2018giz,Kondo:2019ywt}. Similarly, in gravity the conformal mode is a ghost, at least classically. These examples highlight that the question of unitarity and its relation with the positivity of the spectral density are crucially intertwined with the spectrum of asymptotic states: the only dangerous ``minus signs'' that can enter the optical theorem are those associated with negative-norm asymptotic states---those contributing to the sum in Eq.~\eqref{eq:optth}. Going back to the question of QCD, due to confinement the gluon ought not to be part of the Fock space, and this could make the negativity of the gluon spectral density irrelevant for unitarity in QCD. Likewise, for QG the conformal mode is not an on-shell mode.

On top of this, in the context of QG one might even fail to find a suitable definition of unitarity~\cite{Platania:2020knd}. For instance, if gravity is interacting in the UV, the standard concept of an asymptotic state would cease to make sense. In addition, the definition of asymptotic states requires the knowledge of the (true) vacuum of the theory and determining it is expected to be highly non-trivial~\cite{Bonanno:2013dja}. Finally, if fluctuations of the spacetime signature can take place, unitarity itself would become meaningless. 

Addressing these issues goes well beyond the scope of this work, and we will refrain from discussing them again in the remainder of this paper. 
In the following we will limit ourselves to a discussion of the constraints on the transverse-traceless (TT) part of the graviton propagator, assuming that its poles define the asymptotic states of the theory.\\

\noindent\fbox{\parbox{\linewidth}{ Based on the considerations of this section, \textit{(non-perturbative) unitarity is better studied at the level of the effective action}. To preserve unitarity \textit{the dressed (graviton) propagator should not have real poles with negative residue.}}}

\section{Fictitious ghosts in ``truncated'' field theories and the residue conjecture}\label{sect:polology}

In this section we show that truncating the derivative expansion of an effective action to a finite  order generally produces fake ghosts, i.e., ghosts that are  artifacts of the approximations and that do not appear in the full theory. The fake ghosts decouple dynamically for sufficiently large truncations, as their residues vanish.

As a toy model for the effective action, we shall consider a non-local effective action interpolating between QED and Lee-Wick QED. The reason is twofold: first, unitarity in QED is better under control than in QG, where even the precise definition of unitarity is unclear. Second, as we shall see explicitly in Sect.~\ref{sect:complexpoles}, the propagators of Lee-Wick QED and one-loop QG share very similar properties (cf. Fig.~\ref{fig:polesleewickqedandqg}).

\subsection{A toy model for QED and Lee-Wick QED}

Starting from the classical QED action 
\begin{equation}
	S_{c}=-\frac{1}{4}\int d^{4}xF_{\mu\nu}F^{\mu\nu}\,,
\end{equation}
and integrating out one or more massive degrees of freedom,
one typically obtains a non-local effective action whose quadratic part reads
\begin{equation}
	\Gamma_{QED}[A_{\mu}]=-\frac{1}{4}\int d^{4}x\left\{ F_{\mu\nu}P(\square)F^{\mu\nu}\right\} \,,\label{eq:toyQED}
\end{equation}
with $\square\equiv \partial_\mu \partial^\mu$.
The latter has to be complemented with a gauge fixing term,
\begin{equation}
	S_{\text{gf}}=-\frac{1}{2\xi}\int d^{4}x\left\{ \partial_{\mu}A^{\mu}Q(\square)\partial_{\nu}A^{\nu}\right\} \,,
\end{equation}
and the resulting propagator on Minkowski spacetime reads
\begin{equation}
	\bold{\Delta}_{\alpha\beta}(q^{2})=-\frac{i}{q^{2}P(q^{2})}\left\{ \eta_{\alpha\beta}-\left(1-\xi\frac{P(q^{2})}{Q(q^{2})}\right)\frac{q_{\alpha}q_{\beta}}{q^{2}}\right\} \,. \label{eq:propagator}
\end{equation}
In what follows we shall fix $\xi=0$. 
Following \cite{LandauD,Donoghue:2015nba,Donoghue:2015xla}, in the one-loop approximation, the function $P(q^2)$ for QED can be obtained by integrating out a scalar or fermionic degree of freedom, and it reads \cite{LandauD}
\begin{equation}
	P(q^2)=1-\frac{\alpha}{3\pi}\log\left(\frac{-q^2+4m^{2}}{4m^{2}}\right)\,,
\end{equation}
where $\alpha$ is the fine structure constant and $m$ is the mass of the degree of freedom which has been integrated out. 
In the case of Lee-Wick QED, the form of the function $P(q^2)$ is instead \cite{Boulware:1983vw,Donoghue:2018lmc}
\begin{align}
	P(q^{2}) 
	=1-\frac{\alpha}{3\pi}\mathrm{log}\left(\frac{m_{th}^{2}-q^{2}}{m_{th}^{2}}\right)-\frac{q^{2}}{M^{2}}\,,
\end{align}
where $m_{th}^{2}=2m^{2}$ corresponds to the threshold of production of a fermion-antifermion pair (or a pair of scalars) and $M$ is a cutoff scale. The momentum-space propagator can be written as\footnote{A detailed explanation of the origin of this equation and its regime of validity are reported in Sect.~\ref{sect:complexpoles}, where we study logarithmic interactions of the Lee-Wick and one-loop quantum-gravity type, cf.~Eq.~\eqref{eq:logdec}.}
\begin{equation}\label{eq:logwiththeta}
	{\bold{\Delta}}_{\alpha\beta}(q^{2})=-\frac{i}{q^{2}-\frac{\alpha}{3\pi}q^{2}\log\left|\frac{-q^{2}+m_{th}^{2}}{m_{th}^{2}}\right|-\frac{q^{4}}{M^{2}}+q^{2}\frac{i\alpha}{3}\theta(q^{2}-m_{th}^{2})}\left\{ \eta_{\alpha\beta}-\frac{q_{\alpha}q_{\beta}}{q^{2}}\right\} \,.
\end{equation}
The absorptive part of the scalar part of the propagator thus reads
\begin{equation}
	2\mathrm{Im}\left[i\Delta(q^{2})\theta(q_{0})\right]=2\pi\,\delta(q^{2})\theta(q_{0})+\Delta(q^2)\left(\frac{2\alpha}{3}q^{2}\,\theta(q^{2}-m_{th}^{2})\right)\Delta^\ast(q^2),
\end{equation}
with
\begin{equation}
	\Delta(q^2)\equiv\frac{-i}{\left(q^{2}+i\epsilon\right)\left(1-\frac{\alpha}{3\pi}\log\left|\frac{-q^{2}+m_{th}^{2}}{m_{th}^{2}}\right|-\frac{q^{2}}{M^{2}}+\frac{i\alpha}{3}\theta(q^{2}-m_{th}^{2})\right)}
\end{equation}
Its structure is similar to that in Eqs.~\eqref{eq:absprop1}-\eqref{eq:absprop2}. The first term corresponds to the standard photon propagator, the second describes the propagation of an unstable ghost, includes fermionic/scalar loops, and contributes when~$q^{2}>m_{th}^{2}$. Diagrammatically
\begin{center}
	\includegraphics[scale=0.25]{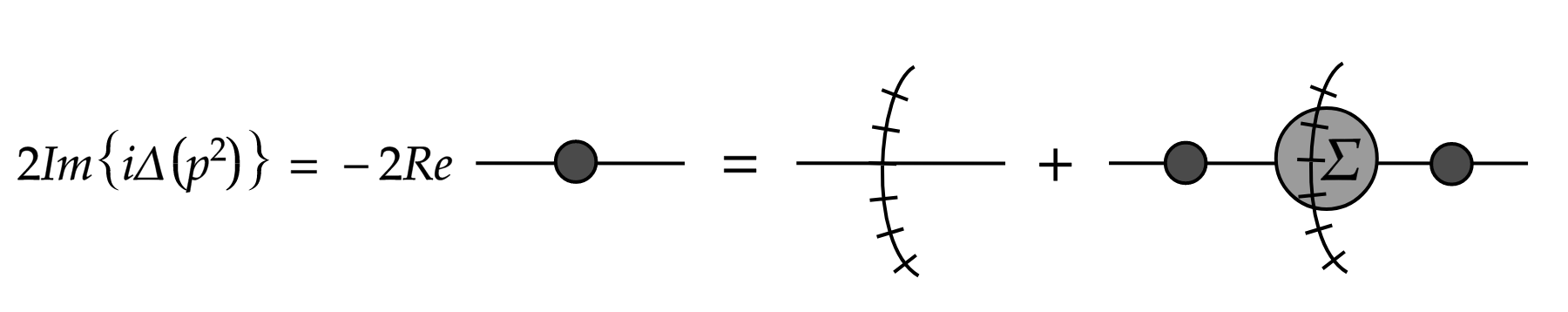}
\end{center}
For a dressed propagator with one single pole, only one of the two terms on the right-hand side is realized. The first term arises in the case of a theory with one single, stable degree of freedom, the second one describes an unstable degree of freedom. We note here that if the dressed propagator is computed from a one-loop effective action, the gray bubble in the right-hand-side is the sum of all possible processes involving one loop of the integrated-out degrees of freedom. For theories with multiple poles both types of terms on the right-hand side could arise.

\begin{figure}
	\begin{centering}
		\includegraphics[scale=0.55]{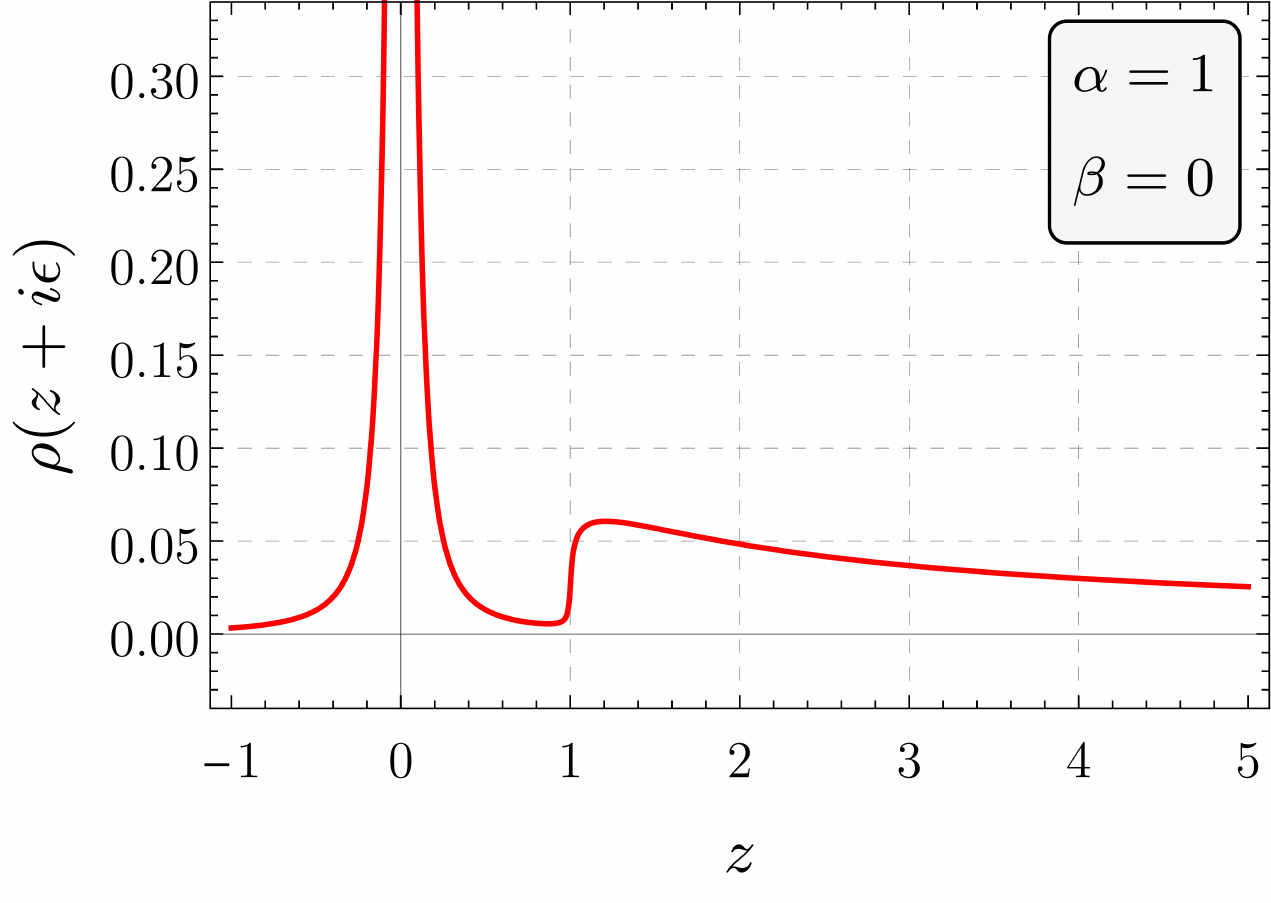}\\
	\end{centering}
	\begin{centering}
		\includegraphics[scale=0.55]{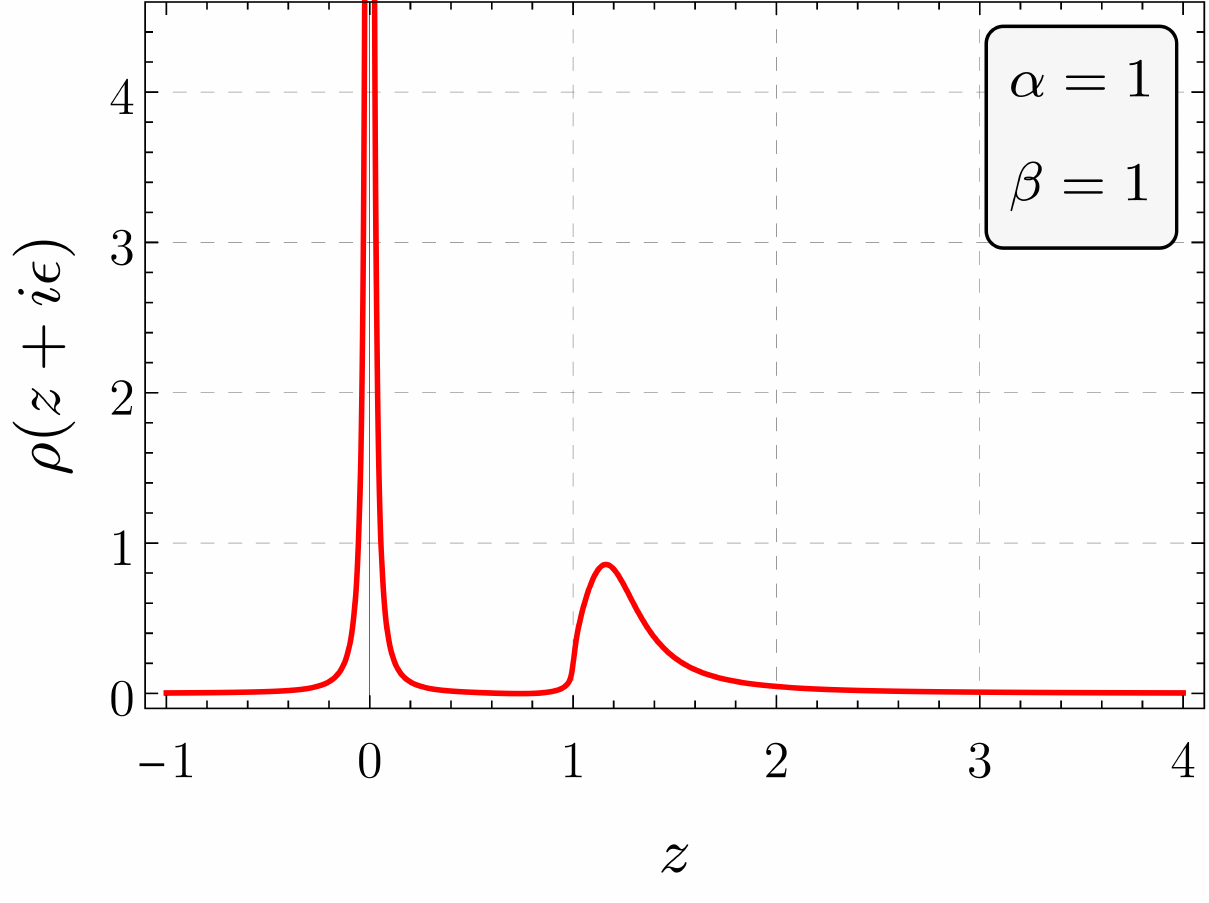}$\quad$ \includegraphics[scale=0.55]{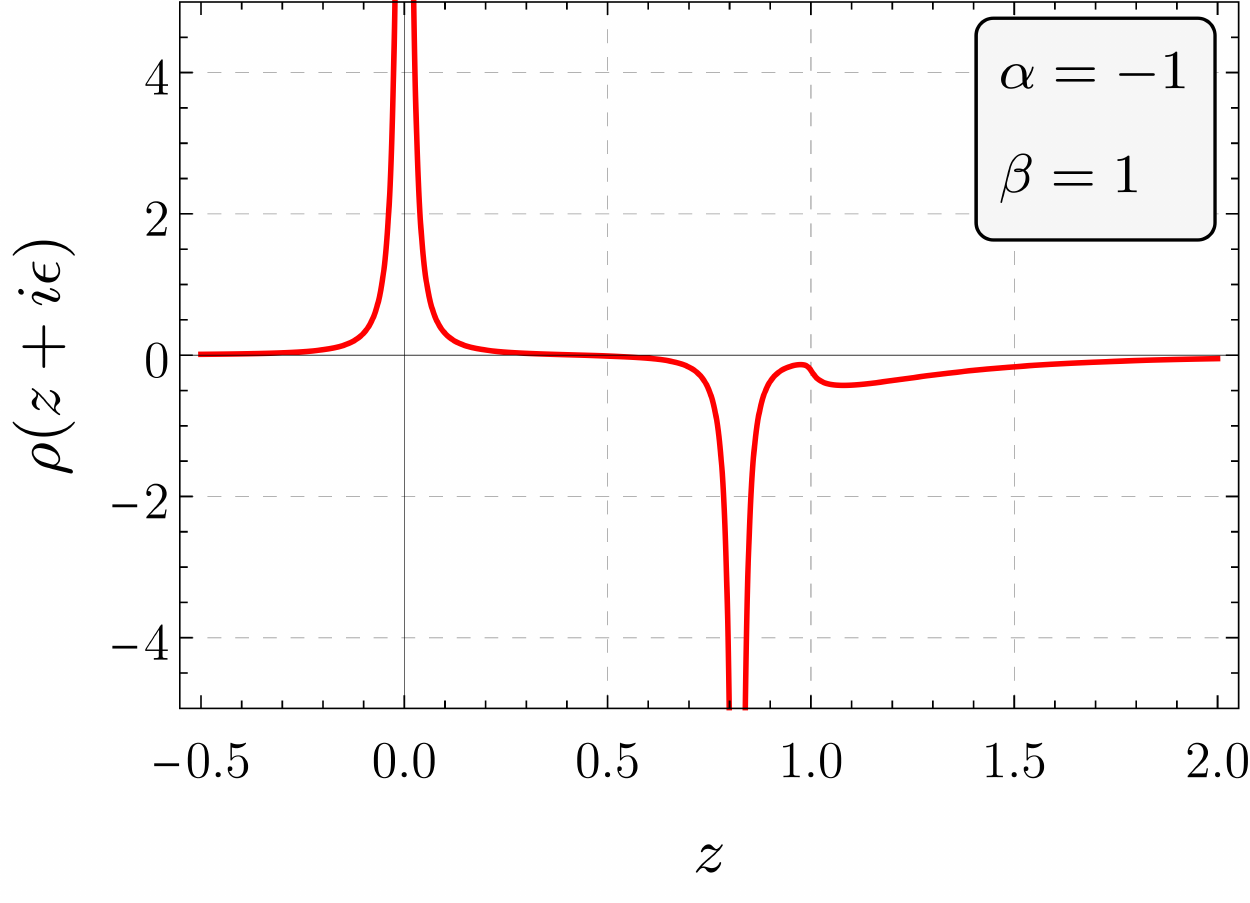}
	\end{centering}
	\caption{Spectral densities for the models summarized in Table~\ref{tab:action}, for $\epsilon=0.01$. In the case of QED (top panel), the spectral density is positive-definite, within the regime of validity of the theory. At the level of the spectral density, the branch cut in the propagator generates a continuum part, corresponding to a multi-particle state. In the case of Lee-Wick QED (bottom-left panel), the branch cut is associated with a ghost resonance, characterized by a negative width. This makes the spectral density positive-definite. If $\alpha<0$ (bottom-right panel), the propagator in Eq.~\eqref{eq:propQED-1} has a pole within its principal branch, corresponding to a stable ghost pole. In this case the spectral density can also be negative and unitarity is violated. \label{fig:rhomodels}}
\end{figure}

In the following we shall employ the generalized propagator
\begin{equation}
	D_{QED}(q^{2})=q^{-2}\left(1-\frac{\alpha}{3\pi}\log\left(-\frac{q^{2}-m_{th}^{2}}{m_{th}^{2}}\right)-\beta\frac{q^{2}}{M^{2}}\right)^{-1}\label{eq:propQED-1}\,\,,
\end{equation}
which reproduces the propagator of one-loop QED for $\beta=0$ and the case of Lee-Wick QED for $\beta=1$ (cf. Tab.~\ref{tab:action}). The one-loop QED propagator has no poles beyond the massless one for small $q^{2}$. The branch cut along $q^2>m_{th}^2$ is related to the threshold of production of lighter particles. However, there is also a \textit{tachyonic ghost pole} at $q^{2}\simeq10^{280}m_{th}^{2}$ (the Landau pole), beyond which the theory breaks down. In the case of Lee-Wick QED, as explained in Sect.~\ref{sect:complexpoles}, there is an unstable ghost pole, whose real part lies in the branch cut region, and a pair of complex-conjugate poles. The pole structure is thus very similar to that of one-loop QG, cf.~Sect.~\ref{sect:complexpoles}. Finally, if $\alpha$ is taken to be negative, the ghost pole is shifted out of the branch cut region, where the logarithmic propagator is real. This makes the ghost stable, and entails a violation of unitarity. The spectral densities corresponding to these three cases are depicted in Fig.~\ref{fig:rhomodels}.
\begin{table}[t!]
	\begin{center}
		\begin{tabular}{|l|l|l|}
			\hline
			 \textbf{Theory}   &         \textbf{Couplings}         &                  \textbf{Real poles}                   \tabularnewline \hline\hline
			One-loop QED           & $\alpha>0$, $\beta=0$ &                         Landau pole \cite{LandauD}               \tabularnewline \hline
			Standard Lee-Wick QED      & $\alpha>0$, $\beta=1$ &       Unstable ghost \cite{Boulware:1983vw,Donoghue:2018lmc} \tabularnewline \hline
			Lee-Wick QED, non-standard sign & $\alpha<0$, $\beta=1$ &                       Stable ghost                      \tabularnewline \hline\hline
		\end{tabular}\caption{The table summarizes the existence and type of (real) massive poles for the generalized QED propagator in Eq.~\eqref{eq:propQED-1}. The values and positivity of the couplings $\alpha$ and $\beta$ determine the specific theory, and the corresponding massive pole.  \label{tab:action}}
	\end{center}
\end{table}

Taking $\eqref{eq:toyQED}$ as a \textit{toy model} for the effective action, with
\begin{equation}
	P(q^{2})=1-\frac{\alpha}{3\pi}\log\left(\frac{-q^{2}+m_{th}^{2}}{m_{th}^{2}}\right)-\beta \frac{q^{2}}{M^{2}}\,\,,\label{eq:Pfunction}
\end{equation}
the propagator in Eq.~$\eqref{eq:propQED-1}$ provides us with a toy model for  the fully dressed photon propagator. Along the lines of~\cite{Platania:2020knd}, in the next subsection we will study the pole structure of this propagator varying the truncation order of the effective action in a derivative expansion.

\subsection{Polology of the expanded one-loop effective action}

\begin{figure}
	\hspace{-0.0cm}\includegraphics[scale=0.39]{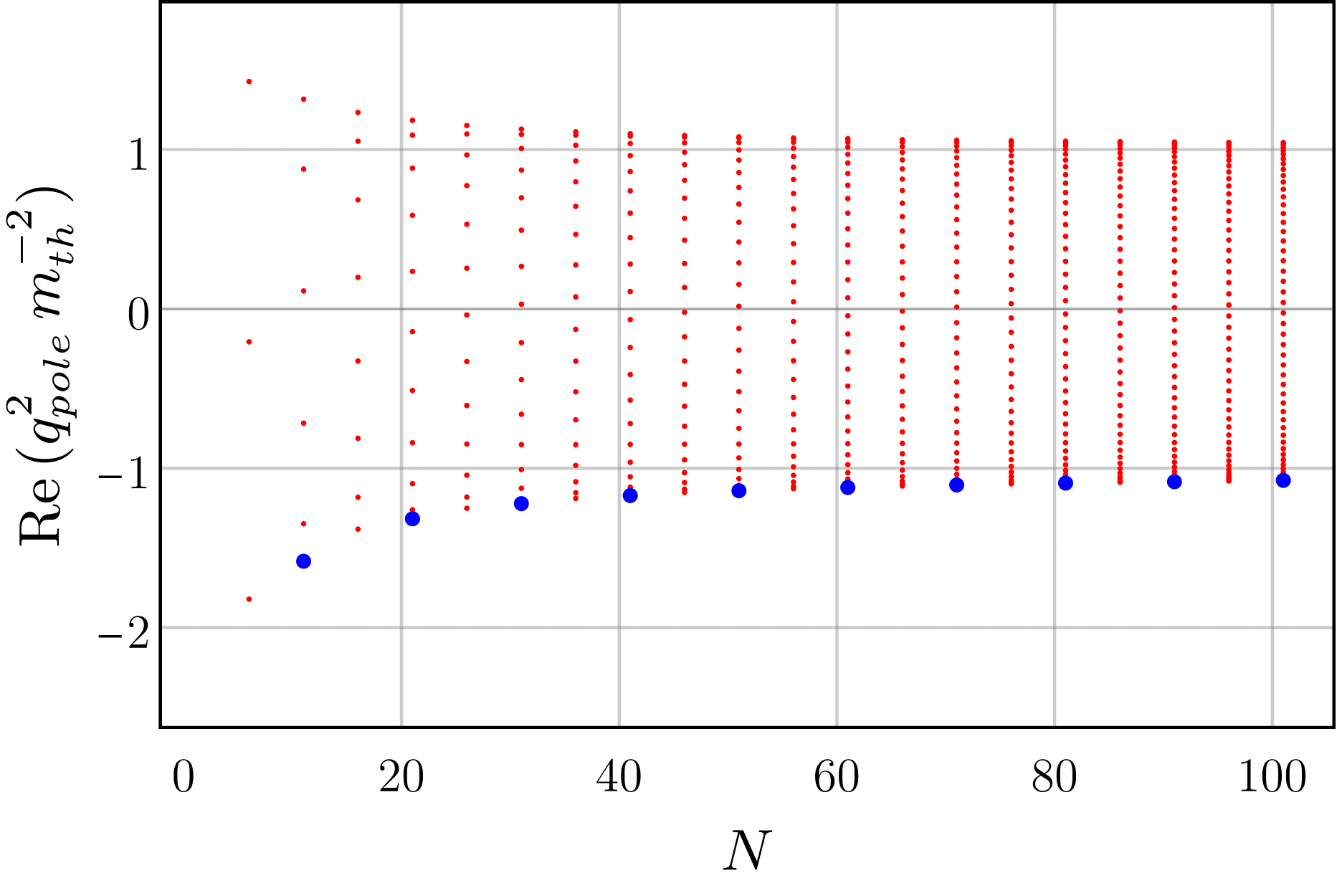}\hspace{0.35cm}\includegraphics[scale=0.39]{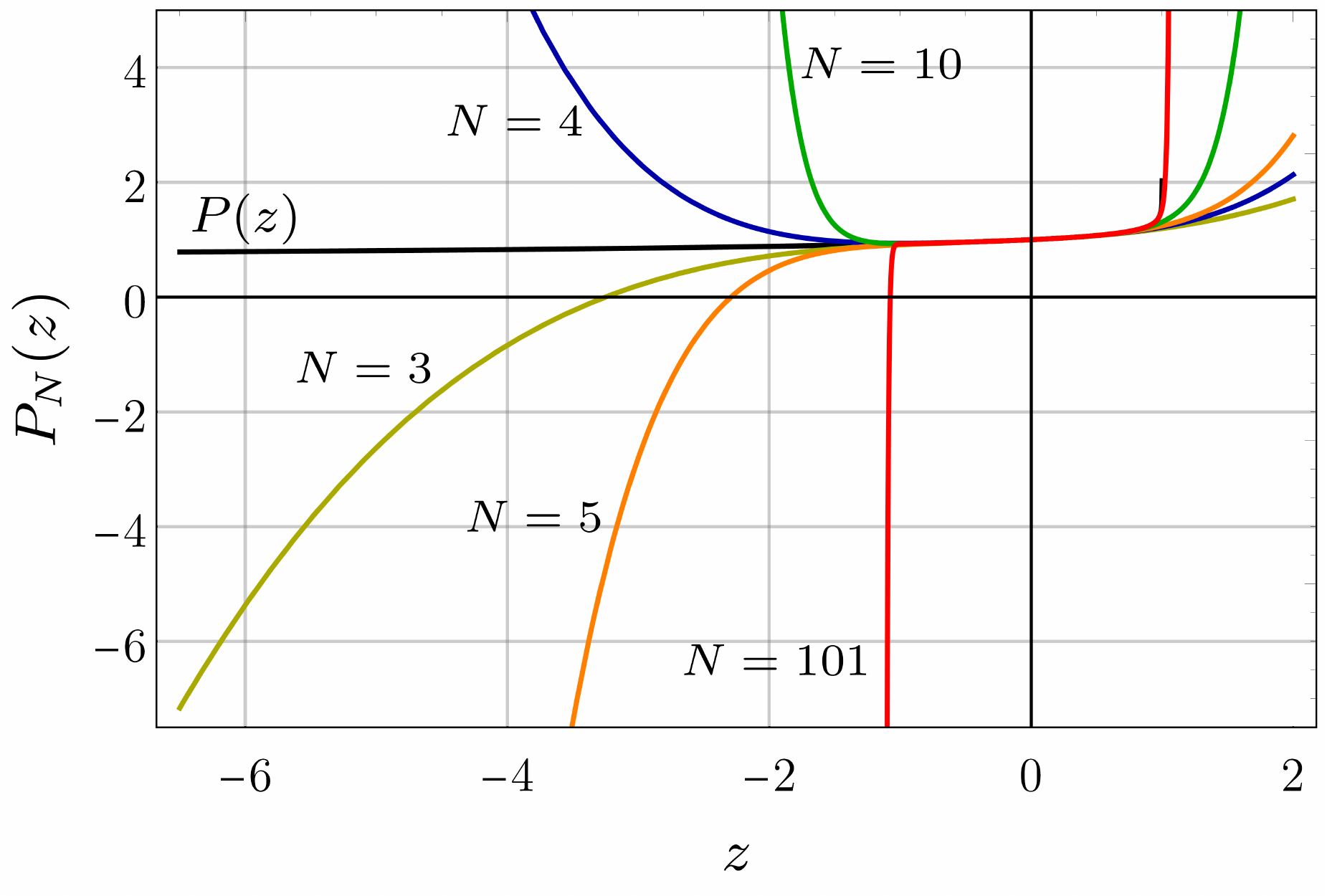}
	\hspace{-0.35cm}\includegraphics[scale=0.39]{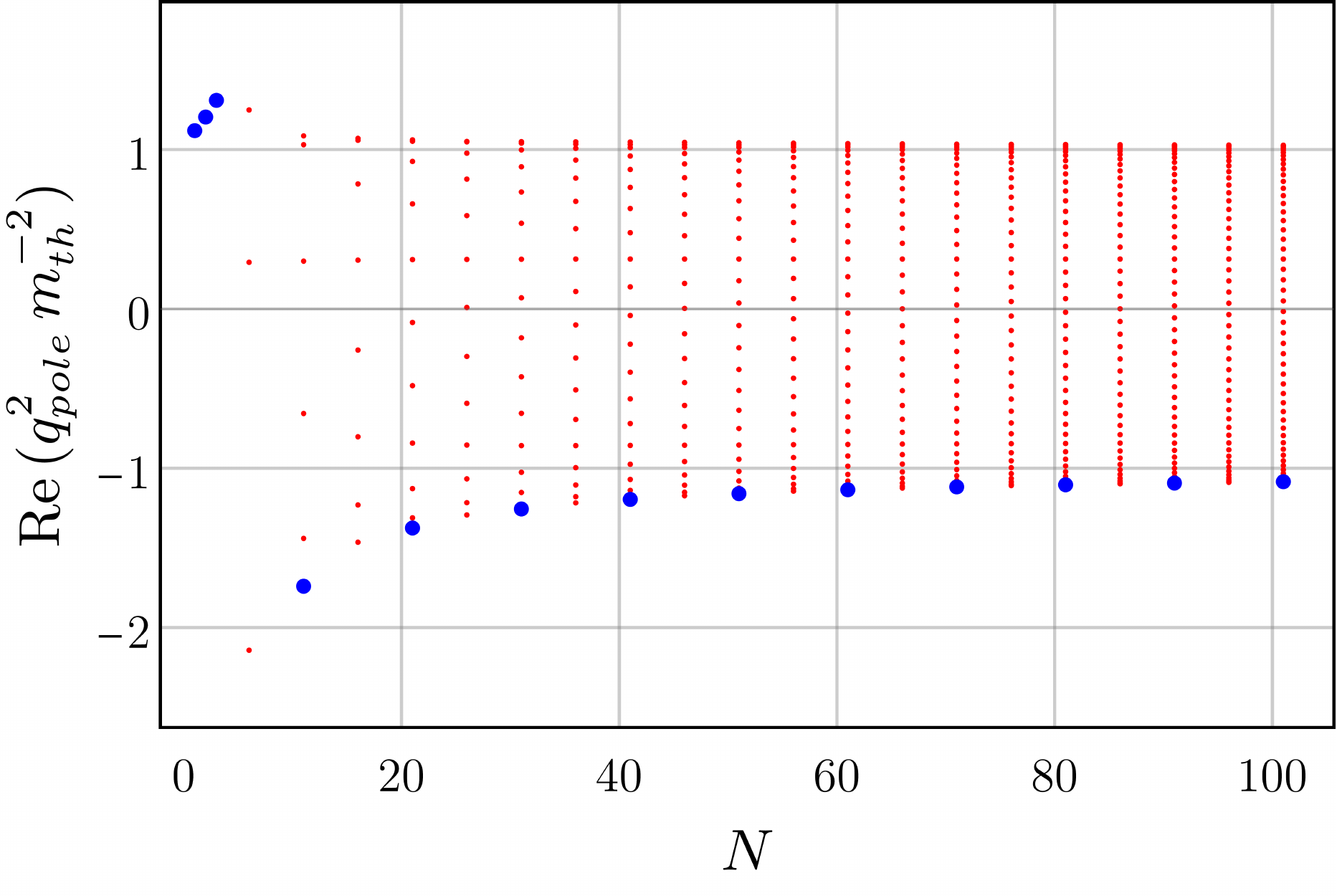}\hspace{0.2cm}\includegraphics[scale=0.54]{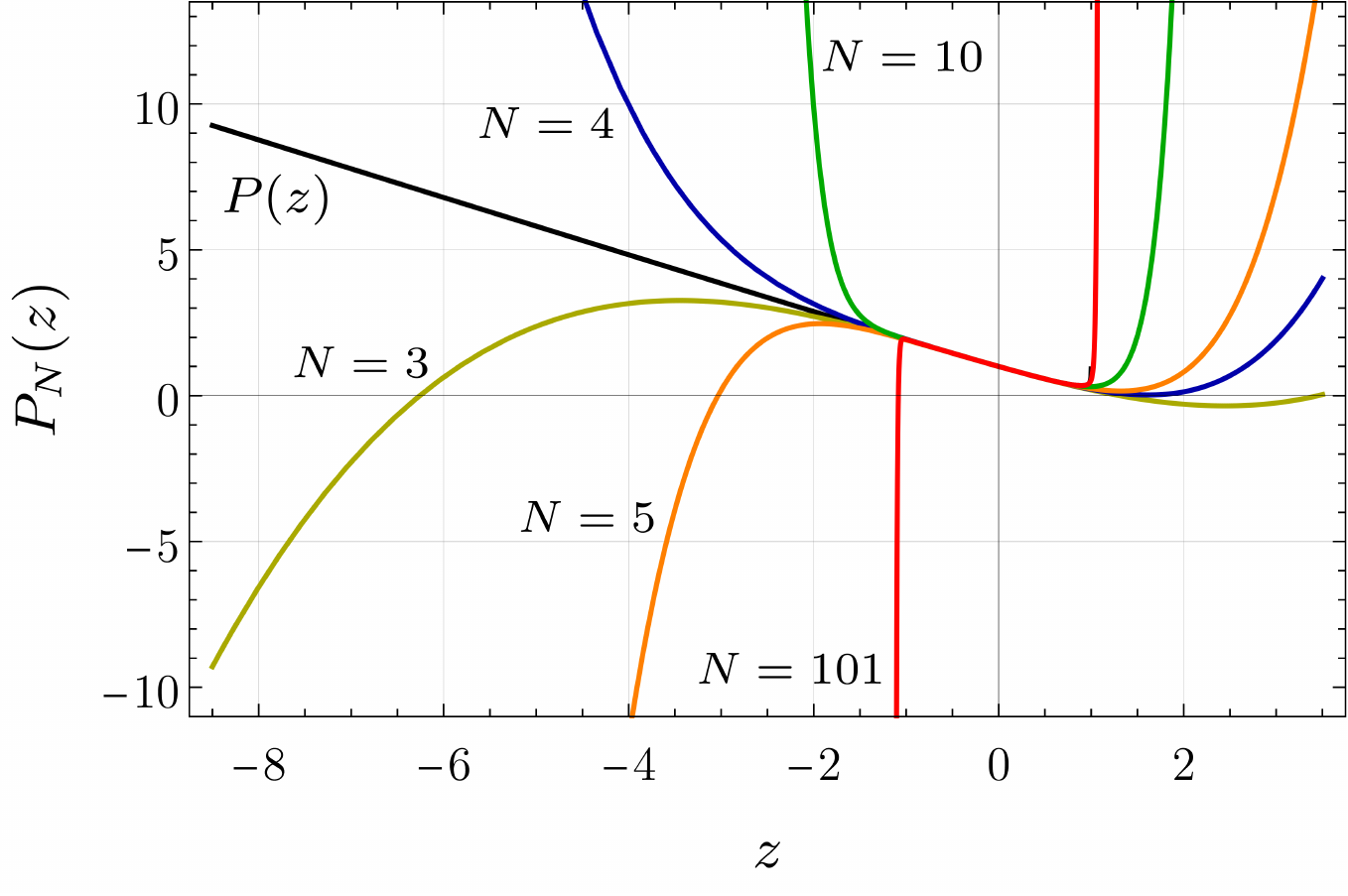}
	\hspace{-0.35cm}\includegraphics[scale=0.39]{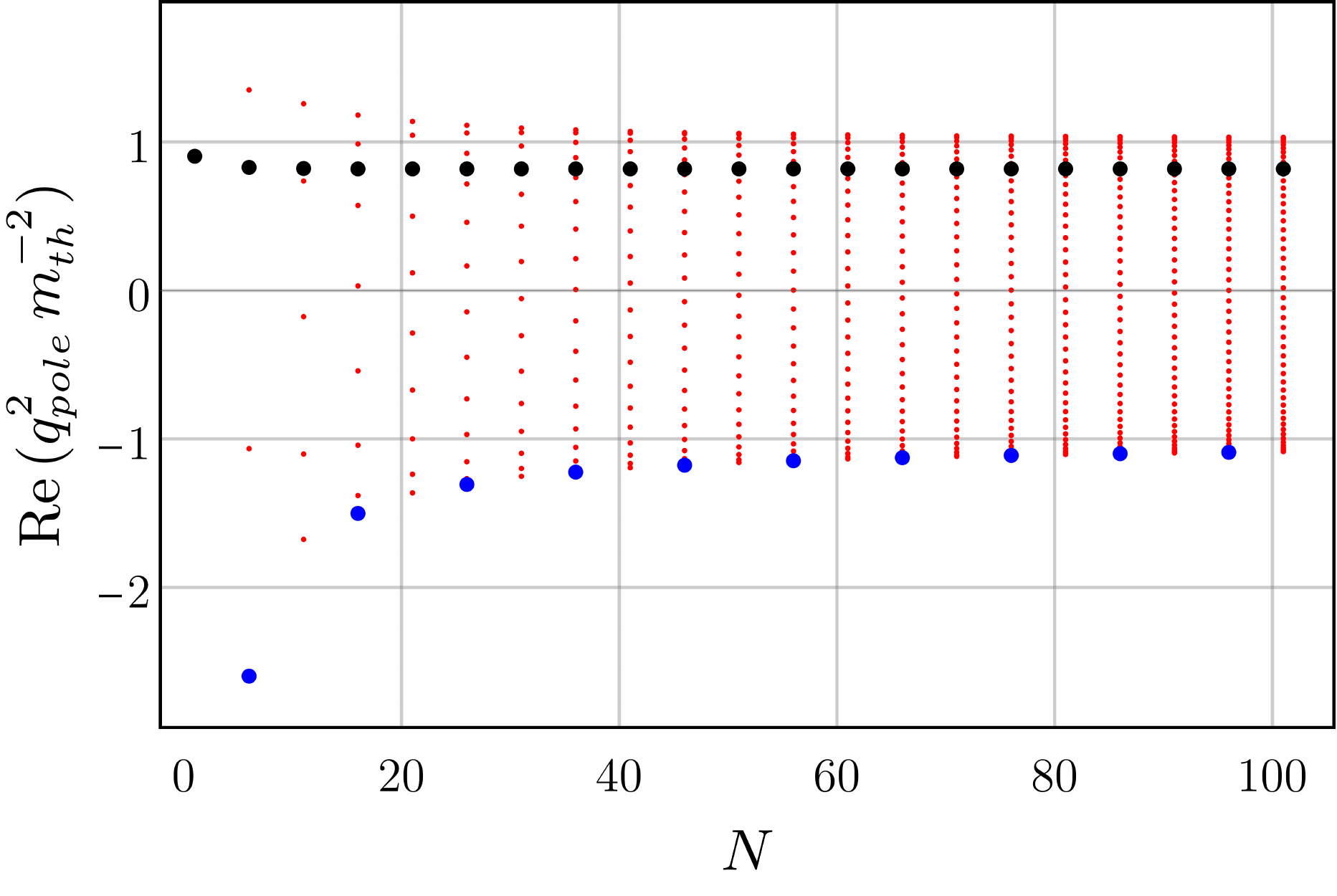}\hspace{0.2cm}\includegraphics[scale=0.39]{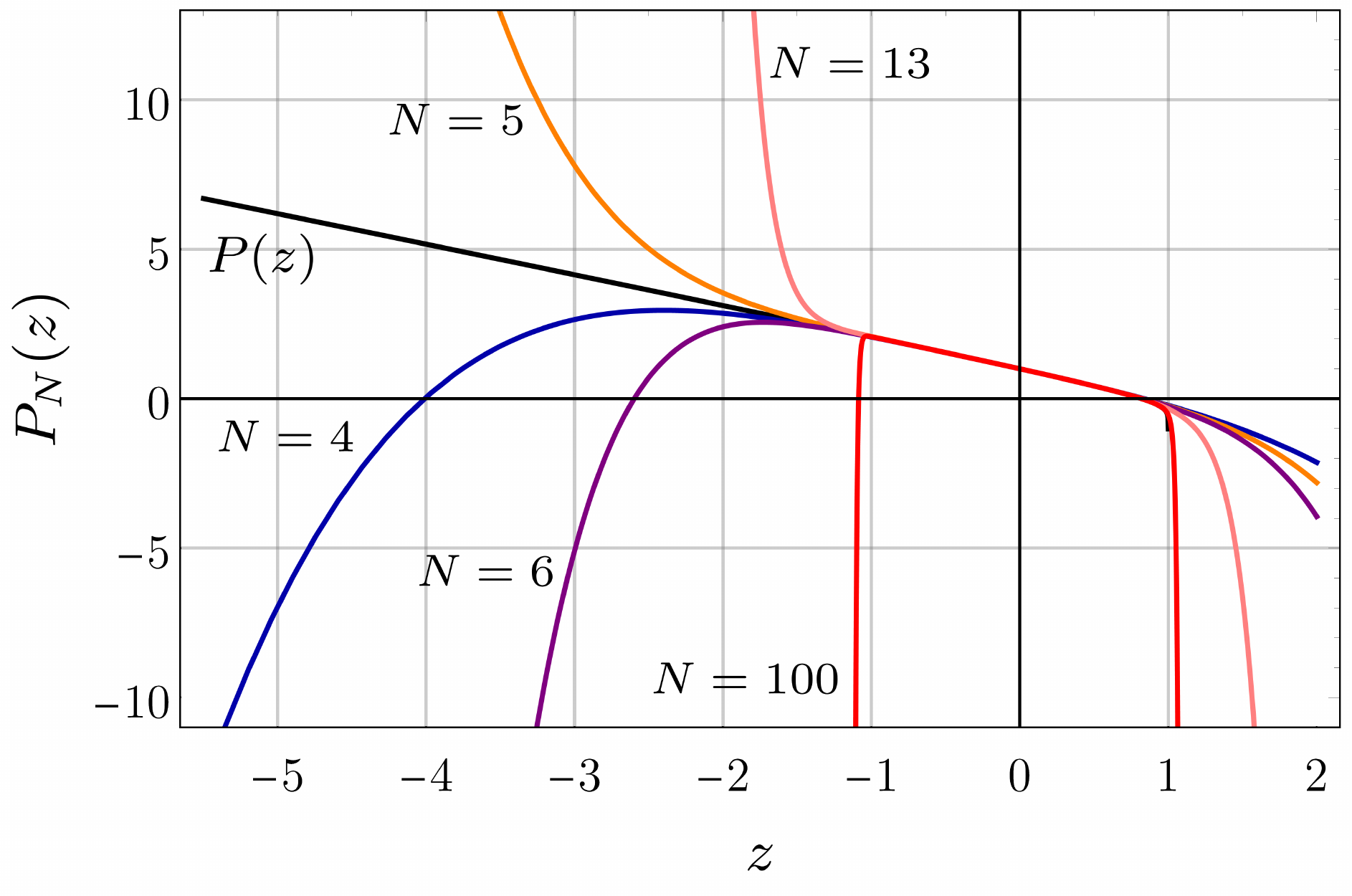}
	\caption{Pole structure of the truncated propagator~\eqref{eq:trunpropagator} as function of the truncation order~$N$ (figures in the left column) and corresponding truncated $P$-functions $P_N(z)$ (figures in the right column) for various $N$ (colorful lines) and in the limiting case $N\to\infty$ (black line). The figures refer to the cases of QED (top panel), Lee-Wick QED (center) and Lee-Wick QED with non-standard coupling (bottom panel). Complex-conjugate poles of the truncated propagator are depicted as small red dots. Blue dots denote real fictitious poles and, in the specific cases analyzed, are generated whenever the corresponding $P_N$-function diverges negatively and crosses the $z$-axis. In the case of Lee-Wick QED with non-standard coupling (bottom panel) there is an additional real pole (black dot) corresponding to the true (non-fictitious) ghost of the theory. The corresponding (untruncated) $P(z)$-function has indeed a zero at $z>0$ in this case. \label{fig:polesanddiv}}
\end{figure}

Truncating the derivative expansion of the effective action~\eqref{eq:toyQED}, one obtains a ``truncated'' inverse propagator of the form~\eqref{eq:propagator}, with  
\begin{equation}
	P(z)\to P_{N}(z)=1-\beta z+\frac{\alpha}{3\pi}\sum_{n=1}^{N}\frac{z^{n}}{n}, \label{eq:expansion}
\end{equation}
$z$ being defined as $z\equiv q^{2}/m_{th}^{2}$. Such an expansion of the effective action cannot reproduce all features of the ``full theory''~\eqref{eq:toyQED} since some physical effects, including the possibility of making a particle unstable, rely on non-perturbative features of the theory. In particular, although the function $P(q^2)$ in Eq.~\eqref{eq:Pfunction} has at most one real massive zero (cf. Tab.~\ref{tab:action}), its truncated version $P_N(z)$ can have several. The scalar part of the truncated propagator,
\begin{equation}\label{eq:trunpropagator}
	i D_N(z)=\frac{i}{z \, P_N(z)}\,,
\end{equation}
thus generally has a number of real and complex-conjugate poles (plots in the left column of Fig.~\ref{fig:polesanddiv}). These poles are generated by the convergence properties of the derivative expansion of~$P(z)$ (plots in the right column of Fig.~\ref{fig:polesanddiv}). This can be seen by studying the sequence of approximated functions~$P_N(z)$ for increasing values of $N$. For large and negative~$z$, the behavior of $P_N(z)$ is dominated by the term $c_Nz^N$. Depending on the sign of the coefficient~$c_N$ and on whether $N$ is even or odd, $P_N$ can either diverge negatively or positively as $z\to\infty$. Specifically, due to the structure of the function~$P(z)$, in the region $z<0$ the function~$P_N$ alternates between positive and negative signature divergences, with a certain periodicity in $N$ (see plots in the right column of Fig.~\ref{fig:polesanddiv}). The convergence properties of $P(z)$ can thus lead to accidental zeros for $z<0$, and the corresponding truncated propagator $D_N$ develops a number of fictitious complex-conjugate and real poles. The disappearance of these fictitious zeros of $P_N(z)$ can only be achieved in the limit $N\to\infty$.
Since the function~$P_N(z)$ truncated to order $N$ is polynomial, and
\begin{equation}
	\frac{1}{(q^{2}-m_{0}^{2})(q^{2}-m_{1}^{2})\dots(q^{2}-m_{N}^{2})}=\sum_{i=1}^{N}\left(\prod_{j=1}^{N}\frac{1}{m_{i}^{2}-m_{j}^{2}}\right)\frac{1}{q^{2}-m_{i}^{2}}\,\,,
\end{equation}
some of the aforementioned fictitious degrees of freedom \emph{must} be ghosts. Specifically, in the cases at hand, the truncated derivative expansion generates several complex-conjugate poles and one single fake (tachyonic) ghost. As is apparent from Fig.~\ref{fig:polesanddiv} (left column), the corresponding zero is generated at large negative $q^2$ for low-order truncations, but moves towards $\mathrm{Re}(z)=-1$ for increasing values of $N$. In other words, it seems that fictitious zeros move towards and accumulate on the boundary of the domain of convergence of~$P(q^2)$. It is worth noticing that if the resummation of quantum fluctuations yields a non-local effective action with $P(z)$ given by an entire function, the radius of convergence is infinity and fictitious ghost poles are expected  to approach infinity for increasing values of~$N$. We will come back to this point in the next subsection.

\begin{figure}
	\centering\includegraphics[scale=0.23]{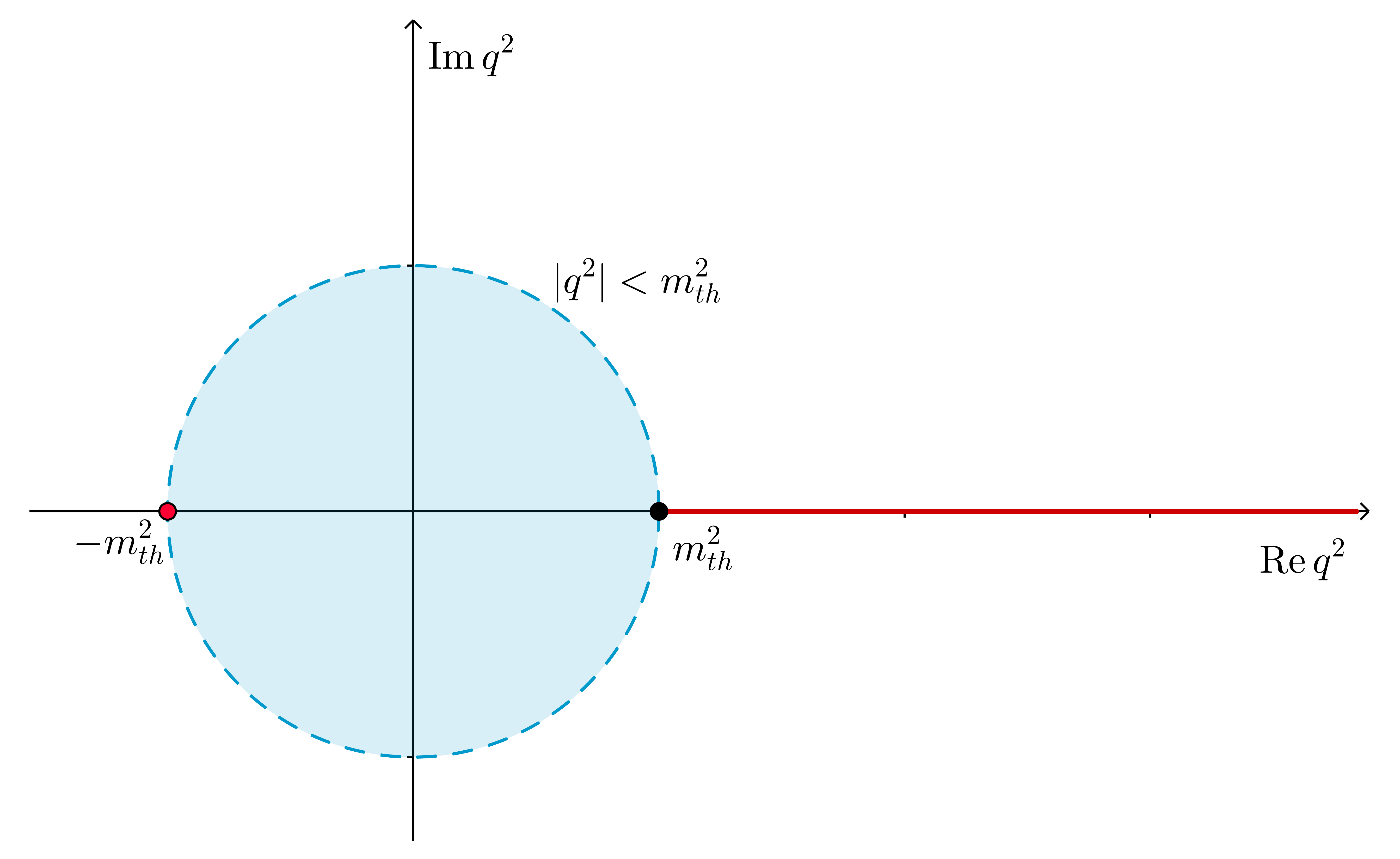}
	\caption{Branch cut (red line) and domain of convergence (blue region) of the truncated propagator~\eqref{eq:trunpropagator}. Unstable ghosts of the full theory can only live in the branch cut region and cannot be detected within truncated derivative expansions of the effective action. Genuine and fake (stable) ghosts live in the principal branch of the logarithm, i.e., in the region $\mathrm{Re}\, q^2<m_{th}^2$. The former stay inside the domain of convergence of the logarithm and are visible within truncations. The latter can only be generated outside of the domain of convergence, are only present within truncations, and they move towards and accumulate on the boundary of the domain of convergence of the form factor characterizing the full theory. \label{fig:domain}}
\end{figure}

\begin{figure}
	\hspace{-0.35cm}\includegraphics[scale=0.55]{Images/DivNegPos2}\hfill\includegraphics[scale=0.55]{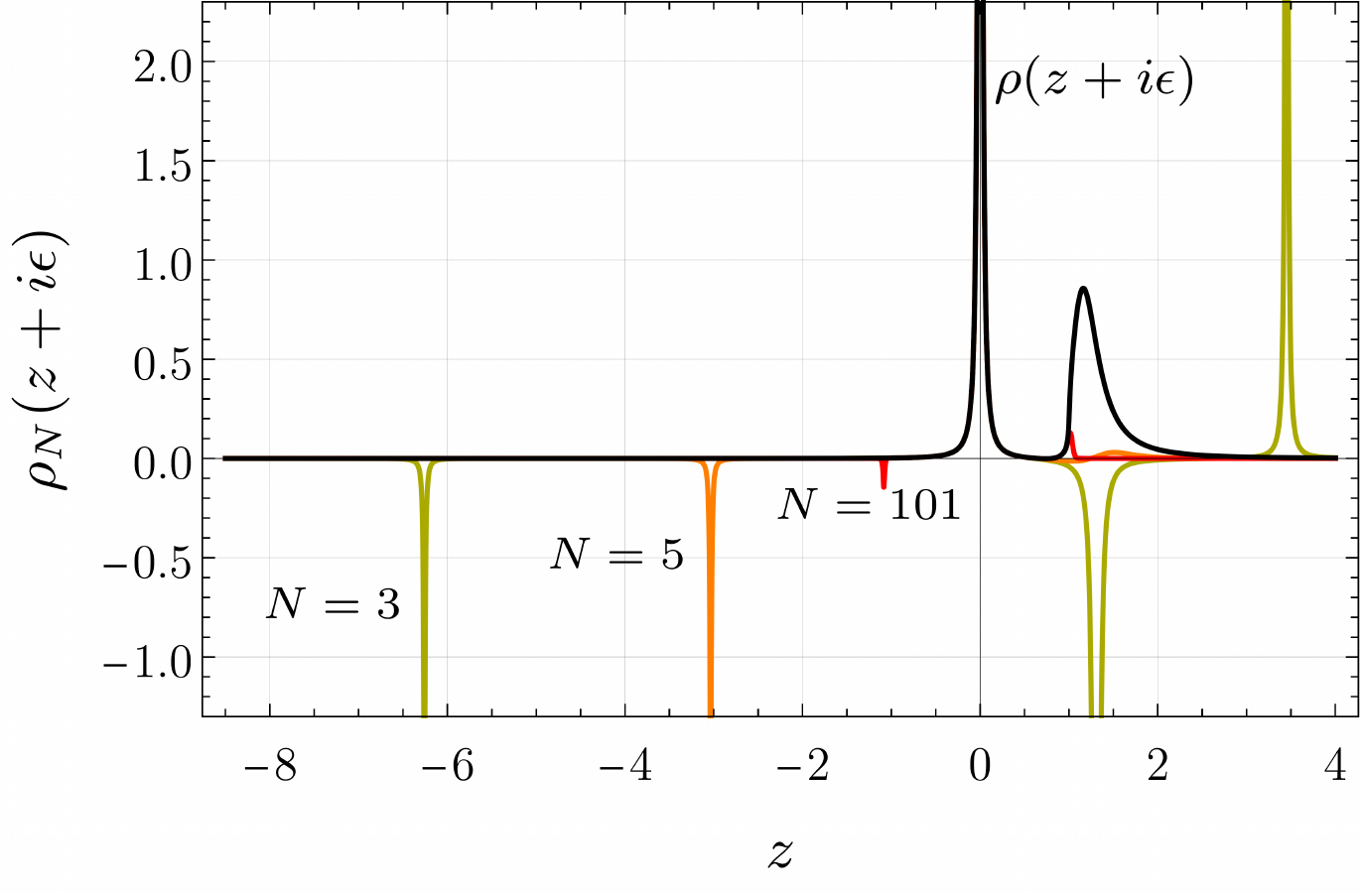}
	\caption{Although Lee-Wick QED is unitary, and thus characterized by a positive spectral density~$\rho(z+i\epsilon)$, its truncated versions have several unphysical ghosts and its spectral density is not positive. The figure on the right panel (obtained by setting $\epsilon=0.01$) shows the spectral density $\rho(z)$ induced by the function $P(z)$ (black line) and by its truncated versions $P_N(z)$ (colorful lines). In the central peak zone, corresponding to the massless pole, all colorful lines overlap, and are hidden behind the black line. The shape of the $P_N(z)$-functions are shown for comparison in the left panel. While the complete spectral function is positive and characterized by a continuum part, its truncated versions $\rho_N(z)$ display several negative-diverging Dirac deltas associated with the fictitious zeros of $P_N(z)$, i.e., with the fake poles of the truncated propagator. Moreover, the truncated spectral function cannot describe the continuum part of $P(z)$, since the resonance necessarily arises as an all-order effect. \label{fig:truncspectral}}
\end{figure}

The domain of convergence and branch cut region of $P(z)$ are shown in Fig.~\ref{fig:domain}.
Fake degrees of freedom live outside the domain of convergence of the $P$-function, but appear in the principal branch of the logarithmic interaction, and approach its boundary for increasing values of $N$. Possible unstable ghosts in the full theory can only live in the branch cut region. The latter region cannot overlap with the domain of converge of the full $P$-function by construction. Thus, unstable ghosts (or particles) cannot be captured using the truncated  expansion~\eqref{eq:expansion} of the inverse propagator (see Fig.~\ref{fig:truncspectral}).
More precisely, the radius of convergence is finite due to the breakdown of perturbation theory near the mass of an unstable particle \cite{Veltman:1963th}. Indeed, this is because the truncated propagator is a finite sum of single-pole propagators, and the corresponding spectral density is thereby a sum of Dirac deltas only. Consequently, the spectral function computed from a truncated version of the theory (cf.~Fig.~\ref{fig:truncspectral}), beyond not being positive due to the presence of fake ghosts, does not have the continuum part associated with the resonance. 

\begin{figure}
	\begin{center}
		\includegraphics[scale=0.4]{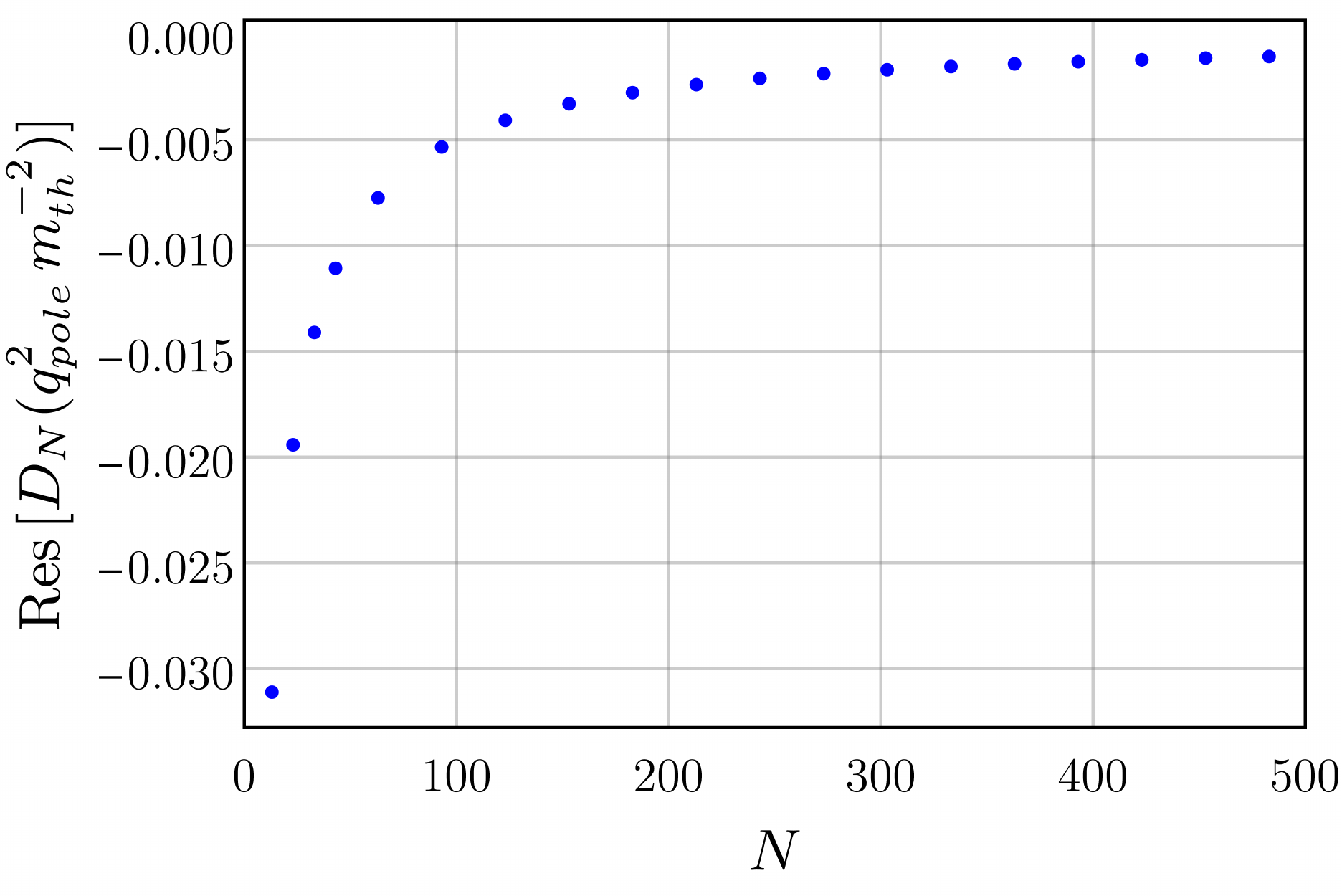}
	\end{center}
	\hspace{-0.3cm}\includegraphics[scale=0.4]{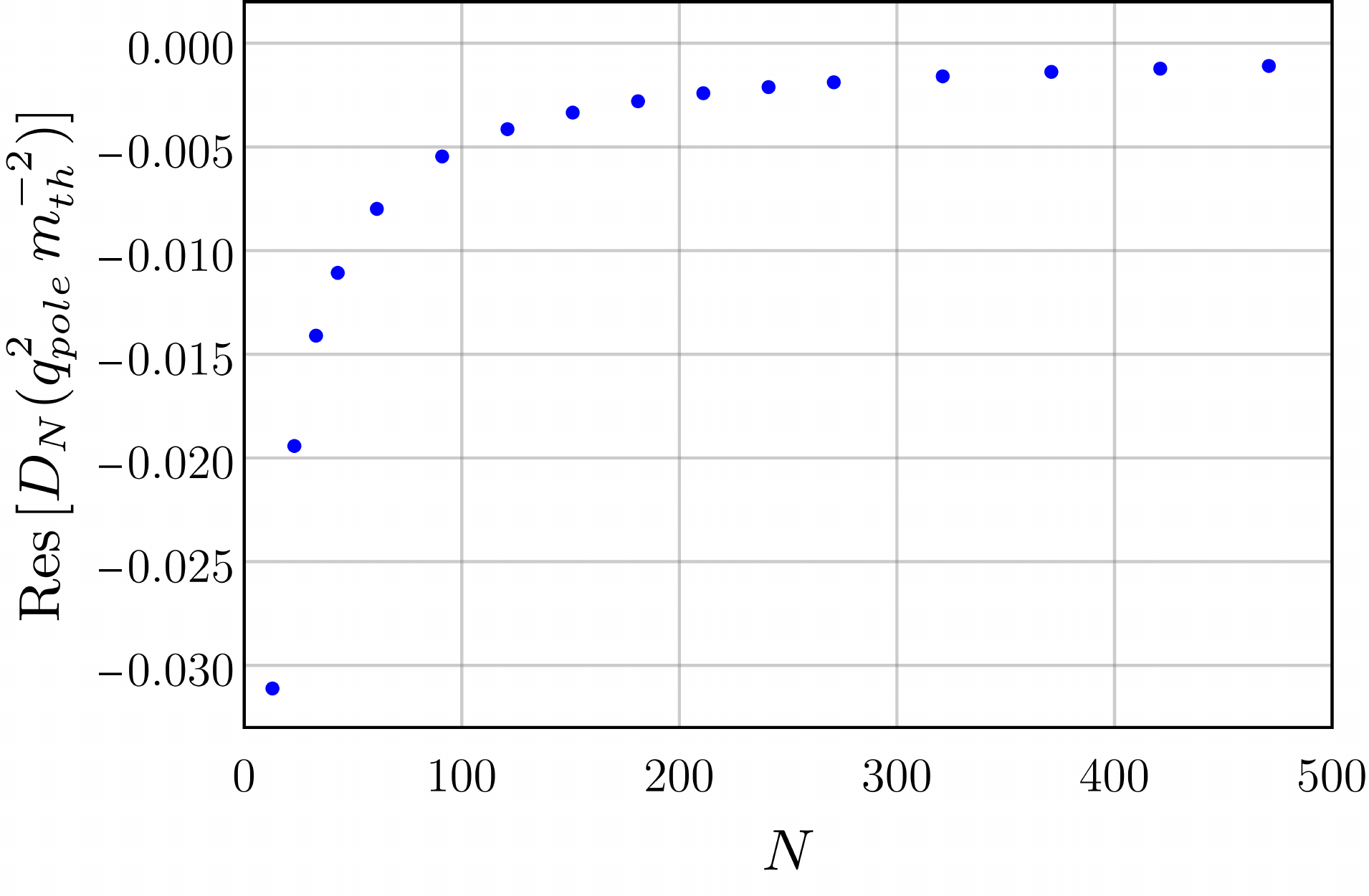}\hspace{0.1cm}
	\includegraphics[scale=0.4]{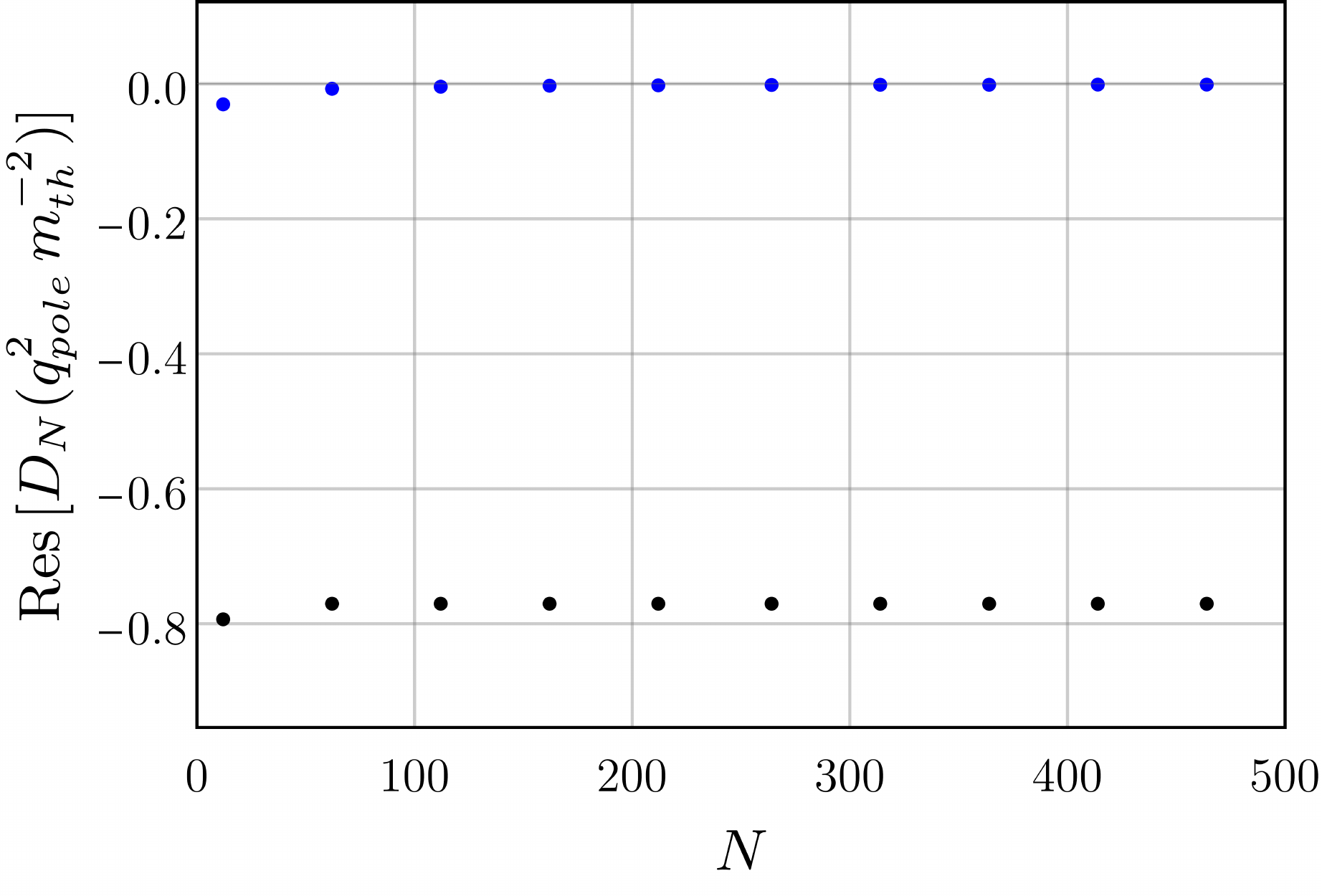}
	\caption{Residues of the truncated propagator~\eqref{eq:trunpropagator} at its poles as a function of the truncation order $N$, for the case of QED (top panel), Lee Wick QED (bottom left panel) and Lee-Wick QED with non-standard coupling (bottom right panel). As in Fig.~\ref{fig:polesanddiv}, blue dots refer to fictitious poles while the black dots describe the true ghost. The residues associated with the fake ghosts tend to zero as $N\to\infty$. The residue of the real ghost is always negative instead.\label{fig:residuesqed}}
\end{figure}

Due to their negative residues, the fake degrees of freedom contribute to the truncated spectral function with negative-diverging Dirac deltas. 
However, since the fake ghosts are not part of the set of asymptotic states of the full theory, they must dynamically decouple by increasing the truncation order $N$. As argued in~\cite{Platania:2020knd}, it might be possible to determine the nature of a ghost in a truncated theory by studying the behavior of the corresponding residue with $N$. In fact, the residue associated with fake ghosts seems to vanish for sufficiently large $N$, at least in the case of QED and Lee-Wick QED with non-standard sign~\cite{Platania:2020knd}, while the residue of a ghost appearing in the full theory is negative and remains negative for any $N$ (cf. Fig.~\ref{fig:residuesqed}).
In the cases at hand, the contribution of fictitious ghost poles to the propagator (and thus to the optical theorem) becomes small for sufficiently large $N$ and vanishes for $N\to\infty$, so that fake ghosts decouple in the limit where no approximation is employed. 
In what follows we will refer to this as the \textit{residue decoupling mechanism} of fake ghosts.

Since quadratic gravity~\cite{Stelle:1977ry} can be viewed as a truncation of a full diffeomorphism-invariant theory of QG, e.g., within the framework of asymptotically safe gravity, the massive ghost of quadratic gravity could be a fake degree of freedom, rather than a problem of the theory. In particular, if the residue decoupling mechanism holds, it can be used to investigate this hypothesis. This provides an important motivation to further explore the validity of the residue decoupling mechanism.

\subsection{Residue decoupling mechanism of fake poles: further numerical evidence}

In the case of logarithmic effective actions, the (absolute value of the) residue of fictitious ghosts decreases as the truncation order is increased, while for genuine ghosts the residue is negative and quickly converges to a finite (negative) value. It is both instructive and useful to investigate the validity of this statement in the case of other form factors~$P(z)$.

\begin{figure}
	\begin{flushleft}
		\hspace{-0.2cm}\includegraphics[scale=0.4]{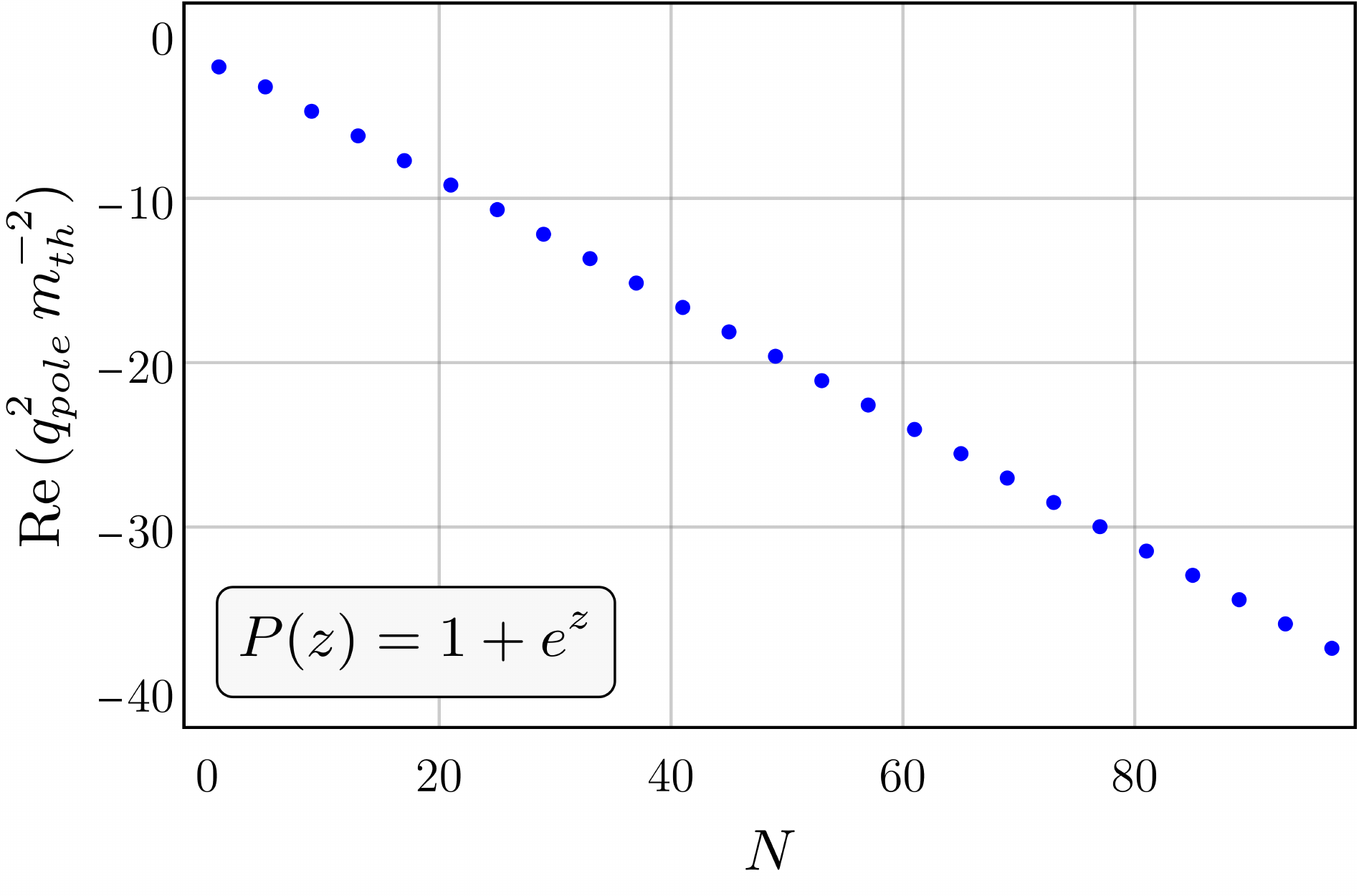}\hspace{0.1cm}\includegraphics[scale=0.4]{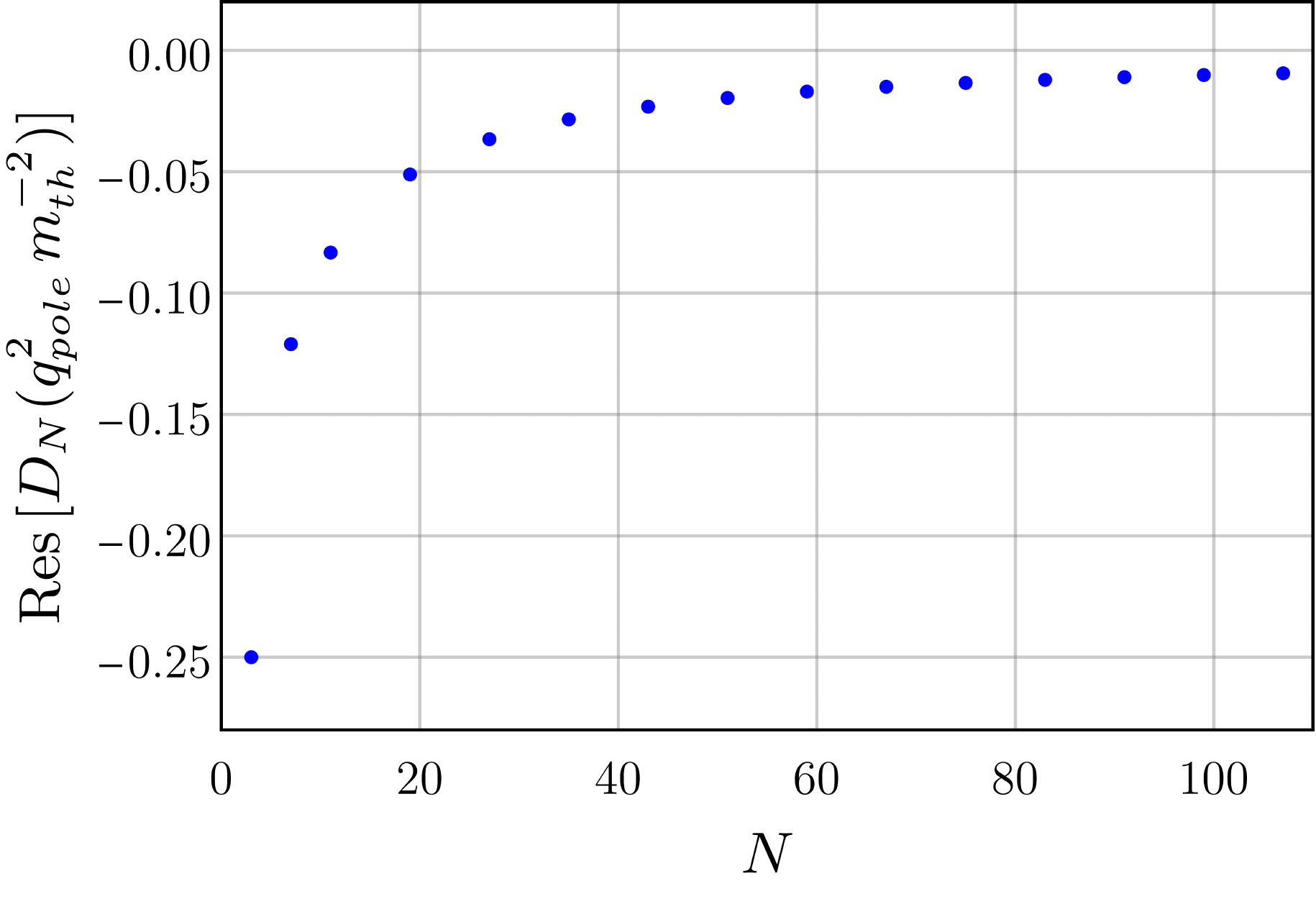}\\
		\hspace{-0.1cm}\includegraphics[scale=0.4]{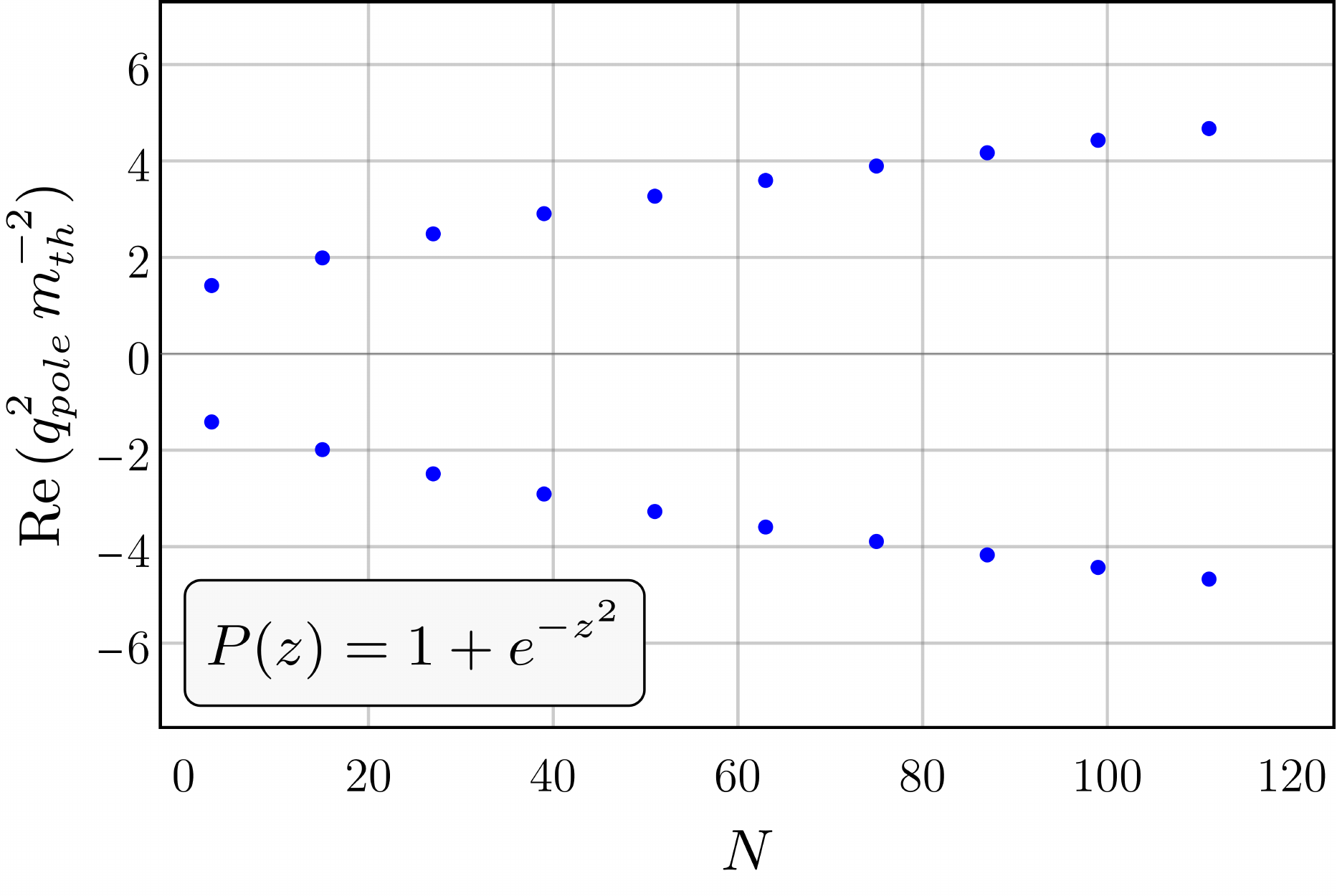}\hspace{0.1cm}\includegraphics[scale=0.4]{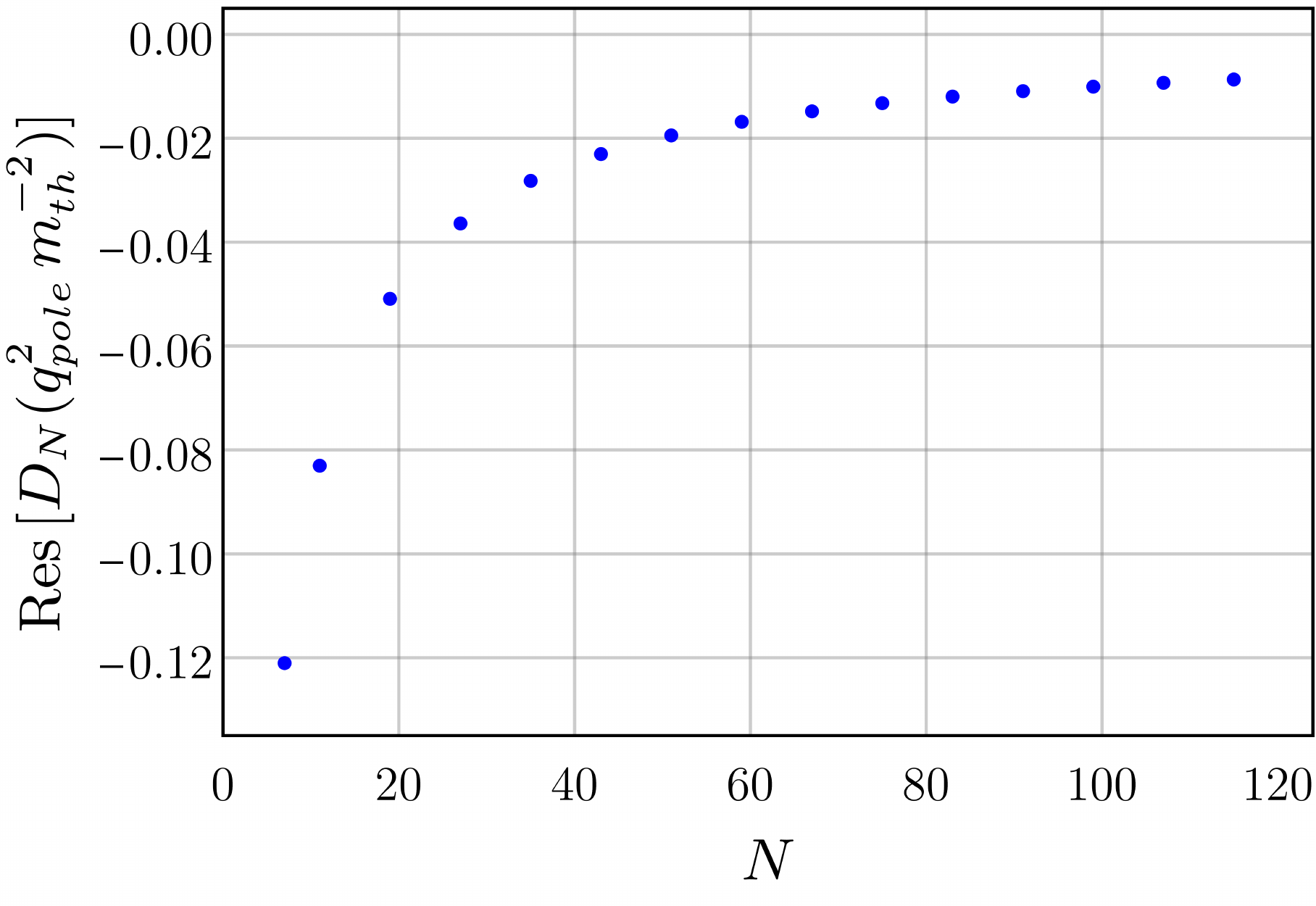}\\
		\hspace{-0.3cm}\includegraphics[scale=0.4]{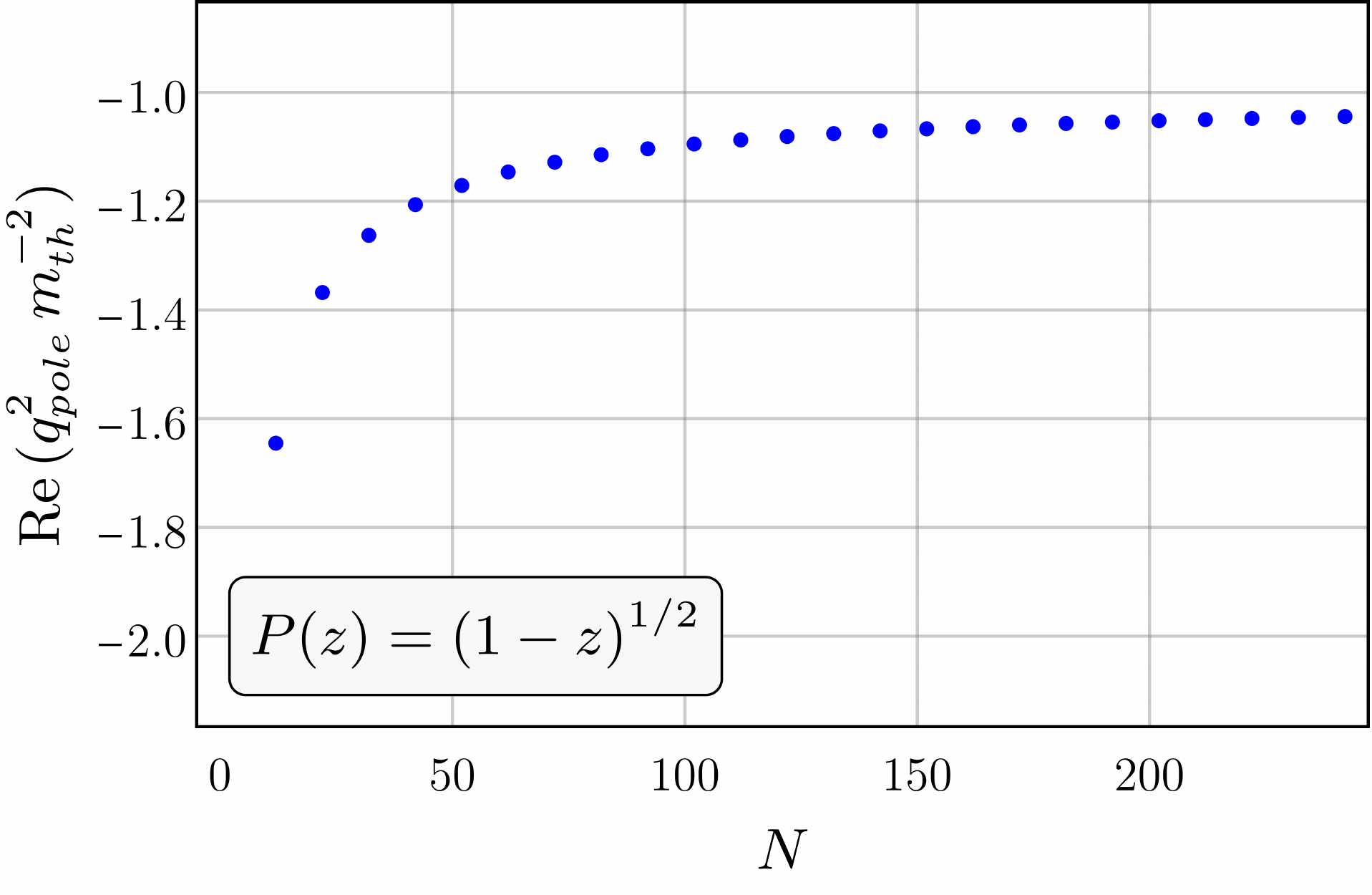}\hspace{0.1cm}\includegraphics[scale=0.4]{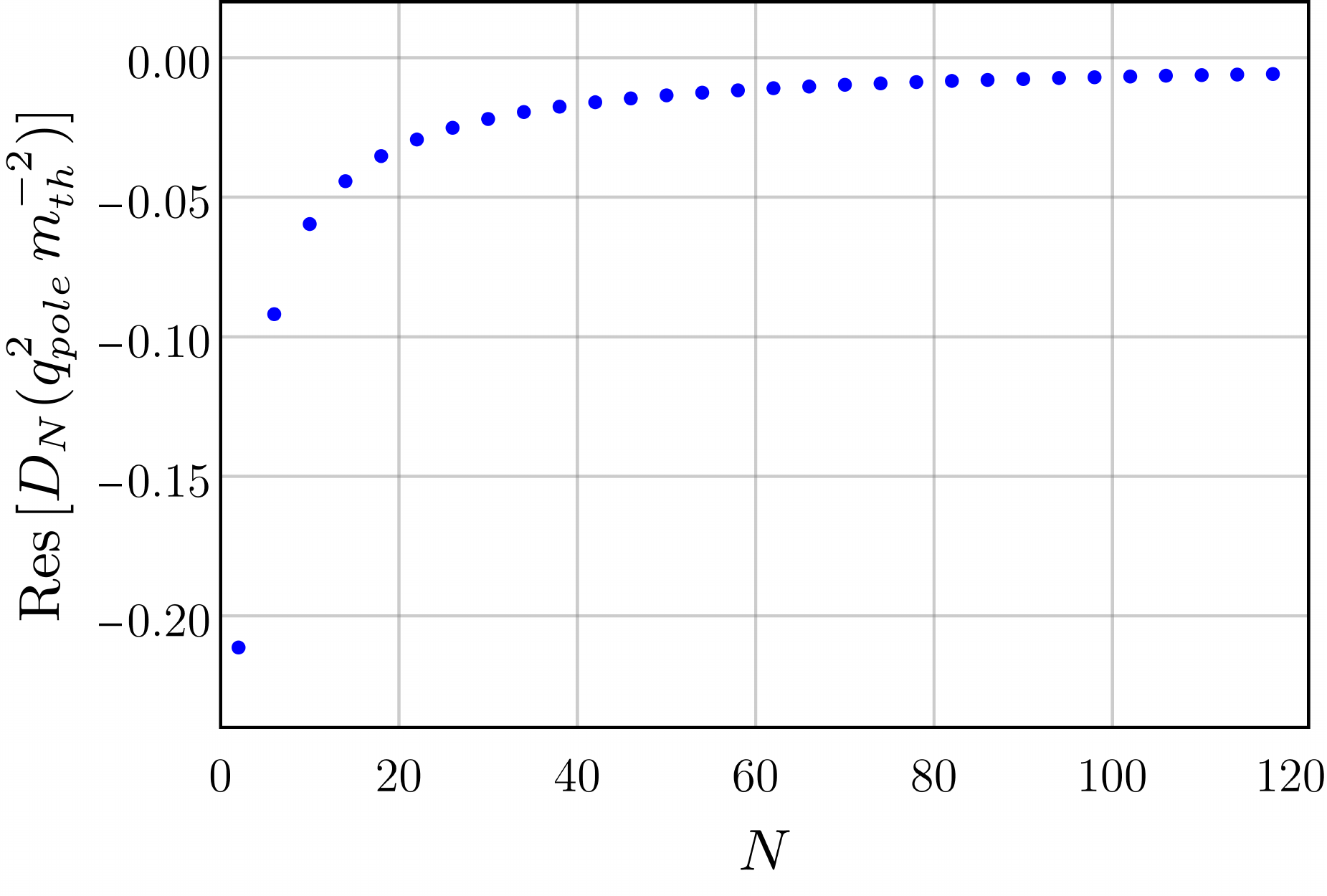}
	\end{flushleft}
	\caption{Evolution of real poles (plots in the left column) and residues (plots in the right column) of truncations of the three test functions: $P(z)=1+e^{z}$ (top panel), $P(z)=1+e^{-z^{2}}$ (central panel) and $P(z)=(1-z)^{1/2}$ (bottom panel). The first function, $P(z)=1+e^{z}$, is entire. Its truncated version~$P_N(z)$ has one single real zero, which moves to infinity as $N$ is increased. At the same time, the corresponding residue approaches zero. The second (entire) function, $P(z)=1+e^{-z^{2}}$, is very similar to the first one, but in this case its truncated version has two real zeros, that evolve symmetrically. The corresponding residues are identical (every blue dot in the figure in the right panel is actually the overlap of two blue dots, corresponding
	to the residues of each pole). The third test function, $P(z)=(1-z)^{1/2}$, has a finite radius of convergence and the single fictitious pole generated by truncations moves
	towards the boundary of the domain of convergence, $|z|\leq1$, while the corresponding 
	residue approaches zero, as expected.}
	\label{fig:res123}
\end{figure}

Our numerical results are summarized in Fig.~\ref{fig:res123} for a sample of $P$-functions. The residue mechanism holds 
in all these examples, which include both entire and non-entire functions. In the case of entire functions, however, the convergence of the sequence of residues is slower than in the case of non-entire functions.
It is straightforward to apply the same analysis to other $P$-functions, such as trigonometric, periodic, hyperbolic and various exponential functions, leading to the tentative conclusion that the residue mechanism holds under very general assumptions. In particular, our numerical studies suggest that its applicability does not depend on the existence of branch cuts and the periodicity properties of the $P$-function, is not limited to even/odd functions, and is influenced neither by the number of zeros of $P$ nor by its divergences.

Summarizing, the numerical results extracted from a sample of functions $P$ seem to point towards the following conclusions:
\begin{itemize}
	\item A stable ghost in the full theory appears in the truncated effective action as a pole in the principal branch of $P(z)$. By increasing the order $N$ of the truncation, the corresponding residue is negative and remains negative for any truncation order $N$. A fake ghost appears in the truncated action for all $N$, but it does not appear in the fully quantum action. 
	\item If the non-local interactions in the effective action are characterized by branch cuts, the fictitious zeros of $P_N(z)$ (corresponding to the fake ghost poles) move towards  and accumulate on the boundary of the domain of convergence of $P(q^{2})$, and disappear only in the limit $N\to\infty$. The absolute value of the corresponding residues decreases with~$N$ and vanishes in the limit $N\to\infty$. The ghost degrees of freedom associated with these fictitious zeros are ``fake'', as they are truncation artifacts and not a feature of the theory. 
	\item If a form factor in the effective action is an entire function, its radius of convergence is infinity. There is a single principal branch, and the function is analytic everywhere. By increasing the truncation order $N$, the locations of the fictitious zeros of $P_N(z)$ approach infinity, while their residues tend to zero. The corresponding ghost degrees of freedom are fake, as in the preceding case, and decouple from the theory for sufficiently large~$N$.
\end{itemize}
A proof of the residue decoupling mechanism of fictitious ghosts will be presented elsewhere~\cite{Platania:2022upcoming}. 

This concludes our discussion on non-perturbative unitarity and the role of truncations. In the next section we will discuss other two fundamental properties of consistent QFTs: causality and stability.\\

\noindent\fbox{\parbox{\linewidth}{Based on the considerations of this section, \textit{when discussing unitarity, one should not trust truncated versions of the dressed propagator, since fictitious ghosts that are not present in the full theory are unavoidably generated}. The resulting fake ghosts decouple for sufficiently large truncations via the \textit{residue decoupling mechanism} described in this section.}}

\section{Complex poles, notions of causality, and (in)stabilities}\label{sect:causality}

In this section we discuss constraints on the pole structure of the dressed propagator that are to be imposed to preserve unitarity, avoid acausalities and dangerous (tachyonic and vacuum) instabilities, and allow for an analytic Wick rotation. The types of poles and their implications for unitarity, causality and stability are summarized in Tab.~\ref{tab:polesconditions} at the end of this section.

Before starting, we have to clarify what  ``causality'' means in this context. 
Several different notions of causality exist across the literature, from vague definitions that apply in many contexts, to more solid conditions that only hold in specific cases. Given the multiplicity of definitions of causality in relativity and in QFT, a quantum-relativistic theory could be causal in one sense, but not necessarily in another. Therefore, before deriving conditions on the propagator of causal field theories, in Sect.~\ref{subsect:causdefs} we will review various notions and definitions of causality and attempt to clarify their mutual relation. In Sect.~\ref{subsect:instabsdefs} we will highlight analogies and differences between tachyonic and vacuum instabilities, both at the classical and quantum level. This will also be key for the discussion in Sect.~\ref{subsect:propycauscond}, where we shall investigate the relations between violation of causality, instabilities, position of the poles of the propagator in the complex $q^2$- and $q_0$-planes, and the possibility of performing an analytic Wick rotation. 

In the case of field theories with a single massless or massive pole this relation is clear. Specifically, the construction of propagators in QFT is based on the Feynman prescription that, via the replacement $q^2\to q^2+i\epsilon$, moves the real poles of a Green's function off the real axis and towards the second and fourth quadrants of the $q_0$-complex plane. Similar arguments imply that  causality can only be preserved if no complex poles are present in the first and third quadrants of the complex $q_0$-plane. As a key example, unstable ghosts having a (small) negative width -- Merlin modes -- entail a violation of causality on microscopic scales~\cite{Donoghue:2019ecz,Donoghue:2019fcb}. This can be seen by studying the Fourier modes of the propagator. In  Sect.~\ref{subsect:propycauscond} we generalize the arguments of~\cite{Donoghue:2019ecz,Donoghue:2019fcb} to the case of generic complex poles with arbitrarily large width, complex-conjugate poles and tachyonic modes. This is key to understand the general conditions under which causality is violated (or preserved), and determine what type of degrees of freedom are compatible with causality, unitarity and stability. We shall conclude this section highlighting a relation between causality and Wick rotation in QFT, and some more caveats about the coexistence of unitarity, causality, and (vacuum and tachyonic) stability at the quantum level. 

\subsection{Avatars of causality and their hierarchy}\label{subsect:causdefs}

An in-depth understanding of the many facets of causality is distinctly important for the construction of field theories that are free from paradoxes. This is even more crucial in the view of QG, where spacetime and its causal structure become dynamical and fluctuating. 

In the literature, many distinct notions of causality have been introduced. Here we list them and discuss their mutual relations, starting from the most general one:
\begin{itemize}
	\item \textit{Causality as ``no backward propagation''}. A rigorous definition of causality that applies to all limiting cases (non-relativistic, ultra-relativistic, field theory, S-matrix, etc.) is to our best knowledge lacking. While na\"{i}ve,
	defining causality as the condition that 
	{information can only propagate forward in time} (for all inertial observers) seems useful to conceptually understand and relate more specific and rigorous notions of causality which apply to particular cases only (e.g., classical limit, axiomatic QFT, scattering amplitudes). As such, it makes their mutual relations and hierarchy clearer. 
	In practical terms, it can be stated as the condition that if a source is activated at a time~$t$, then a two-point correlation function $G(t'-t,|\vec{x}'-\vec{x}|)$ should vanish at any $t'<t$ for all inertial observers (a signal cannot be detected before it is produced by a source). As we will comment in the following, this very minimal condition reduces to the classical notion of causality in the non-relativistic limit, in special relativity it matches Einstein's locality, and in the quantum theory it implies both microcausality (more precisely, microcausality follows from Einstein's locality) and that positive energies must flow forward in time. 
	\item \textit{Causal structure and Lorentz invariance}. The spacetime is endowed with a Lorentzian structure defined by non-degenerate (the speed of light $c$ is finite and non-vanishing) light cones; Lagrangians are written in terms of scalars with respect to Lorentz transformations; the type of spacetime interval (space-, time-, light-like) is preserved under Lorentz transformations.
	\item \textit{Causality in Lorentz-invariant theories: Einstein's locality} (this notion of locality is not be confused with locality of bare Lagrangian densities). There is no action at a distance and all signals propagate subluminally\footnote{On curved spacetimes the propagation could be ``mildly superluminal'' but this would not lead to a  violation of causality or Lorentz invariance~\cite{deRham:2020zyh}.}, i.e., along time-like or light-like directions (within the {future} light cone of the source). Equivalently, spacetime events (in particular, results of measurements) cannot be correlated if separated by space-like intervals. 
	Note that ``causality'' (in the sense of forward propagation as introduced above) and Einstein's locality (in the sense of propagation within the 
	light-cone) are equivalent in Lorentz-invariant theories. In fact, propagation via space-like distances can appear to some inertial observers as a propagation backward in time. Thus, forbidding space-like propagation is equivalent to forbidding backward propagation. The violation of causality on macroscopic scales or, equivalently, the violation of Einstein's locality would lead to causal paradoxes due to the appearance of closed time-like curves~\cite{Bilaniuk:1962zz,Rolnick:1969sk,Csonka:1970az,Friedman:1990xc,Deutsch:1991nm,Everett:1995nn,Garrison:1998da}.
	\item \textit{Causality in Lorentz-invariant relativistic quantum theories: local commutativity or microcausality} \cite{GellMann:1954db}. Microcausality can be regarded as a direct implication of Einstein's locality in QFT, since ``observables'' (results of measurements) arise as expectation values of Heisenberg operators. It states that all space-like separated operators must commute,
	\begin{equation}
		[\mathcal{O}(x),\mathcal{O}(y)]=0,\qquad \forall (x-y)^2<0\,,
	\end{equation}
	i.e., an event occurring at $x$ can only influence events belonging to its future light cone. Microcausality thus requires the excitations of a field to propagate subluminally (note however that the group velocity of a field can be superluminal, as it happens in the case of tachyonic fields~\cite{Aharonov:1969vu}) and a violation of this property on macroscopic scales would lead to paradoxes, since in this case faster-than-light propagation and closed time-like curves would be allowed~\cite{Bilaniuk:1962zz,Rolnick:1969sk,Csonka:1970az,Friedman:1990xc,Deutsch:1991nm,Everett:1995nn,Garrison:1998da}. 
	
	\item \textit{Classical (non-relativistic) causality}. In  the non-relativistic limit, $c\to\infty$, all geodesics belong to the future ``light''-cone (which is degenerate in this limit), i.e., they are all time-like or null. 
	Thus Einstein's locality and local commutativity reduce to the na\"{i}ve, classically emergent notion of causality -- that we can call \textit{non-relativistic causality} -- that ``effects come after their cause'': due to the degeneracy of the light cone, a causal non-relativistic amplitude comes with a universal $\theta(t-t_0)$ (instead of a time-ordered product combined with a decay of the amplitude outside of the light cone), $t_0$ being the time where a source is turned on. Both microcausality and non-relativistic causality are thus compatible with the definition of causality (no backward propagation) given above.
	\item \textit{Bogoliubov-Efimov macrocausality of the $S$-matrix in axiomatic QFT}~\cite{Bogoljubow1,Bogoljubow2,Efimov:1967pjn} \textit{and  bounds on scattering amplitudes}. Causal ordering realized at the level of scattering amplitudes derives from the Bogoliubov conditions, and requires the S-matrix to satisfy
	\begin{equation}
		\frac{\partial}{\partial \phi(x)}\left(\frac{\partial S}{\partial \phi(y)}S^{-1}\right)=0 \,,\qquad \forall(x-y)^2<0\,,
	\end{equation}
	for any field $\phi$. 
	Alternatively, this can also be stated in terms of the analyticity properties of the S-matrix. In local field theories microcausality implies Lorentz invariance of Green's functions and analyticity of all scattering amplitudes in the upper half-plane~\cite{GellMann:1954db}. Note that this is just a sufficient condition. Finally, the latter conditions on the analyticity of the amplitudes, together with the positivity properties of their absorptive parts, imply a number of bounds on scattering amplitudes~\cite{Cerulus:1964cjb,Giddings:2011xs,Epstein:2019zdn,Draper:2020bop} and the condition of cluster  decomposition~\cite{Giddings:2011xs}.
	\item \textit{Lorentz invariance of the S-matrix}~\cite{Weinberg:1995mt}. The S-matrix can be constructed as sum of integrals of time-ordered products. Time-ordering only makes sense for time-like (or light-like) separated spacetime points since if $(x-y)^2$ is spacelike, the time-ordering of the corresponding events is not Lorentz invariant. Thus, if a Lagrangian is Lorentz invariant and microcausality holds, then the S-matrix is manifestly Lorentz invariant. Microcausality in particular is needed to ensure that the time-ordering defining S-matrix elements is Lorentz invariant. This is guaranteed if commutation relations of fields at space-like separated points vanish. Thus, Lorentz invariance (and analyticity) of scattering amplitudes are sufficient conditions for microcausality to hold.
	\item \textit{Arrow of causality}~\cite{Cline:2003gs,Donoghue:2019ecz}. Microcausality does not account for the direction of the arrow of causality, which is instead defined by the propagation of the positive-energy flow of stable particles~\cite{Donoghue:2019ecz}. Microscopic violations of causality (in the sense of forward propagation) can thus arise in the presence of modes propagating against the macroscopic arrow of causality defined by the stable modes of the theory~\cite{Donoghue:2019ecz,Grinstein:2008bg}. This concept is a microscopic realization of causality in the sense of forward propagation of signals, and is strictly related to one of the axioms of QFT: two-point functions have to be analytic in the cut $q^2$-complex-plane (singularities are allowed on the time-like real axis only). Violating this condition typically implies a breakdown of local commutativity~\cite{Habel:1989aq,Alkofer:2000wg}, while causality of the S-matrix can still be preserved~\cite{Grinstein:2008bg}.
\end{itemize}
Not all these conditions are equivalent, but a violation of at least one of them ought to imply that Einstein's locality is violated in at least some regimes, aka, the theory is not causal. In particular,
the conditions on causality of the S-matrix are only sufficient, so that their validity does not imply that a theory be causal in the sense of forward propagation and/or in the sense of Einstein's locality (which would imply local commutativity). Importantly, causality (no backward propagation) of a QFT on microscopic scales (not to be confused with microcausality) is the condition described in~\cite{Donoghue:2019ecz} that modes with positive energies flow forward in time. We shall study the latter in more detail in the next sections.

It is important to remark that causality violations might not constitute a serious problem, as long as the violation is confined to short distances and no paradoxes are realized macroscopically. Lee-Wick theories constitute an emblematic example of theories which violate causality (backward propagation and microcausality) on microscopic scales~\cite{Coleman:1969xz,Cutkosky:1969fq,Lee:1971ix}, but produce causal scattering amplitudes (Bogoliubov conditions and causality bounds)~\cite{Grinstein:2008bg} and cause no paradoxes on macroscopic distances.

A microscopic violation of causality occurs, for instance, in $2\to2$ scattering amplitudes of the large-$N$ Lee-Wick $O(N)$ model due to the acausal propagation of virtual Lee-Wick particles~\cite{Cutkosky:1969fq,Lee:1971ix,Grinstein:2008bg}. More generally, in these causality-violating theories scattering processes can involve outgoing particles being created before the particles in the initial state interact~\cite{vanTonder:1133972,Grinstein:2008bg}, possibly leading to observable signatures such as wrong vertex displacements~\cite{Alvarez:2009af}, outgoing wave packets emerging before the actual collision of the incoming signal~\cite{Cutkosky:1969fq,Grinstein:2008bg}, reversed Wigner resonance time delay due to the backward propagation~\cite{Donoghue:2019ecz}, or ``strange'' interference effects~\cite{Cutkosky:1969fq,Rizzo:2007ae,Rizzo:2007nf}. The latter is subject of a line of research in the context of QCD, aimed at assessing whether the gluon propagator could display pair(s) of complex-conjugate poles~\cite{Hayashi:2021nnj}, potentially explaining confinement~\cite{Baulieu:2009ha,Hayashi:2018giz,Binosi:2019ecz} (in this case reflection positivity of the euclidean theory would be violated~\cite{Hayashi:2018giz,Kondo:2019ywt}) or the negativity of the spectral density~\cite{Hayashi:2018giz,Kondo:2019ywt}.

We conclude this subsection with a word of caution. In addition to the possibility of defining causality in a plethora of different inequivalent ways, turning gravity on makes the concept of causality even more subtle: the propagation on a curved spacetime can be ``mildly superluminal''~\cite{deRham:2020zyh}, and the causal structure can display some uncertainties---even at low energies, rendering the concept of a lightcone ill-defined~\cite{Donoghue:2021meq}. Moreover, causality of the S-matrix can only be shown to follow from microcausality and Lorentz invariance in the case of local theories defined on flat spacetimes, especially since the concept of microcausality is based on the definition of local operators. Finally, in the context of  gravitational effective field theories causality constraints may be even more stringent thanks to considerations stemming from black-hole physics~\cite{deRham:2021bll}. As this work focuses on the properties of the (dressed) graviton propagator on a Minkowski background, we will not further discuss these points. 

\subsection{Classical and quantum (in)stabilities: ghosts vs tachyons} \label{subsect:instabsdefs}

Unitarity, stability and causality are subtly related, both at the classical and at the quantum level. Similarly to the case of causality, ``stability'' can refer to different concepts, and in particular ghost and tachyonic instabilities have distinct causes and features. The notions of tachyonic and ghost instabilities, their differences, as well as their relation with unitarity and causality will enter our discussion on the Fourier modes of the propagator. In this section we review these concepts and their relation.

\subsubsection{Classical (in)stabilities}

\begin{figure}[t]
	\hspace{-0.5cm}\includegraphics[scale=0.41]{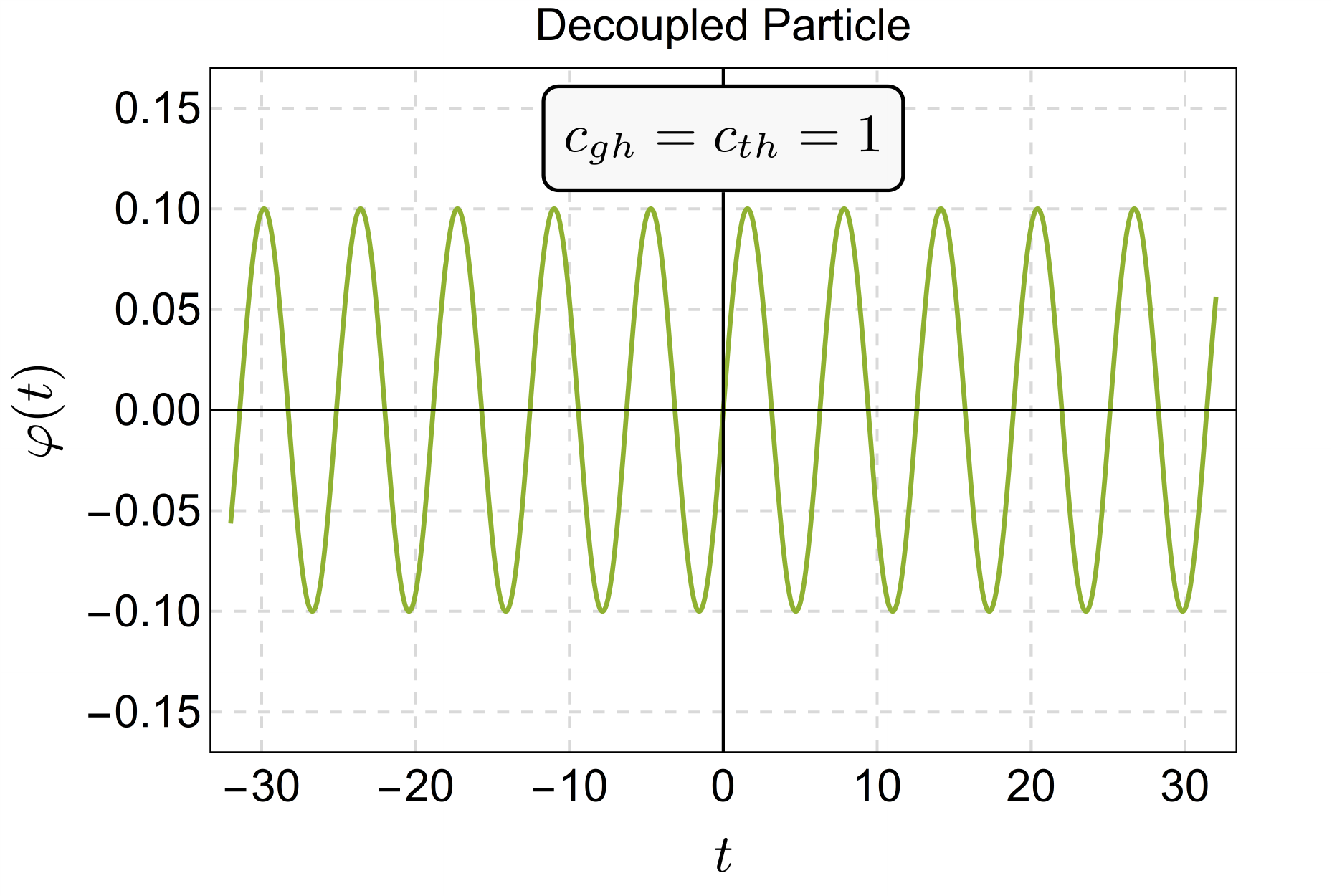}\includegraphics[scale=0.41]{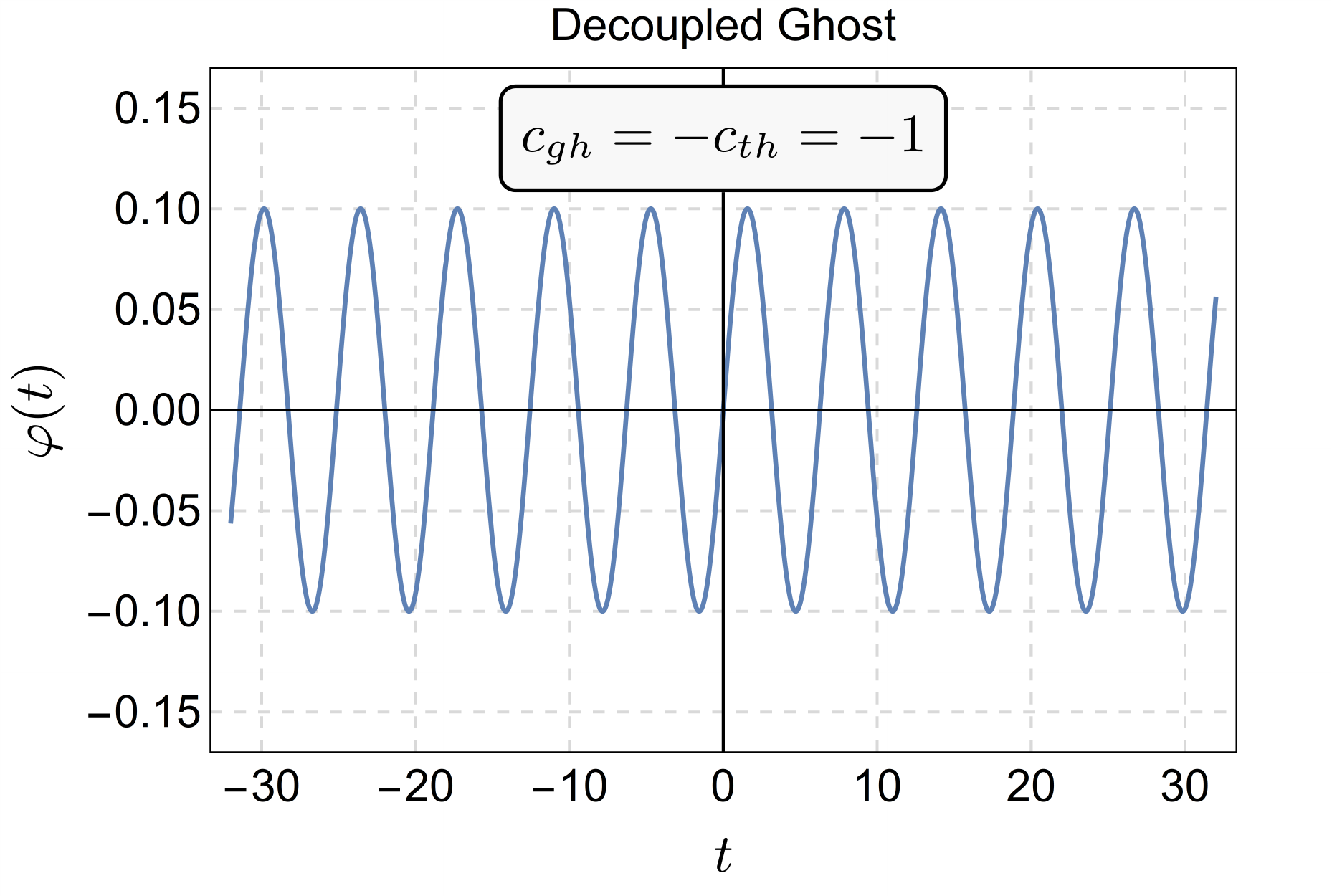}\\
	\centering{\includegraphics[scale=0.41]{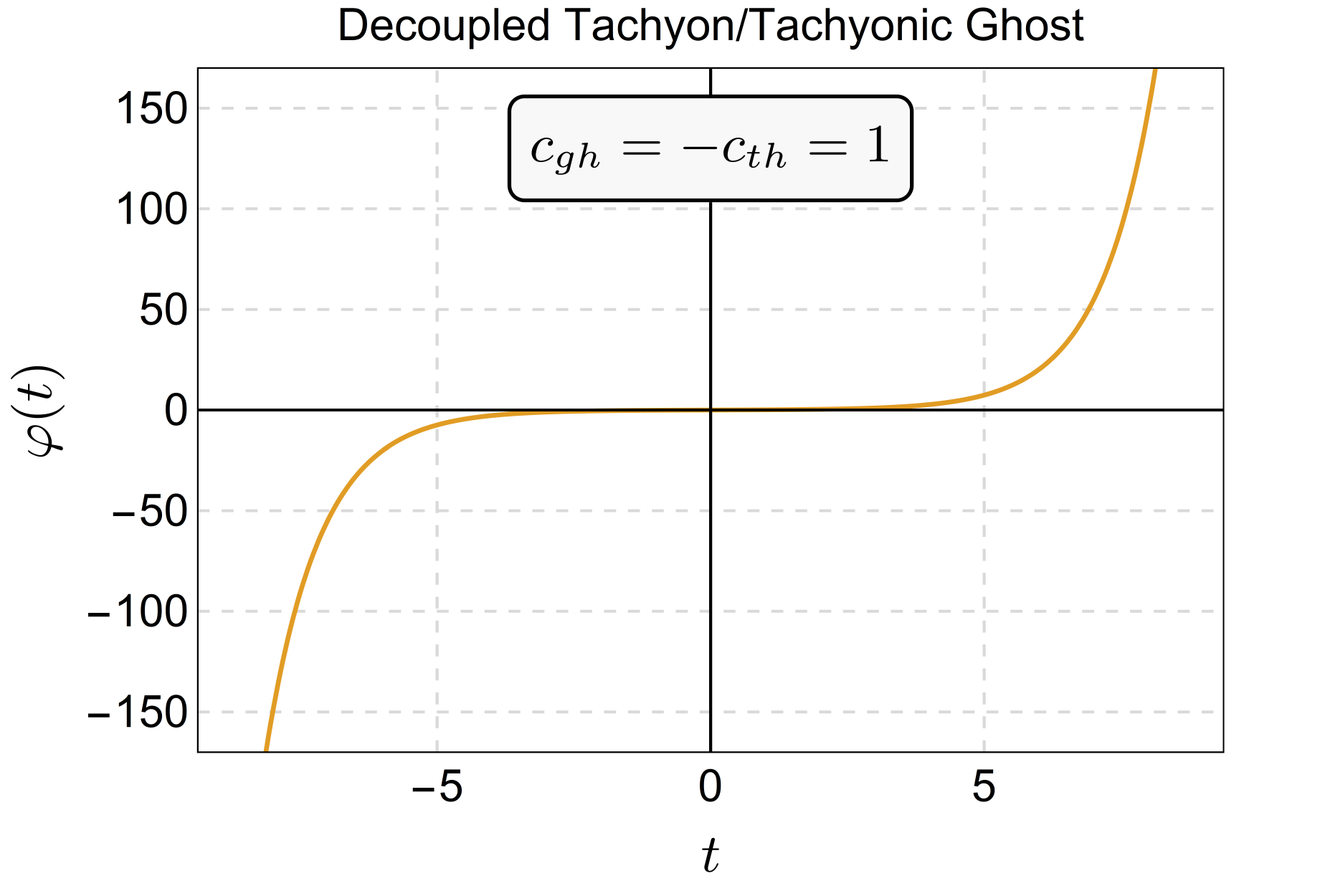}}\\
	\includegraphics[scale=0.41]{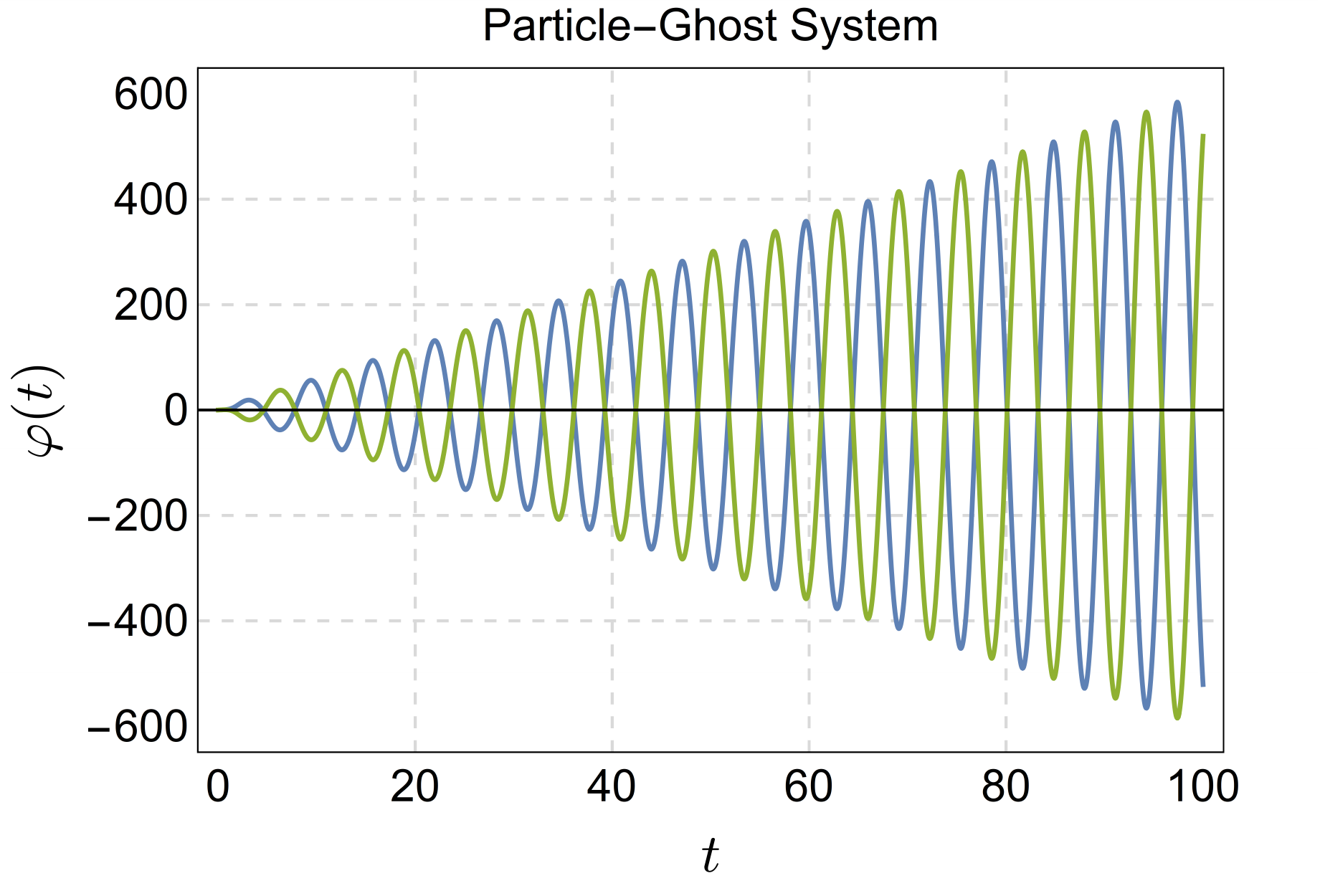}\includegraphics[scale=0.41]{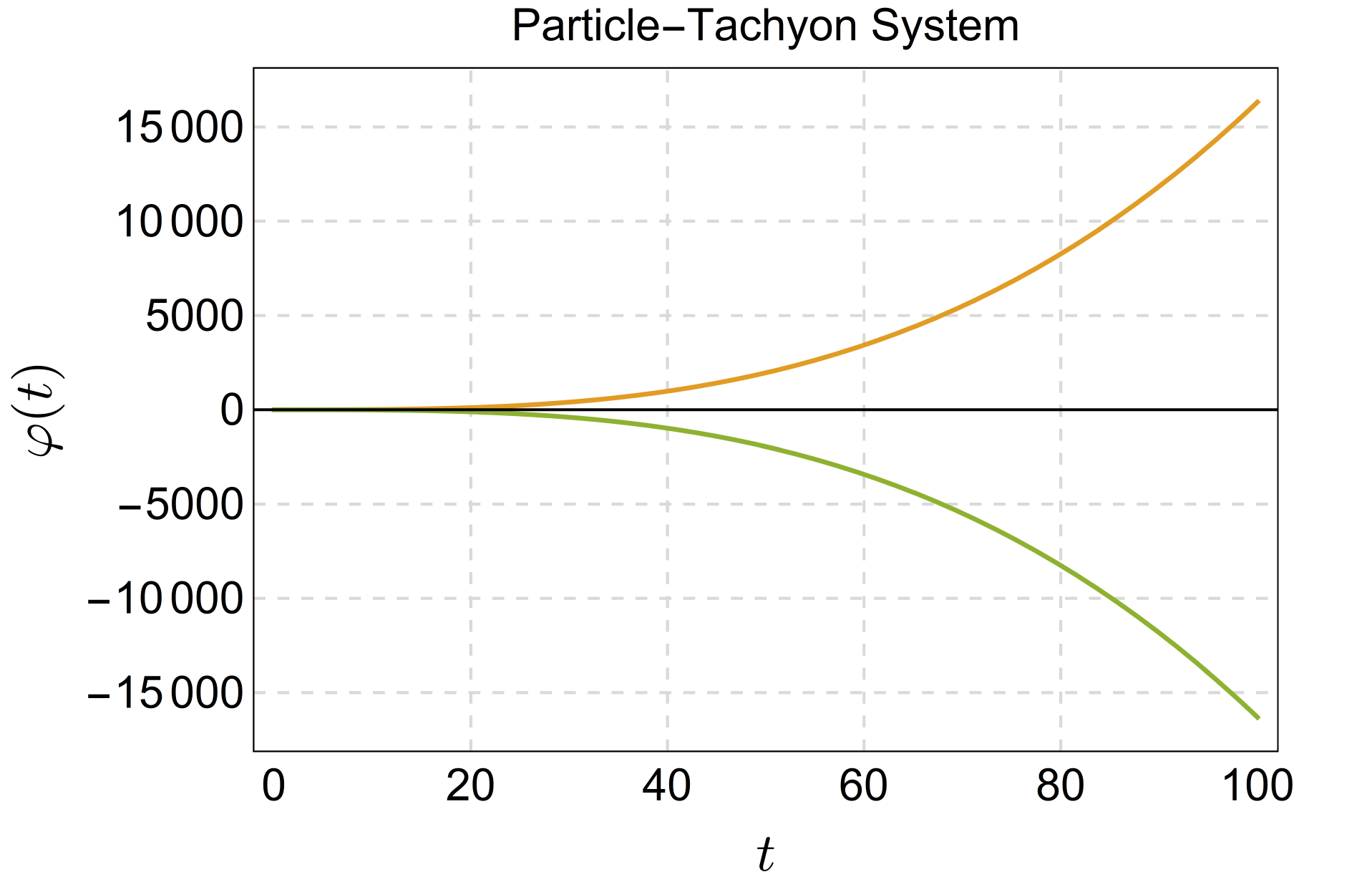}
	\caption{Comparison of classical ghost and tachyonic instabilities in the solutions $	\varphi(t)\equiv\varphi(t,\vec{q}=0)$ to the field equations. The green, blue, and orange lines depict the Fourier modes of  particle, ghost and tachyonic fields, respectively. Non-interacting particles and ghosts (top-left and top-right panel, respectively) do not show exponentially growing modes, rather their fields oscillate periodically in time with constant amplitudes. In the case of free tachyons, there exist regimes (shown in the central panel) where the characteristic frequency $\omega_{\vec{q}}$ is imaginary and the fields modes are exponentially growing. The figures on the bottom refer to the case of a standard particle-field $\phi$ coupled with a ghost (left-bottom panel) and with a tachyon (right-bottom panel) via the Lagrangian~\eqref{eq:coupledlagrangian}. The plots have been obtained using a quartic interaction, $\mathcal{L}_I=-(\phi+\varphi)^4$. Turning the ghost-particle interaction on triggers the ghost instability, resulting in oscillatory and exponentially growing Fourier modes for both the particle and the ghost fields. Similarly, a tachyonic instability propagates into other sectors of the theory.  \label{fig:class-instab}}
\end{figure}

It is convenient to start from a classical scalar field theory with the Lagrangian
\begin{equation}\label{eq:lagra}
\mathcal{L}=-c_{gh}\frac{1}{2}\varphi(\square+c_{th}m^{2})\varphi\,,
\end{equation}
where $c_{gh}=-1$ in the case of ghosts and, keeping $m^{2}>0$, $c_{th}=-1$ in the case of tachyons. The corresponding field equation reads
\begin{equation}\label{eq:LagraSingleGhTh}
(\square+c_{th}m^{2})\varphi=(-q^{2}+c_{th}m^{2})\varphi=0\,,
\end{equation}
and, if the field $\varphi$ does not interact with other fields, it is independent of $c_{gh}$.
The real-space solutions to the field equation at a fixed $\vec{x}$ read 
\begin{equation}
	\varphi(t,\vec{q})=c_1 e^{-i\omega_{\vec{q}} t}+c_2 e^{i\omega_{\vec{q}} t}\,,
\end{equation}
and are characterized by a frequency $\omega_{\vec{q}}=\sqrt{|\vec{q}|^{2}+c_{th}m^{2}}$.
The fact that the solutions are independent of the sign of $c_{gh}$ implies that a non-interacting ghost behaves as a standard particle and no instability arises (see top panels of Fig.~\ref{fig:class-instab}). In particular, the modes of a non-interacting ghost are oscillatory, while the modes of a non-interacting tachyon in the regime where $\omega_{\vec{q}}$ is imaginary are non-oscillatory and exponentially growing (see central panel of Fig.~\ref{fig:class-instab}). 
If one instead couples the degree of freedom in Eq.~\eqref{eq:LagraSingleGhTh} with a standard ($c_{gh}=c_{th}=1$) field, 
\begin{equation}\label{eq:coupledlagrangian}
\mathcal{L}=-\frac{1}{2}\phi(\square+m^{2})\phi-c_{gh}\frac{1}{2}\varphi(\square+c_{th}m^{2})\varphi+\mathcal{L}_I(\phi,\varphi)\,,
\end{equation}
$\phi$ being the field of a standard particle,
the ghost-particle interaction would trigger an instability of the type shown in the bottom-left panel of Fig.~\ref{fig:class-instab}. As for the case of particle-tachyon couplings, the tachyonic instability would also induce an instability in the other sectors of the theory, as shown in the bottom-right panel of Fig.~\ref{fig:class-instab}.

Summarizing:
\begin{itemize}
	\item \textit{Classical tachyonic instability} is characterized by non-oscillating exponentially growing modes due to an \textit{imaginary energy spectrum}, and is a problem on its own.
	\item \textit{Classical ghost instability} is provoked by the \textit{energy spectrum being unbounded from below}. This is problematic only when the ghost is coupled with non-ghost degrees of freedom. Indeed, in the case of non-interacting ghosts one could unambiguously flip the sign of the Lagrangian, making their Hamiltonian bounded from below. In fact in this case the energy of the two degrees of freedom is not separately conserved and thus the two fields can carry arbitrarily large energies. As a result, the field configurations display oscillating exponentially growing modes. At variance with the tachyonic instability, which is generated by the interaction ``potential'' (e.g., the mass term in the case of a free tachyon), the ghost instability is a kinetic instability.
\end{itemize}

\subsubsection{Canonical quantization, ambiguities in the tachyonic case, and quantum (in)stabilities}\label{sect:ambitachy}

Following the standard quantization procedure for~\eqref{eq:lagra} results in the modified commutation relations $[a_{k},a_{p}^{\dagger}]=c_{gh}(2\pi)^{3}\delta^{3}(\vec{p}-\vec{k})$. Some more algebra yields the following free-field expansion for $\varphi$
\begin{equation}\label{eq:freefield}
	\varphi(\vec{x},t)=\int\frac{d^{3}q}{(2\pi)^{3}}\frac{1}{\sqrt{2\omega_{\vec{q}}}}\left\{ a_{\vec{q}}\,e^{i(\vec{p}\cdot\vec{x}-\sqrt{\omega_{\vec{q}}^{2}}\,t)}+a_{\vec{q}}^{\dagger}\,e^{-i(\vec{q}\cdot\vec{x}-\sqrt{\omega_{\vec{q}}^{2}}\,t)}\right\} \,,
\end{equation}
where $\sqrt{\omega_{\vec{q}}^{2}}=|\omega_{\vec{q}}|$ only if $\omega_{\vec{q}}^{2}>0$. For standard particles, the classical statement that the energy spectrum is bounded from below translates in the condition that $a_{\vec{q}}|0\rangle=0$.  Many-particle states are instead created using the creation operator $a_{\vec{q}}^{\dagger}$ and they carry positive energy $E=+\omega_{\vec{q}}$. 

The case of tachyons requires more attention, as several ambiguities can emerge in the quantization procedure. First, let us distinguish two different types of tachyons: superluminal tachyons (or simply tachyons~\cite{Feinberg:1967zza}) -- faster-than-light particles with a real energy spectrum -- and subluminal tachyons (dubbed bradyons~\cite{Cawley:1970us,Recami:1985jb}), characterized by an imaginary energy spectrum. This difference can be straightforwardly seen by inspecting the relativistic expression of the energy and momentum of free tachyons
\begin{equation}
	E_{\vec{q}}=\frac{\sqrt{c_{th}m^2}}{\sqrt{1-v^2}}=\frac{m}{\sqrt{v^2-1}}\,,\qquad |\vec{q}|=\frac{m|\vec{v}|}{\sqrt{v^2-1}}\,,
\end{equation}
in natural units, where $c=1$. For subluminal tachyons $v^2<1$, $|\vec{q}|\in[0,\infty]$, and the energy spectrum is imaginary, since $E_{\vec{q}}^2<0$. On the other hand, for superluminal tachyons $|\vec{q}|\in[m,\infty]$, so that $E_{\vec{q}}^2>0$ for all momenta. The case of tachyons offers a clear example for the inequivalence of subluminality, causality, and stability~\cite{Aharonov:1969vu}.

A comprehensive review of problems and solutions attached with the quantization of tachyons is reported in~\cite{Perepelitsa:2014pva}. Here we want to focus on two particular aspects that will be crucial in the following discussions.

An important difference between the tachyonic and non-tachyonic case arises from the dispersion relation $E^2-\vec{q}^2=c_{th} m^2$. For $c_{th}=1$ the mass-shell relation is a double-sheeted hyperboloid of revolution, one corresponding to  $E\geq m$, and one to $E\leq-m$. In the quantization procedure the corresponding two sets of plane-wave solutions with positive and negative energy are associated with creation and annihilation operators, respectively. Importantly, any proper Lorentz transformation cannot change the sign of the energy. The situation is very different in the case of tachyons, since the mass-shell relation describes a one-sheeted hyperboloid, and the sign of $q_0=E$ is no longer Lorentz-invariant, since a Lorentz transformation can connect different points of the single-surface hyperboloid with energies of opposite signs~\cite{Arons:1968smp,Dhar:1968hkz,Schwartz:2016usj}. Hence, in the case of tachyons the plane-wave expansion~\eqref{eq:freefield} cannot be used, as there is no clear distinction between negative- and positive-energy solutions, and no corresponding unambiguous assignment of a branch with  creation or annihilation operators. This is crucial, since if one would naively use Eq.~\eqref{eq:freefield} and replace $|E_{\vec{q}}|=\sqrt{-m^2+\vec{q}^2}$ in the regime $E_{\vec{q}}^2<0$, one would conclude that, even at a quantum level, tachyons lead to exponentially growing modes. This might still be the case, at least in principle, but such a conclusion cannot be drawn directly from the free-field expansion~\eqref{eq:freefield}, since it must be modified in the case of tachyons~\cite{Arons:1968smp,Dhar:1968hkz}\footnote{Let us remark that the ``reinterpretation principle'' advocated in~\cite{Arons:1968smp} and attached with the different free-field expansion for tachyons seems to solve the paradoxes typically associated with tachyons~\cite{Parmentola:1971auf}, e.g., the Toolman, Bohm, and Pirani paradoxes~\cite{Perepelitsa:2014pva}.}. On the other hand, from a path-integral perspective, once all quantum fluctuations are integrated out, one is left with a fully quantum action, i.e., the effective action $\Gamma_0$ discussed in Sect.~\ref{subsect:effactfrg}. The quantum solutions are obtained by solving the corresponding field equations. If these field equations look like the classical ones in Eq.~\eqref{eq:LagraSingleGhTh}, with $c_{th}=-1$, then exponentially growing modes of the type encountered in the classical case are expected to arise. Notwithstanding quantum fluctuations are expected to correct the simple classical Lagrangian~\eqref{eq:LagraSingleGhTh} by many more interaction operators, and this can lead to the appearance of additional vacua, with respect to which the theory can be stable. This will be the topic of the next subsection, Sect.~\ref{sect:interactionssavetachyons}.

We can now proceed by summarizing how ghost and tachyonic instabilities can occur at the quantum level:
\begin{itemize}
	\item \textit{Quantum tachyonic instability}: In the case of tachyons the quantum instability arises because in some regions of the momentum space $\omega_{\vec{q}}^{2}$ is imaginary. In these regions the field $\varphi$ can potentially display \textit{non-oscillating runaway solutions}, in analogy to the classical case, even though this cannot be directly inferred from Eq.~\eqref{eq:freefield}, as explained above. In addition, the energy spectrum is complex and the states of the theory have vanishing norm. However, one could cure the instability by choosing a more appropriate vacuum, if it exists \footnote{Note that this is not possible in the case of a free tachyon, since the potential is a concave parabola  and thus one cannot tunnel from one vacuum (the unstable one) to another one (the true vacuum). In other words, there is no available vacuum with respect to which the theory would be stable.}. This will be discussed in more detail in the next subsection.
	\item \textit{Quantum ghost instability}: In the case of ghosts, the classical instability turns into either a problem of \textit{unitarity} or \textit{vacuum stability}:
	if one imposes the energy spectrum to be bounded from below, with a ground state identified by the condition $a_{\vec{q}}|0\rangle=0$, the norm of one-particle states  $\langle\vec{q}|\vec{p}\rangle=2E_{\vec{p}}c_{gh}(2\pi)^{3}\delta^{3}(\vec{p}-\vec{q})$
	is negative. Alternatively, one could exchange the role of the creation and annihilation operators, using $a_{\vec{q}}$ to define multi-particle states and $a_{\vec{q}}$ to define the vacuum,  $a_{\vec{q}}^{\dagger}|0\rangle$. In this case one would have states with positive norm, but the one-particle states would carry negative energies,  $E=-E_{\vec{q}}$, rendering the combined particle-ghost vacuum unstable.
\end{itemize}
This is a choice that one also encounters using Feynman quantization: if a ghost is present and the standard Feynman prescription $q^2\to q^2+i\epsilon$ is used, the theory is not unitary, in the sense that there are negative-norm states and the probabilistic interpretation of the quantum theory ceases to make sense. But one could also trade non-unitarity with vacuum instability by quantizing the theory with an opposite Feynman prescription $q^2\to q^2-i\epsilon$\footnote{Let us remark that this type of vacuum instability is not strictly related to ghosts, since in principle one could quantize other degrees of freedom having $c_{gh}=1$ with the inverted Feynman prescription. In this case one would obtain a theory that has both negative-norm states (violation of unitarity) and vacuum instability.}. This makes the theory unitary (no negative-norm states) but the spectrum is no longer bounded from below and the vacuum is unstable~\cite{Cline:2003gs,Sbisa:2014pzo}. As we shall see in the next section, this possibility is also accompanied by a microscopic violation of causality in the sense of backward propagation\footnote{The effect of the inverted Feynman prescription on ghosts is very similar to that of a negative width with the standard Feynman quantization, since the negative width would flip the sign of the imaginary part of the inverse propagator. Thus, the inverted arrow of causality described in~\cite{Donoghue:2019ecz} is strictly related to the vacuum instability discussed in~\cite{Cline:2003gs}.}.

\subsubsection{Path integral quantization, effective actions, and tachyons: are tachyons really a problem?}\label{sect:interactionssavetachyons}

In the last subsection we discussed how tachyonic and ghost instabilities can emerge at a quantum level. 

The problem of ghosts is related to the ``kinetic part'' of the action, and therefore its cure is to be sought in the momentum dependence of the propagator arising from the physical flow in the quadratic part of the action. The tachyonic instability is instead due to the ``potential part'' of the action, and thus its resolution relies on the interactions generated at the level of the effective action. 

In this subsection we provide an explicit example of how interaction terms (or, the field dependence) in the effective action can cure tachyonic instabilities. The way a non-trivial momentum dependence can cure ghost instabilities will instead be the focus of Sect.~\ref{sect:complexpoles} and~\ref{sect:goodpropy}.

\begin{figure}[t!]
	\centering\includegraphics[scale=0.59]{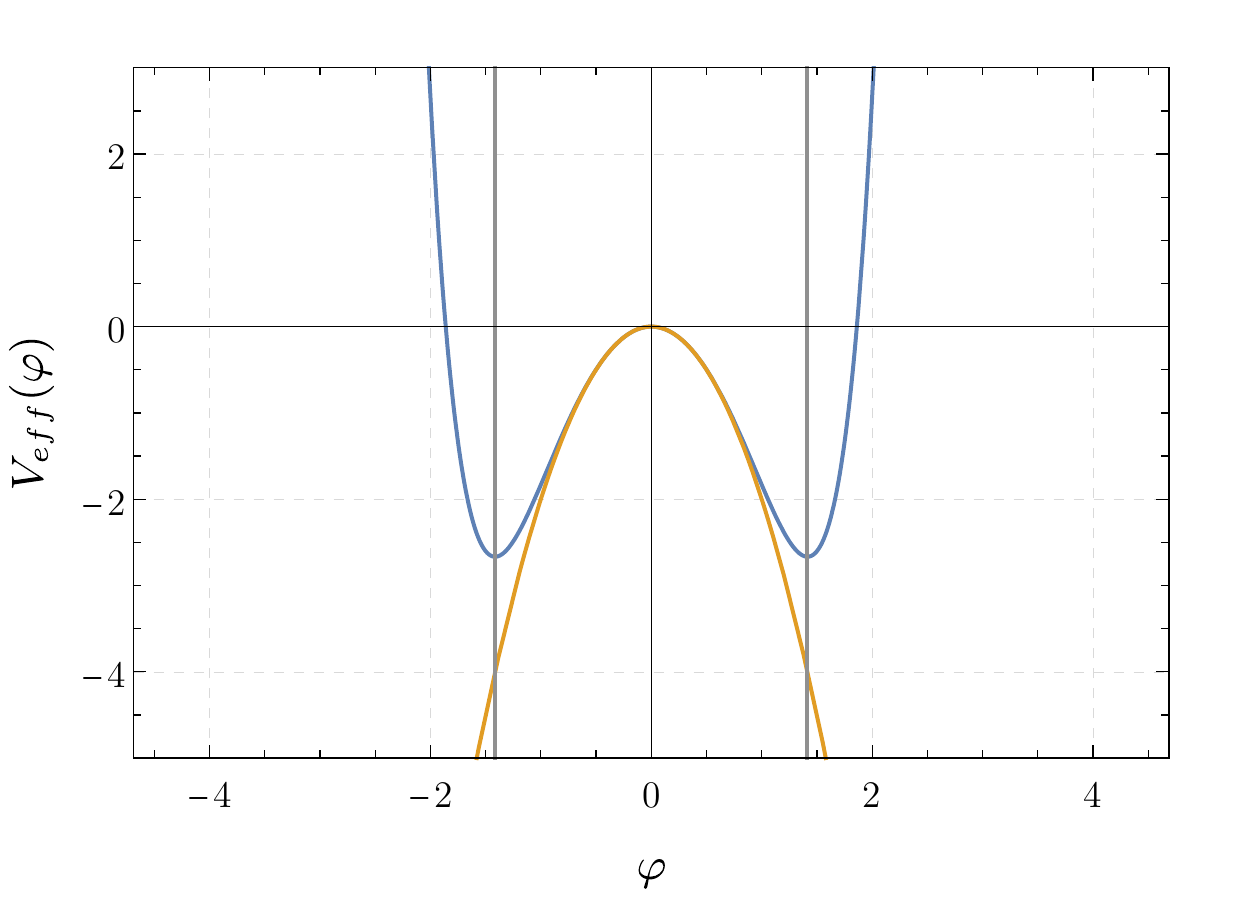}\\\includegraphics[scale=0.59]{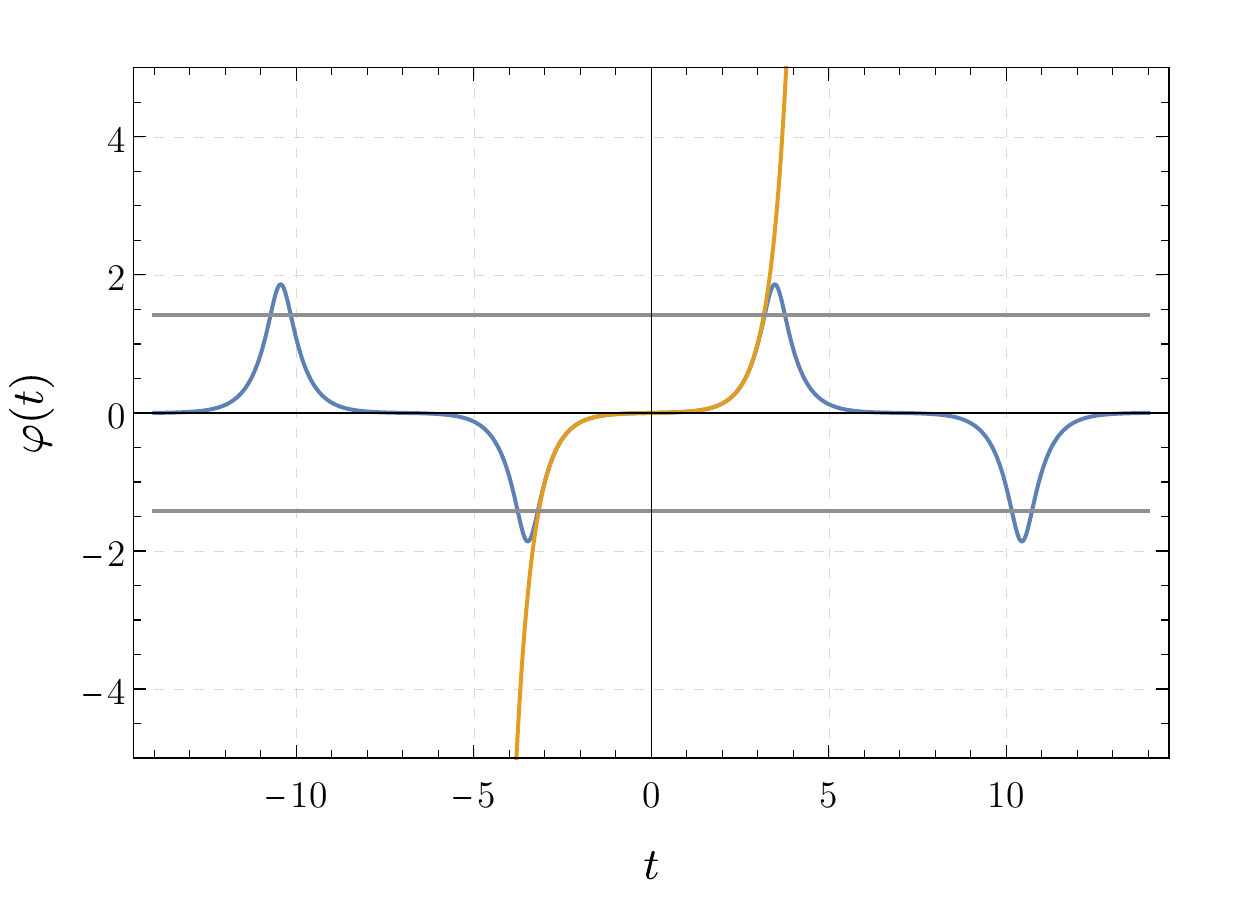}\hfill\includegraphics[scale=0.59]{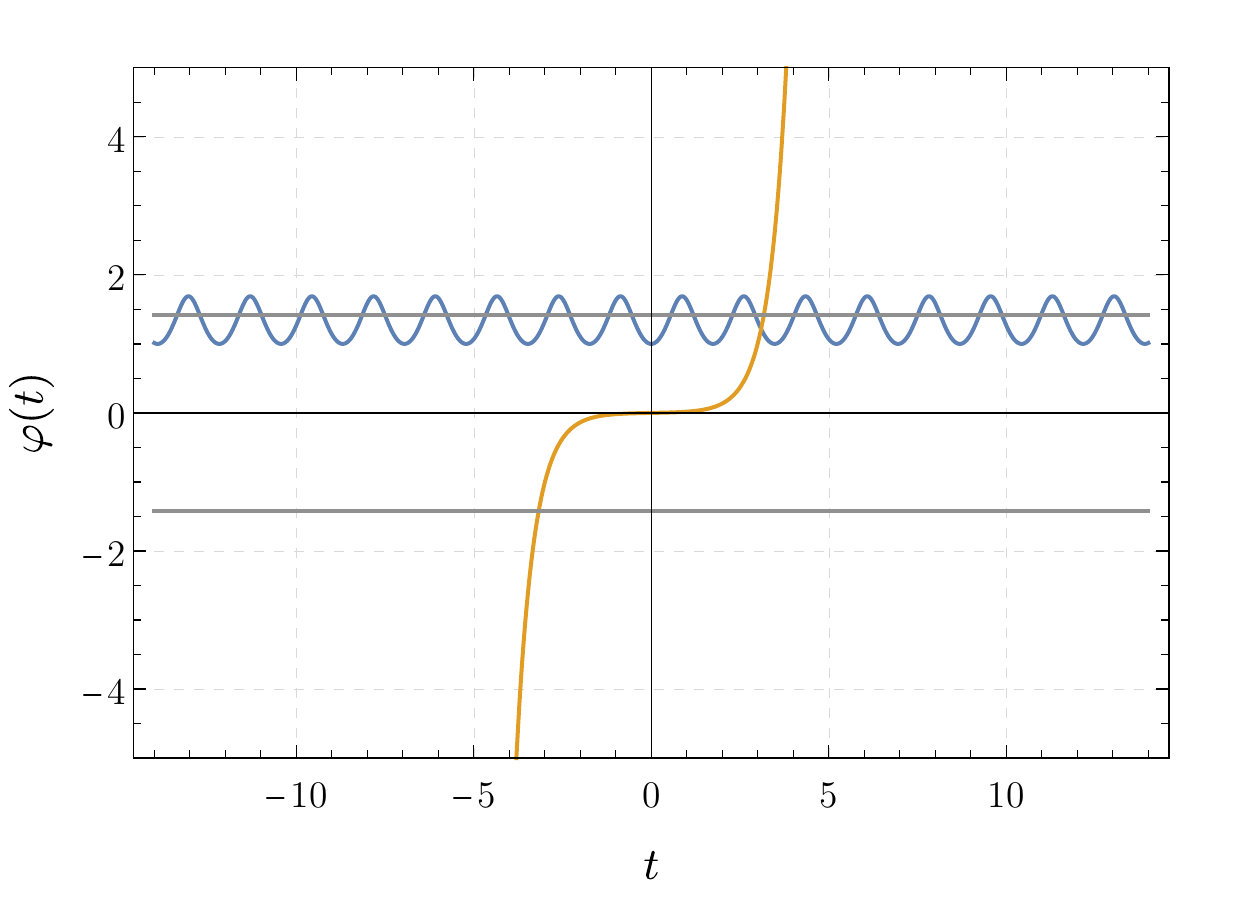}
	\caption{Effective potential (top panel) and solutions to the field equations (bottom panel) in the case of a free tachyon (orange lines) and a tachyonic field interacting via a toy model for the effective potential, $V_{eff}(\phi)=c_{th}m^2\varphi^2+\varphi^6/6$ (blue lines). The parameters $m^2=2 \text{GeV}^2$ and $c_{th}=1$ are used to generate the plots. Two straight gray lines at $\varphi=\pm \sqrt{2}$ are drawn to facilitate identifying the position of the two minima in all plots. The solution $\varphi(t)$ in the bottom-left panel is obtained by using the initial conditions $\varphi(0)=0$ and $\varphi'(0)=0.1$, while the one on the right panel is generated for $\varphi(0)=1$ and $\varphi'(0)=0$. The motion starts close to the unstable configuration associated with the tachyonic modes, and moves away from it. In the case of a free tachyonic field the potential is unbounded from below, since $c_{th}m^2<0$, and this generates the non-oscillating exponentially growing modes discussed in the previous section. Once an interaction is turned on, the potential becomes bounded from below. Correspondingly, the motion is oscillatory, and bounded both from above and from below, similarly to the case of standard particles. Specifically, the figure on the bottom-left panel describes the situation where the field's kinetic energy overcomes the potential energy, and the field oscillates between the two stable vacua. In the second case, depicted in the bottom-right figure, the field oscillates about one of the two possible stable vacua, yielding an oscillatory motion very similar to that of stable particles. \label{fig:interct-tach-instab}}
\end{figure}

In the case of a free tachyon $c_{th}m^2<0$, the potential is a convex parabola (orange line in the top panel of Fig.~\ref{fig:interct-tach-instab}) and the solutions $\varphi(t)\equiv\varphi(t,\vec{q}=0)$ to the field equations display non-oscillatory exponentially growing modes (orange lines in the bottom panel of Fig.~\ref{fig:interct-tach-instab}).
Integrating out quantum fluctuations typically generates a plethora of terms at the level of the effective action and, as anticipated, this might yield a globally stable potential. To illustrate this idea, we can consider a toy model where 
\begin{equation}
V_{eff}(\phi)=c_{th}m^2\varphi^2+\varphi^6/6\,.
\end{equation}
As depicted in the top panel of Fig.~\ref{fig:interct-tach-instab}, the interaction term $\varphi^6/6$ renders the effective potential $V_{eff}$ (blue line) bounded from below. At this point one can solve the field equations with different initial conditions. Two examples of such solutions are plotted in the bottom panel of Fig.~\ref{fig:interct-tach-instab}
(blue lines). As one can easily realize, interaction in this case has removed all tachyonic instabilities, as the solutions to the field equations are oscillating and bounded both from below and above.

Summarizing, since free tachyons cannot exist in nature---barring miraculous cancellation that eliminate all interactions in the effective action---tachyons are not necessarily fatal for the theory, and a full analysis in the presence of interactions is in order to establish whether the theory is sick.

\subsection{Fourier modes of the propagator: conditions for microscopic violations of causality and relation with (in)stability} \label{subsect:propycauscond}

Although in the quantum theory it is more natural to work in momentum-space, causality and microcausality cannot be directly examined at the level of momentum-space two-point Green's functions (since they involve four-momenta, and thus cannot be localized in space and time). The causality condition (no backward propagation) is to be spelled out and studied at the level of position-space propagators \cite{Veltman:1963th}. Indeed, it would not make sense to talk about causality of the momentum-space propagator of an on-shell photon or graviton, since the corresponding amplitude is simply a Dirac delta function; nonetheless, one would wish electromagnetic (or gravitational) waves in real space to propagate forward in time.

In this section we determine relations between the poles of the propagator, unitarity, stability and microscopic violation of causality, quantifying the time-scale of the causality violation in terms of the distance of a pole from the real $q_0$-axis. In what follows we will assume that the inverse (dressed) propagator has no essential singularities at infinity (this is not the case for exponential form factors such as those studied in the context of non-local gravity~\cite{Tomboulis:1997gg,Modesto:2011kw,Biswas:2011ar,Tomboulis:2015esa}, in which case one has to apply the procedure outlined in~\cite{Tomboulis:2015gfa}), and that each of its poles has multiplicity one, such that one can ``close the contour'' and apply the Cauchy integral  formula~\eqref{eq:fullydressed} (which in the case of polynomial inverse propagators gives the same result as the partial fraction decomposition). Under this assumption, the full propagator can be decomposed into a sum of single-pole propagators, and it is thus sufficient to study the Fourier modes of a propagator associated with a single degree of freedom (i.e., an isolated pole). Its most general form reads
\begin{equation}\label{eq:simplepropy4causality}
iD(q^2_0)=c_{gh}\tilde{R}\frac{i}{q^{2}-c_{th}m^2+i\gamma+i\epsilon}\,,
\end{equation}
where $R=c_{gh}\tilde{R}$ is the residue of the complete propagator at the corresponding pole, $\tilde{R}$ being real and positive for real poles and complex in the case of complex-conjugate poles. As before, $c_{gh}=c_{th}=1$ for particles, $c_{gh}=-c_{th}=-1$ for ghosts,  $c_{gh}=-c_{th}=1$ for tachyons, and $c_{gh}=c_{th}=-1$ for tachyonic ghosts.
The physical mass is defined by the real part of $m^{2}$, as this is the only quantity that can be measured, while $\gamma$ is related to the decay width of the particle. 
In particular, if $\gamma\ll |m|$ and $m\neq0$ one can approximate
\begin{equation}
q_{pole}=\pm\sqrt{m^{2}-i\gamma}\simeq\pm\left(|m|-i\frac{\gamma}{2|m|}\right)\qquad\Rightarrow\qquad\frac{1}{q^{2}-m^{2}-i\gamma}\simeq\frac{1}{q^{2}-\left(|m|-i\frac{\gamma}{2|m|}\right)^{2}}\,,
\end{equation}
and realize that the corresponding spectral density resembles a Breit-Wigner distribution for a resonance with mass $m$ and decay width
\begin{equation}\label{eq:gammamassive}
\Gamma\equiv\frac{\gamma}{|m|}\,.
\end{equation}
In the case of massless particles the above derivation fails, as it is no longer possible to expand about $\gamma=0$. If the inverse propagator is of the form $(q^{2}\pm i\gamma)$, then the massless complex poles will be located at
\begin{equation}
q_{pole}=\pm\sqrt{\mp i\gamma}=\pm\sqrt{|\gamma|e^{\mp i\pi/2}}=\pm\sqrt{|\gamma|}e^{\mp i\pi/4}=\pm\sqrt{|\gamma|}\frac{1\mp i}{\sqrt{2}}\,.
\end{equation}
Thus, the inverse propagator can be written as
\begin{equation}
q^{2}\pm i\gamma= q^{2}-\left(\sqrt{\frac{|\gamma|}{2}}\mp i\sqrt{\frac{|\gamma|}{2}}\right)^{2}\,,
\end{equation}
and one can identify the resonance width with
\begin{equation}
\Gamma\equiv\pm\sqrt{2|\gamma|}\,.
\end{equation}
Therefore $\Gamma$ is finite even when $m=0$. Starting from Eq.~\eqref{eq:gammamassive} and taking a na\"{i}ve limit $m\to0$ would instead lead to the misleading conclusion that massless complex poles have infinite decay width (or, vanishing lifetime), and thus that they do not violate causality on any relevant time scale.

Finally, it is important to notice that assuming $\tilde{R}$ is real\footnote{A complex $\tilde{R}$ only arises in the case of complex-conjugate poles, and the contribution of the pair to the spectral density is zero. Thus, it is sufficient for this argument to assume $\tilde{R}$ to be real.}, the contribution of the single-pole propagator in Eq.~\eqref{eq:simplepropy4causality} to the full spectral density is
\begin{equation}
	\rho=\frac{\tilde{R}}{\pi}\frac{c_{gh}\gamma_{tot}}{(q^2-c_{th}\,\mathrm{Re}(m^2))^2+\gamma_{tot}^2}\,,
\end{equation}
where $\gamma_{tot}=\epsilon+\gamma+c_{th}\,\mathrm{Im}(m^{2})$ and, since $\tilde{R}$ is positive by assumption, the positivity of the spectral density requires $c_{gh}\gamma_{tot}>0$. This means that unitarity can only be preserved if ghosts are quantized according to a different Feynman prescription, where one replaces $\epsilon\to-\epsilon$~\cite{Cline:2003gs,Sbisa:2014pzo}, or if they are unstable and come with negative decay width~\cite{Donoghue:2019clr}. In what follows we will neglect the constant $\tilde{R}$, since its specific value only affects the Fourier modes of the propagator by an unimportant, constant multiplicative factor.

Having closed these digressions, we can proceed with the determination of the Fourier modes of the simple propagator~\eqref{eq:simplepropy4causality}. 
In the $q_{0}^{2}$-complex plane, the poles are located at
\begin{equation}
q_{0,pole}^{2}=\underbrace{(\vec{q}^{2}+c_{th}\,\mathrm{Re}(m^{2}))}_{E_{q}^{2}}-i\underbrace{(\epsilon+\gamma+c_{th}\,\mathrm{Im}(m^{2}))}_{\gamma_{tot}}\,.
\end{equation}
Therefore:
\begin{itemize}
\item Standard particles or tachyons ($\mathrm{Im}(m^{2})=0$, $\gamma\geq0$) correspond to poles of the propagator which lie in the fourth quadrant of the $q_{0}^{2}$-complex plane.
\item Poles representing unstable particles or unstable tachyons with negative decay width ($\mathrm{Im}(m^{2})=0$, $\gamma<0$) lie in the first quadrant of the  $q_{0}^{2}$-complex plane.
\item Complex-conjugate degrees of freedom ($\mathrm{Im}(m^{2})\neq0$,
$\gamma=0$) lie in the first and fourth quadrant of the $q_{0}^{2}$-complex plane.
\end{itemize}
In the $q_0$-complex plane, the poles are located at
\begin{equation}
q_{0,pole}^{\pm}=\pm\sqrt{E_{\vec{q}}^{2}-i\gamma_{tot}}\,.
\end{equation}
When $\gamma_{tot}\ll1$ one can expand about $\gamma_{tot}=0$. Here we study the general case where~$\gamma_{tot}$ can take any value and we examine where the poles are located in the complex plane, and the corresponding physical consequences.

\begin{figure}[t]
	\centering{\includegraphics[scale=0.41]{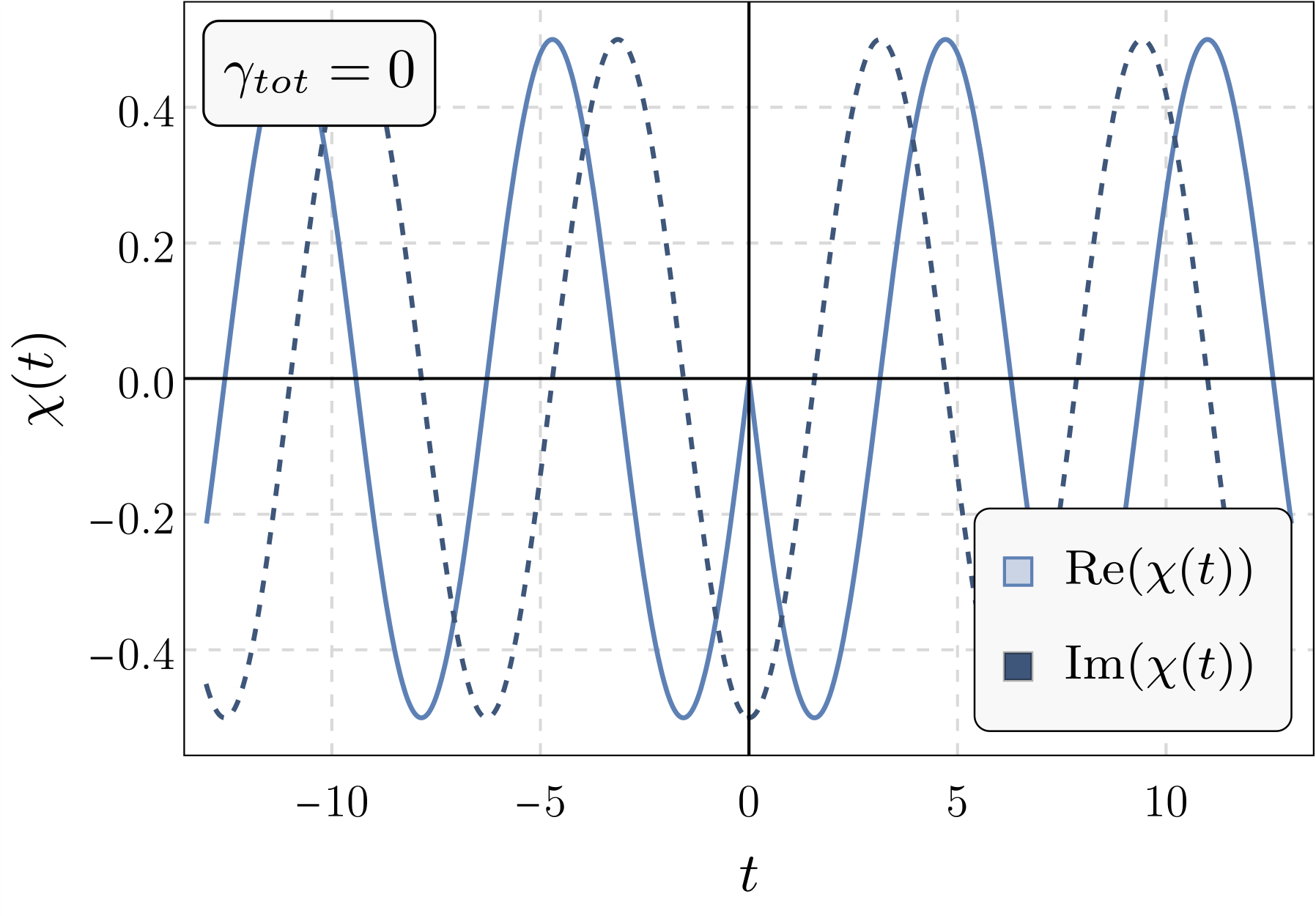}}\\
	\hspace{-0.1cm}\includegraphics[scale=0.41]{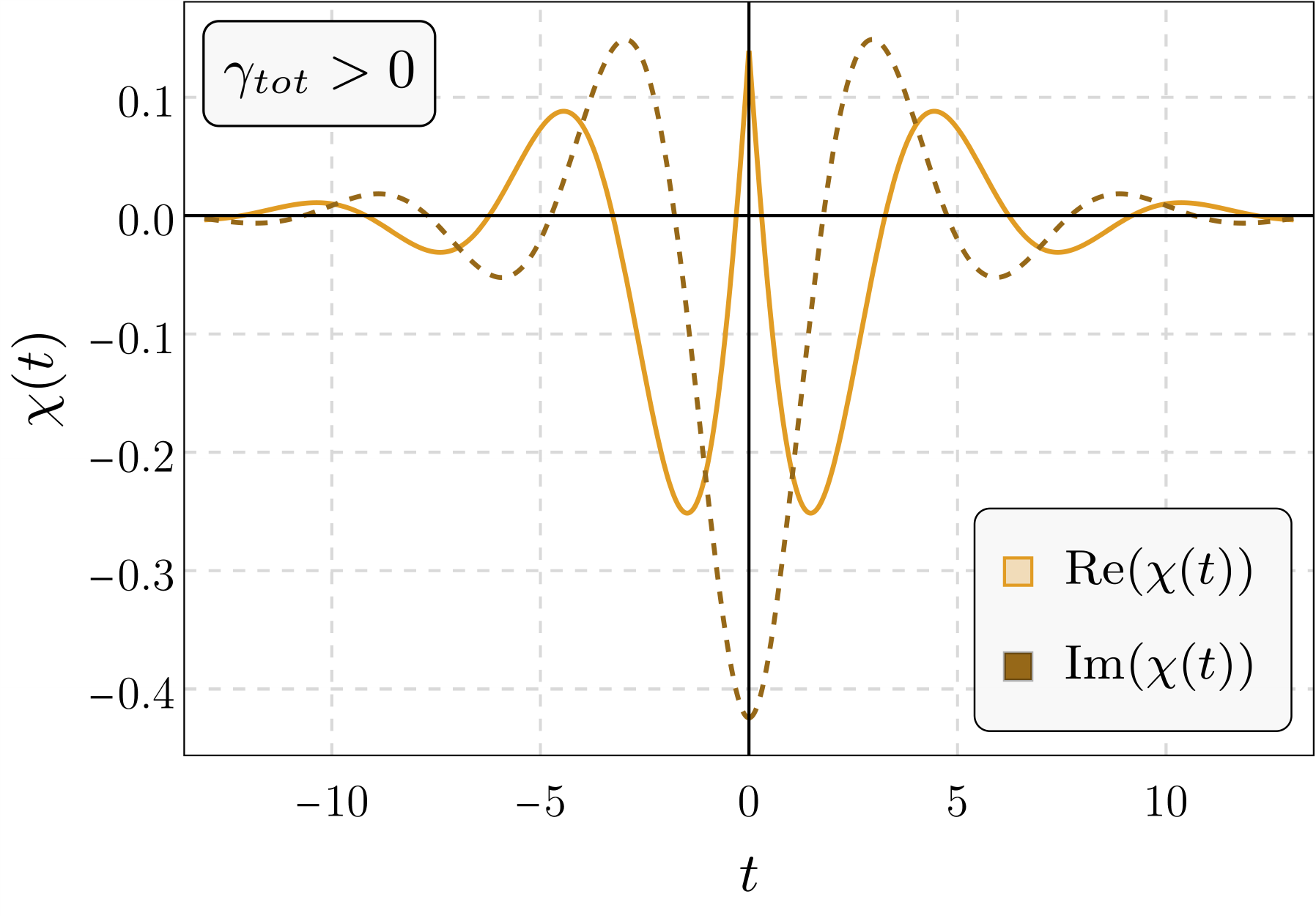}$\,\,$\includegraphics[scale=0.41]{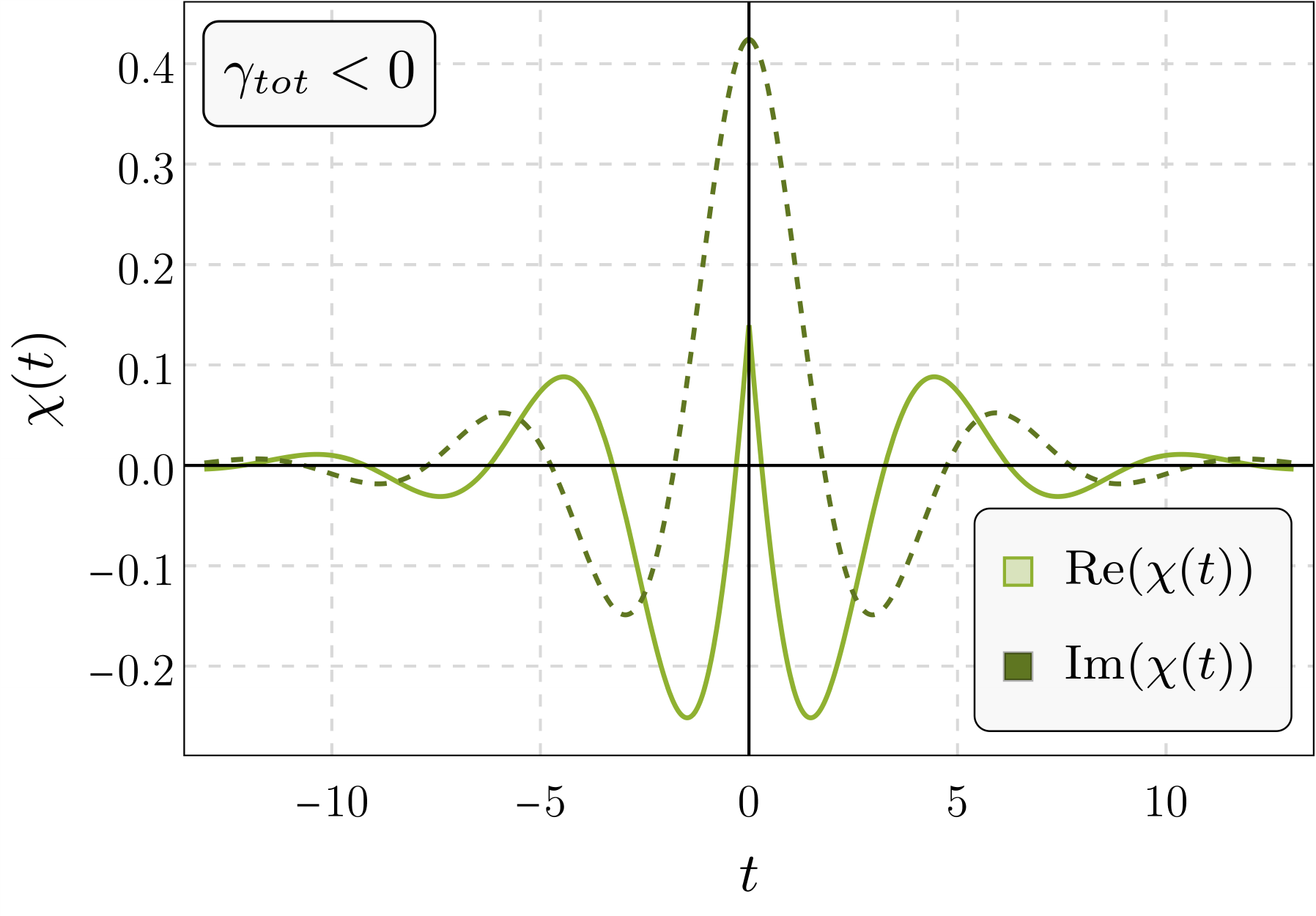}\\
	\hspace{-0.1cm}\includegraphics[scale=0.41]{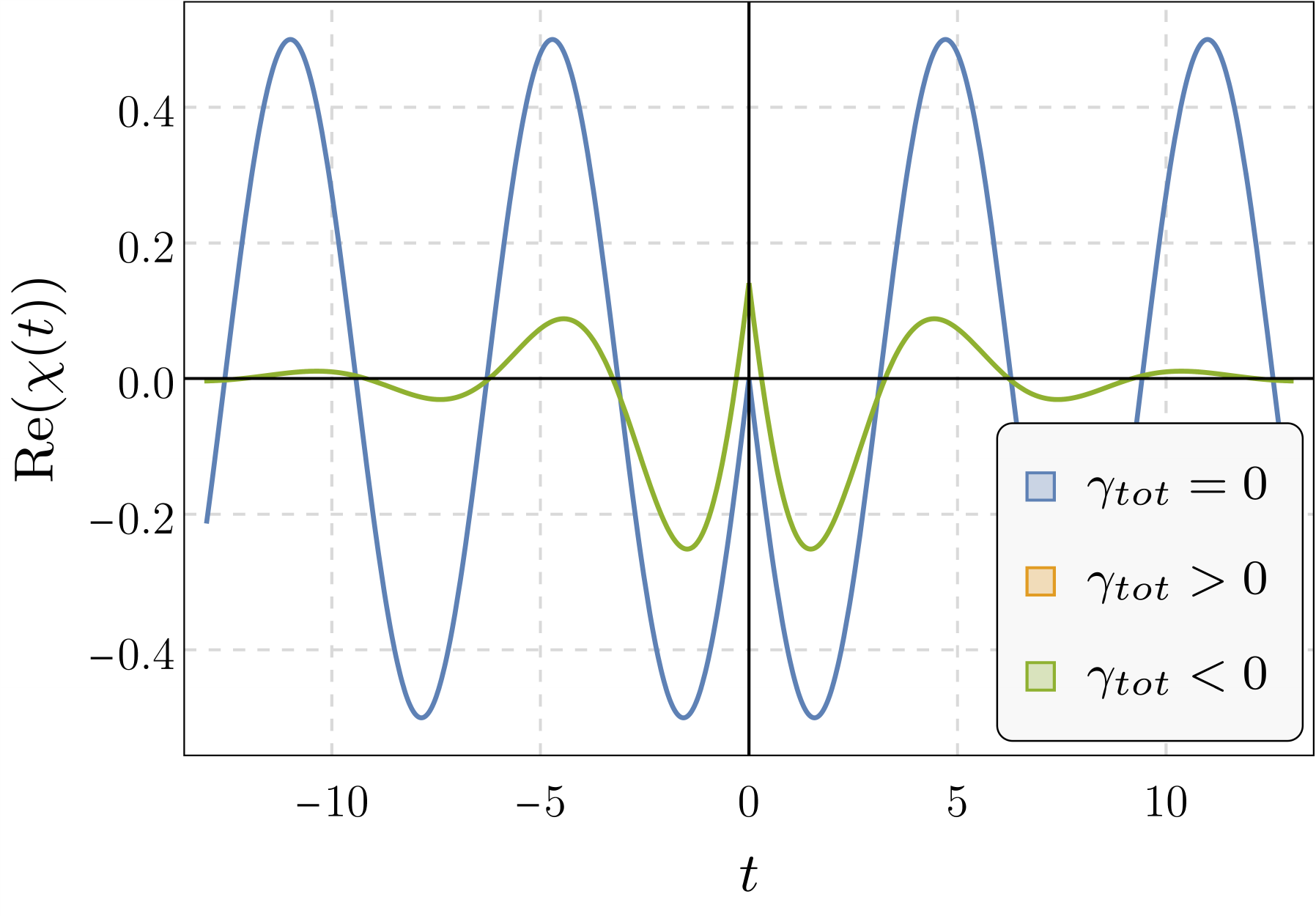}$\,\,$\includegraphics[scale=0.41]{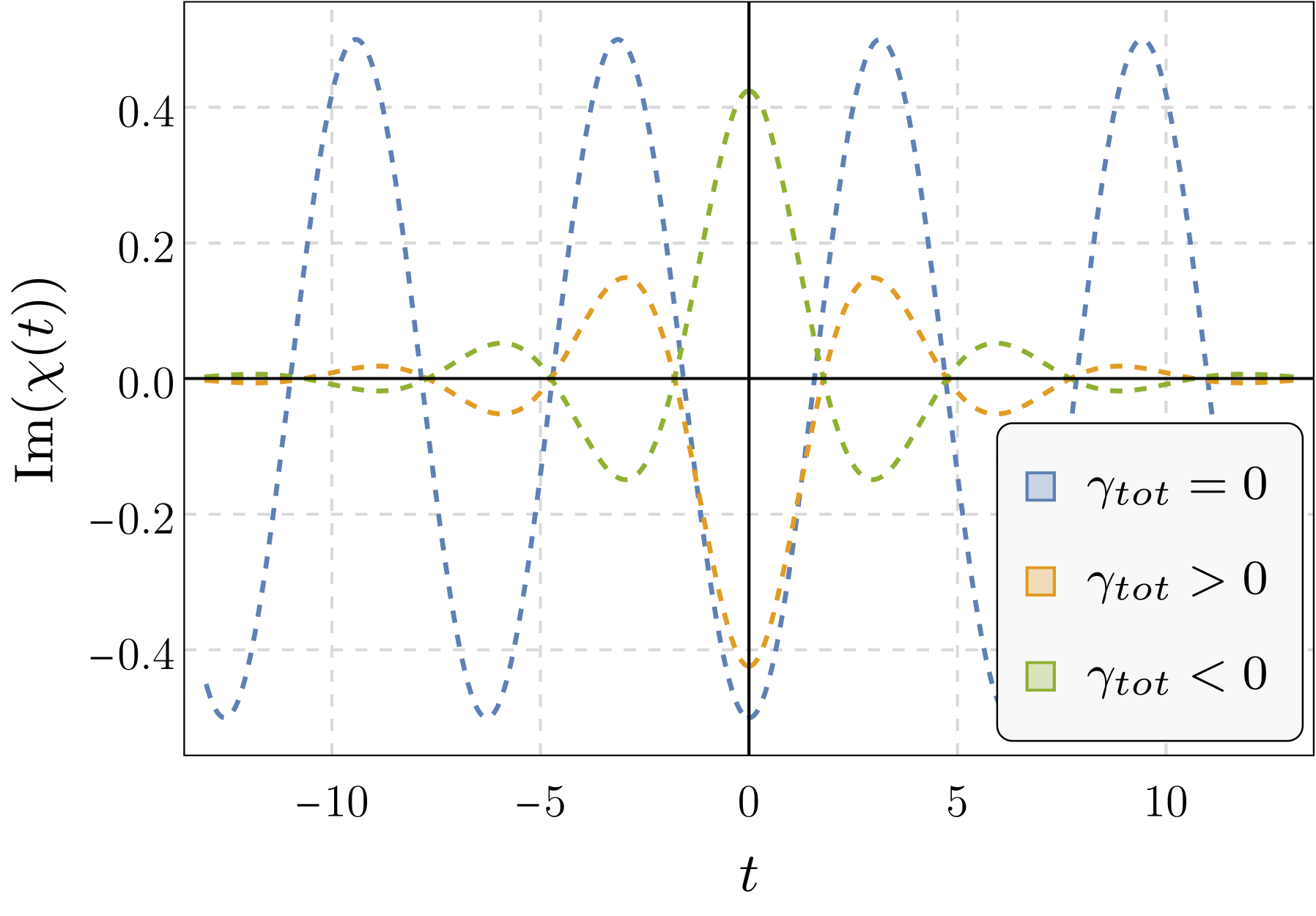}
	\caption{Fourier modes of the propagator (function $\chi(t)$ in Eq.~\eqref{chifunction}) in the case $c_{gh}=1$ and $c_{th}=1$, for various signs of $\gamma_{tot}$. For tachyons, the plots are the same, provided that $E_{\vec{q}}^2>0$ (see Fig.~\ref{fig:Fourier-modes-tachyons} for the case~$E_{\vec{q}}^2<0$). For ghosts, all curves are mirrored with respect to the time axis and no instabilities arise at the level of the propagator. \label{fig:Fourier-modes-particles}}
\end{figure}

In the case of tachyons propagating with $E^2_{\vec{q}}>0$, particles, and ghosts (i.e., for $E_{\vec{q}}^2>0$ and for any $\gamma_{tot}$), or in the case of unstable particles propagating with any energy (i.e., for any sign of $E_{\vec{q}}^2$ and for $\gamma_{tot}\neq0$), the poles $q_{0}^{\pm}$ can be written as
\begin{equation}
\begin{aligned}
q_{0,pole}^{\pm}&=\pm\left(E_{\vec{q}}^{4}+\gamma_{tot}^{2}\right)^{1/4}\left\{ \cos\left[\frac{1}{2}\arg\left(E_{\vec{q}}^{2}-i\gamma_{tot}\right)\right]+i\sin\left[\frac{1}{2}\arg\left(E_{\vec{q}}^{2}-i\gamma_{tot}\right)\right]\right\} \\
&=\pm\left(E_{\vec{q}}^{4}+\gamma_{tot}^{2}\right)^{1/4}\left\{\frac{1-i\mathcal{W}}{\sqrt{1+\mathcal{W}^{2}}} \right\}\,,
\end{aligned}\label{eq:Fpoles}
\end{equation}
where
\begin{equation}
\mathcal{W}=\frac{\gamma_{tot}}{\left(E_{\vec{q}}^{4}+\gamma_{tot}^{2}\right)^{1/2}+E_{\vec{q}}^{2}}\in\mathbb{R}
\end{equation}
has the same sign as $\gamma_{tot}$.
Therefore, in this case (which excludes the case of stable tachyons propagating with $E_{\vec{q}}^2<0$, see below), it is the sign of $\gamma_{tot}=\epsilon+\gamma+\text{Im}(m^{2})$ that determines the position of the poles in the $q_0$-complex plane, independently of its origin ($i\epsilon$-prescription, width of the unstable degree of freedom, imaginary part of the mass square~$m^{2}$ of a complex pole coming with a complex-conjugate partner). In particular:
\begin{itemize}
\item $\gamma_{tot}\geq0$ implies that $q_{0,pole}^{+}$ lies in the forth quadrant and $q_{0,pole}^{-}$ in the second quadrant of the $q_{0}$-complex plane. This is the case of stable particles/tachyons, or standard resonances.
\item $\gamma_{tot}<0$ implies that $q_{0,pole}^{+}$ lies
in the first quadrant and $q_{0,pole}^{-}$ in the third quadrant of the $q_{0}$-complex plane. Since $\epsilon\ll1$, $(\gamma+\mathrm{Im}(m^2))$ should dominate over $\epsilon$ and determines the sign of $\gamma_{tot}$. Therefore, this is the case of unstable degrees of freedom characterized by a negative decay width. 
\item In the case of of one pair complex-conjugate poles, one of the two poles will have~$\gamma_{tot}>0$, while its complex-conjugate partner will have  $\gamma_{tot}<0$. Thus, there will be four poles distributed in all quadrants of the $q_0$-complex plane.
\end{itemize}
At this point, the Fourier modes of the propagator can be computed using:
\begin{equation}
\begin{aligned}
D(x)&=\lim_{\epsilon\to0}\int\frac{d^{3}q}{(2\pi)^{3}}e^{-i\vec{q}\cdot\vec{x}}\int_{-\infty}^{+\infty}\frac{dq_{0}}{2\pi}\,\,D(q^2_0-\vec{q}^2+i\epsilon)\,e^{itq_{0}}=\int\frac{d^{3}q}{(2\pi)^{3}}e^{-i\vec{q}\cdot\vec{x}}\chi(t)\,\,,
\end{aligned}
\end{equation}
where 
\begin{equation}\label{chifunction}
\chi(t)\equiv i \sum_{j\in\mathcal{C}_{+}}\mathcal{R}[D(q^2_{0,j})e^{itq_{0,j}}]\theta(+t)- i \sum_{j\in\mathcal{C}_{-}}\mathcal{R}[D(q^2_{0,j})e^{itq_{0,j}}]\theta(-t) \,\,.
\end{equation}
Here $\mathcal{C}_{+}$ and $\mathcal{C}_{-}$ denote the standard integration contours closed in the upper- and lower-half parts of the complex
plane, respectively, the sums run over the poles $j$ inside $\mathcal{C}_{+}$ and~$\mathcal{C}_{-}$, and $\mathcal{R}[\cdot]$ is the residue of the integrand function evaluated at the pole $j$. Note that the above derivation can only be straightforwardly applied when $D(q^2)$ has \textit{no essential singularities at infinity} (as we assumed). 
Owed to the previous result, and after some manipulation, in the case $\gamma_{tot}\geq0$ one finds
\begin{equation}
\begin{aligned}
\chi(t)&=\frac{-ic_{gh}}{2\sqrt{E_{\vec{q}}^{2}-i\gamma_{tot}}}\left(e^{-i\sqrt{E_{\vec{q}}^{2}-i\gamma_{tot}}\,|t|}\right)=\frac{-ic_{gh}e^{-i \frac{\sqrt{|E_{\vec{q}}^{2}-i \gamma_{tot}|}}{\sqrt{1+\mathcal{W}^{2}}} \,|t|} }{2\sqrt{E_{\vec{q}}^{2}-i\gamma_{tot}}}\,\,e^{-|t|\frac{\sqrt{|E_{\vec{q}}^{2}-i \gamma_{tot}|}}{\sqrt{1+\mathcal{W}^{2}}}\,\mathcal{W}}\,\,,
\end{aligned}
\end{equation}
while in the case $\gamma_{tot}<0$, it is the pole at negative energy that contributes for $t>0$, and the propagator in coordinate space is determined by
\begin{equation}\label{eq:propyfouriermodes}
\chi(t)=\left(ic_{gh}\right)\left({\frac{e^{-i\frac{\sqrt{|E_{\vec{q}}^{2}-i \gamma_{tot}|}}{\sqrt{1+\mathcal{W}^{2}}}\,t}}{2\sqrt{E_{\vec{q}}^{2}-i\gamma_{tot}}}\theta(-t)}+{\frac{e^{+i\frac{\sqrt{|E_{\vec{q}}^{2}-i \gamma_{tot}|}}{\sqrt{1+\mathcal{W}^{2}}}\,t}}{2\sqrt{E_{\vec{q}}^{2}-i\gamma_{tot}}}\theta(t)}\right)e^{+|t|\frac{\sqrt{|E_{\vec{q}}^{2}-i \gamma_{tot}|}}{\sqrt{1+\mathcal{W}^{2}}}\,\mathcal{W}}\,.
\end{equation}
The function $\chi(t)$ is shown in Fig.~\ref{fig:Fourier-modes-particles} for various values of $\gamma_{tot}$.

This result is reminiscent of the way the Feynman prescription is constructed. The position-space Feynman propagator, describing the causal propagation of a particle between two different space points, can be decomposed into a forward- and backward-(on shell) propagating parts. The first is associated with the flow of positive energy, the former that of negative energy. In the case $\gamma_{tot}<0$, modes with negative energies are propagating forward in time, thus entailing a violation of causality on microscopic scales~\cite{Cline:2003gs,Donoghue:2019fcb,Donoghue:2019ecz}. 

\begin{figure}[t]
\centering\includegraphics[scale=0.5]{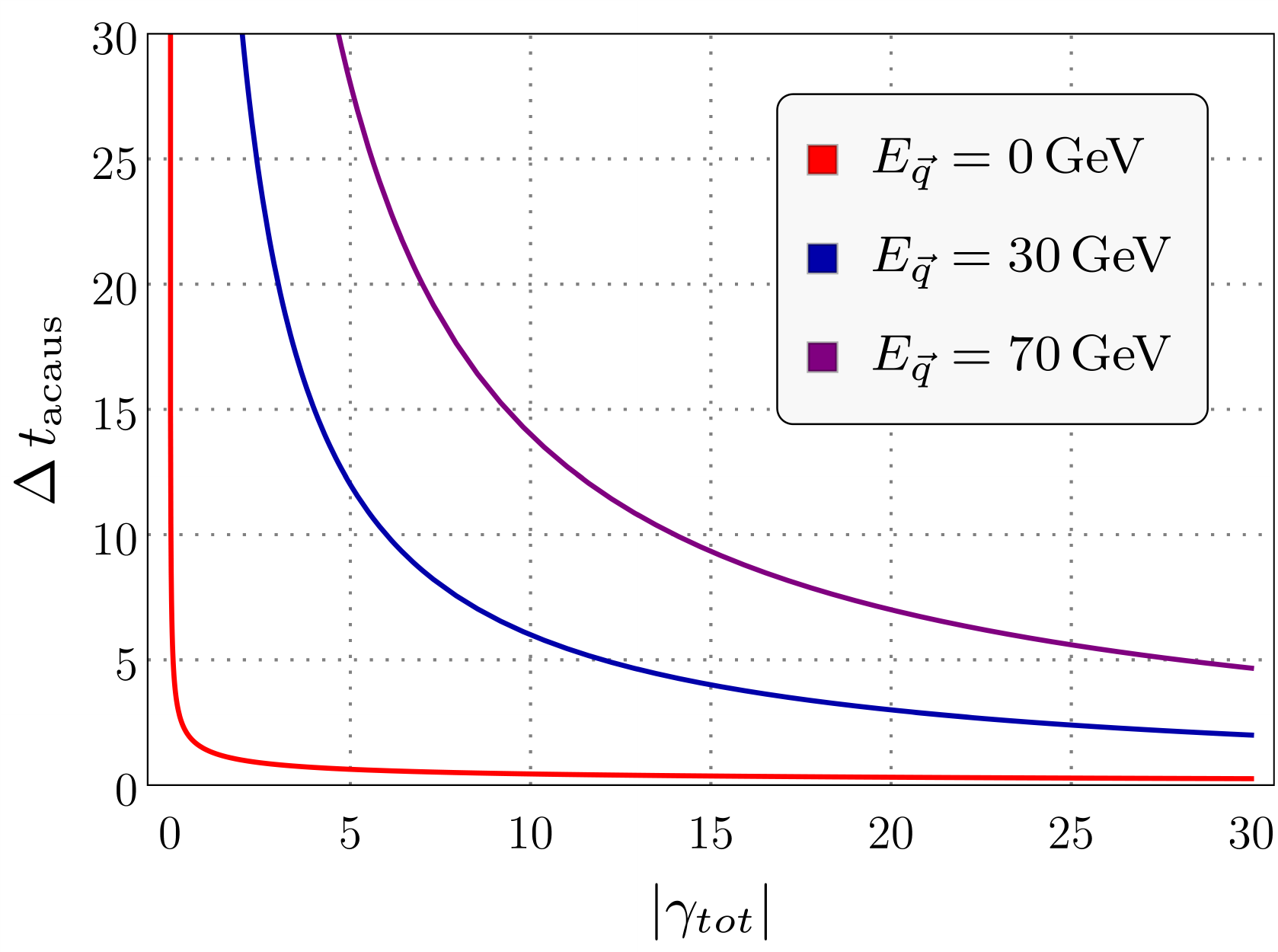}
\caption{Time scale of acausality $\Delta t_{acaus}$ (measured in $\text{GeV}^{-1}$) in the presence of a complex poles $q_{0,pole}^2=E_{\vec{q}}^2-i\gamma_{tot}$ as a function of $|\gamma_{tot}|$ (in units of $\text{GeV}^2$)  for various energies $E_{\vec{q}}$.\label{fig:decaylifetime}}
\end{figure}

The causality violation occurs at energies $E_{\vec{q}}\geq m$ and on time scales comparable with the lifetime of the degree of freedom inducing the violation
\begin{equation}\label{eq:tauviolation}
\Delta t_{acaus}^{-1}\simeq\tau^{-1}=\frac{\sqrt{|E_{\vec{q}}^{2}-i \gamma_{tot}|}}{\sqrt{1+\mathcal{W}^{2}}}\,\mathcal{W}\equiv \frac{\Gamma}{2}\,\,.
\end{equation}
The relation between the scale of acausality and the distance $\gamma_{tot}$ of a complex pole from the real axis is shown in Fig.~\ref{fig:decaylifetime}.
In particular, when $\gamma_{tot}\ll m^2$, our result reduces to that in~\cite{Cline:2003gs,Donoghue:2019fcb,Donoghue:2019ecz}. In fact, expanding about $\gamma_{tot}=0$ yields
\begin{equation}
\Delta t_{acaus}^{-1}\simeq\tau^{-1}\simeq \frac{\gamma_{tot}}{2E_{\vec{q}}}\,,
\end{equation}
which, in the rest frame of a massive particle, reduces to the approximate formula (cf. Eq.~\eqref{eq:gammamassive})
\begin{equation}
\tau^{-1}\simeq\frac{\gamma_{tot}}{2|m|}\equiv \frac{\Gamma}{2}\,.
\end{equation}
These results are independent of the signs of $c_{gh}$ (which would only flip the sign of the corresponding $\chi(t)$-function) and $c_{th}$, and only rely on the sign of $\gamma_{tot}$.

\begin{figure}[t]
	\centering{\includegraphics[scale=0.41]{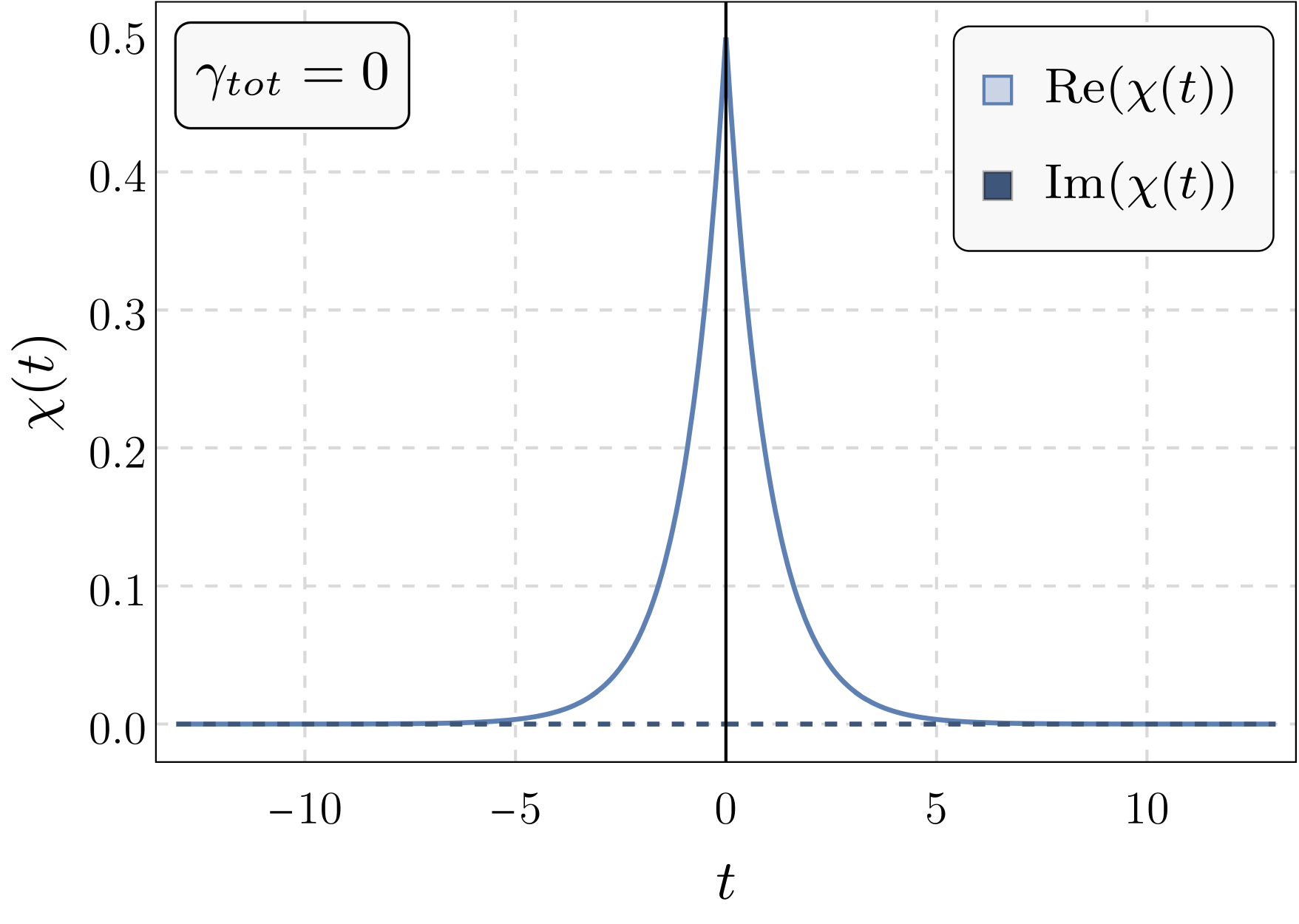}}\\
	\hspace{-0.1cm}\includegraphics[scale=0.41]{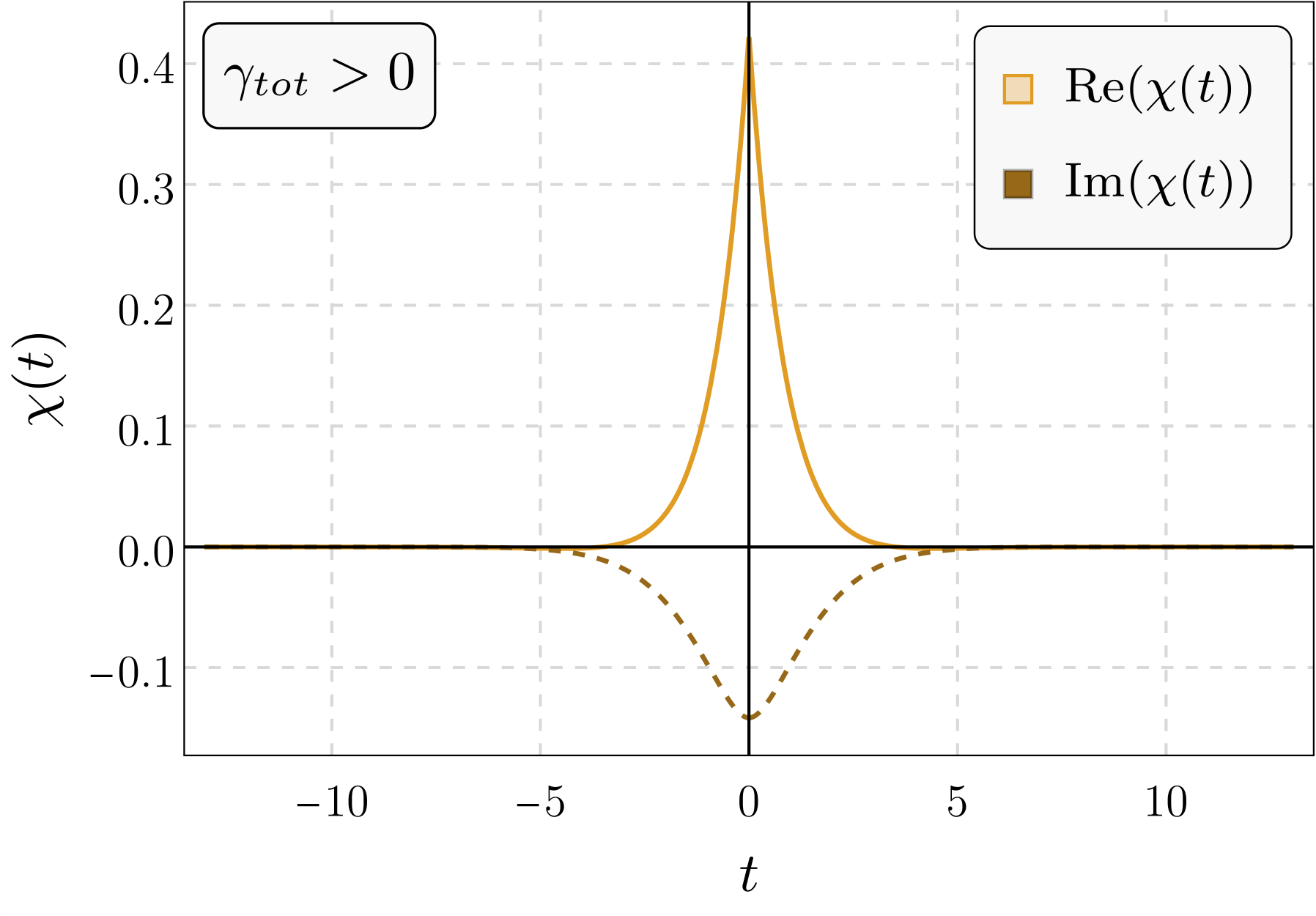}$\,\,$\includegraphics[scale=0.41]{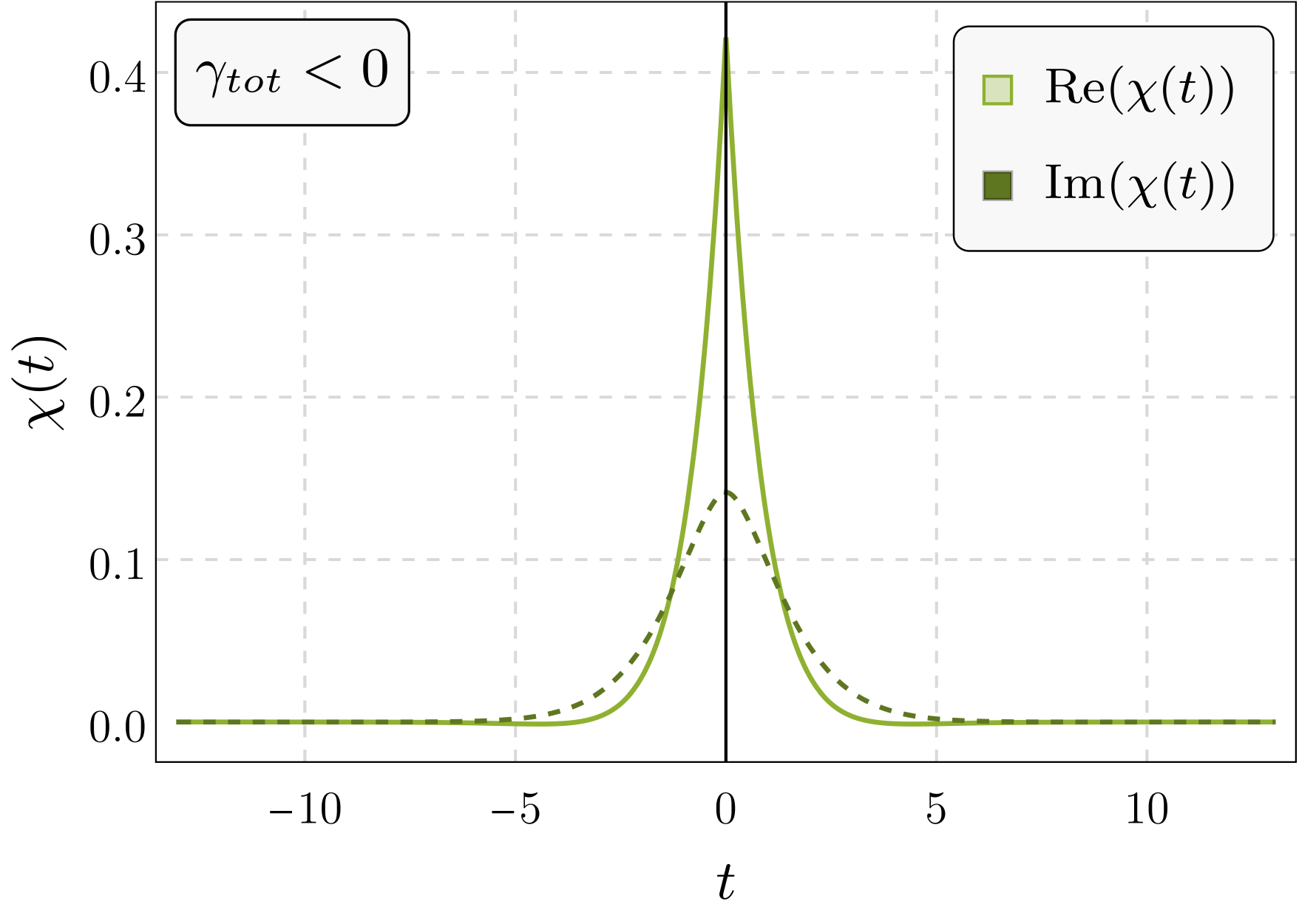}\\
	\hspace{-0.1cm}\includegraphics[scale=0.41]{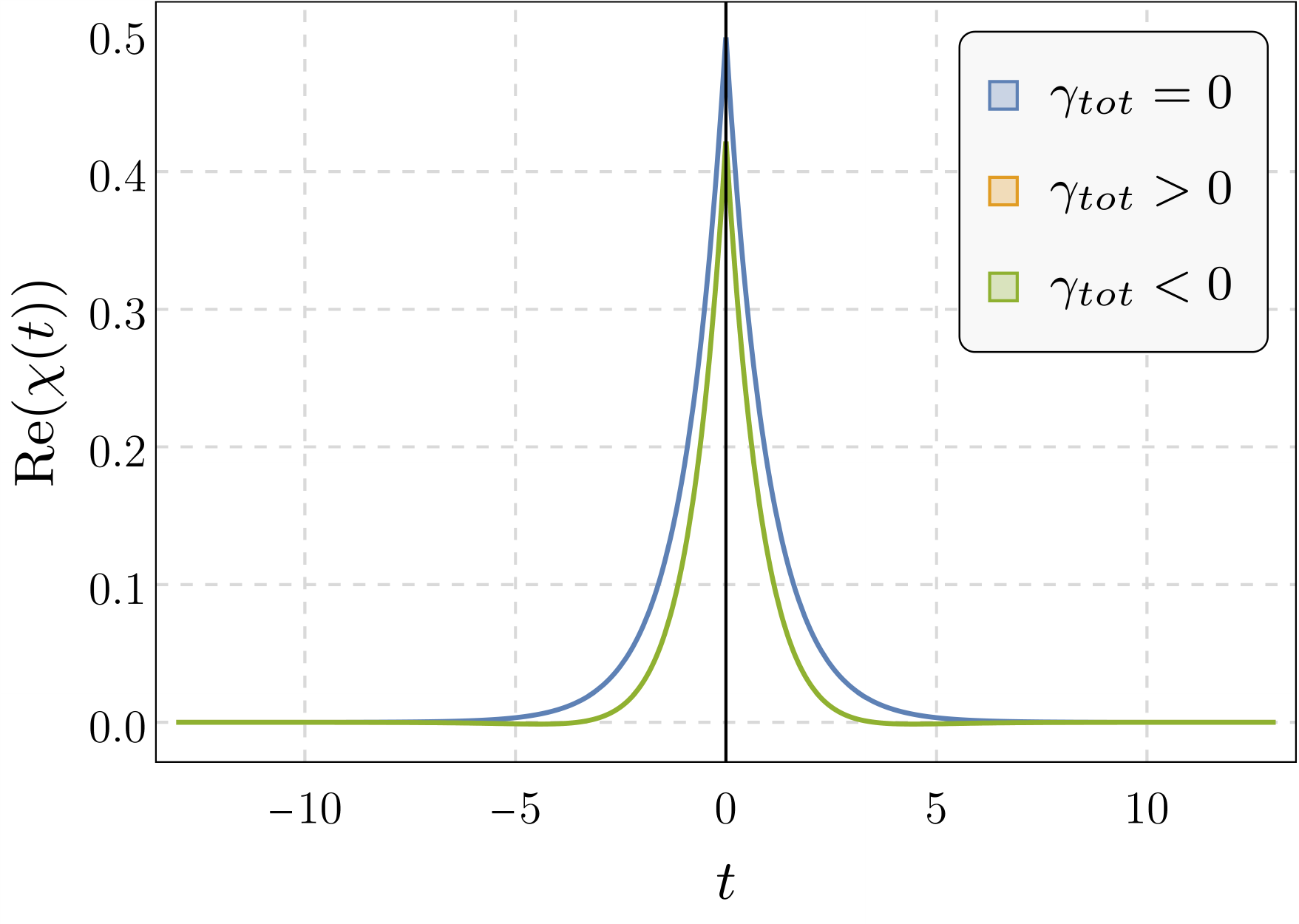}$\,\,$\includegraphics[scale=0.41]{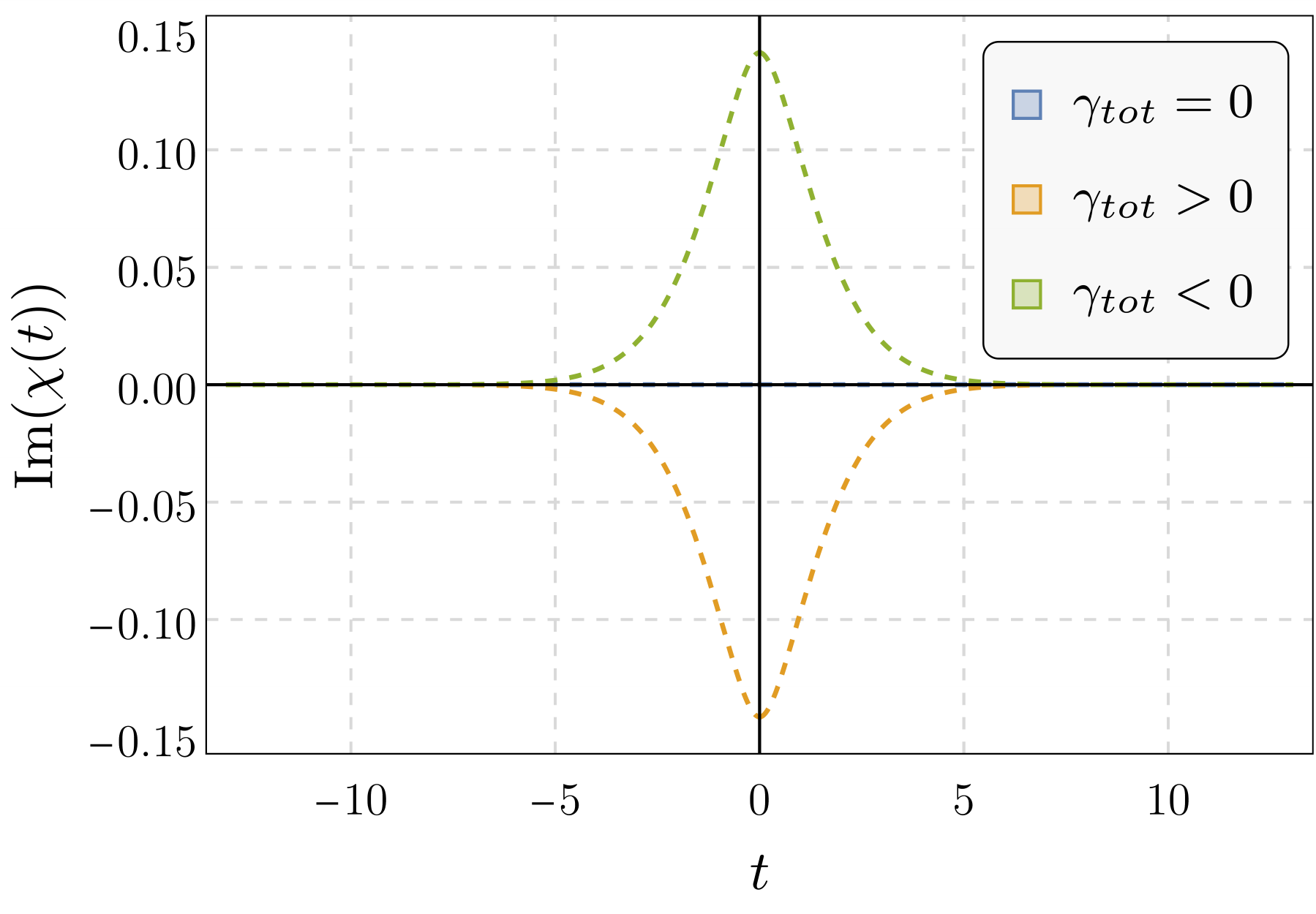}
	\caption{Fourier modes of the propagator (function $\chi(t)$ in Eq.~\eqref{chifunction}) in the case $c_{gh}=1$, $c_{th}=-1$, and $E_{\vec{q}}^2<0$, for various signs of $\gamma_{tot}$. \label{fig:Fourier-modes-tachyons}}
\end{figure}

Let us now analyze the case of stable ($\gamma=0$) tachyons propagating with $E_{\vec{q}}^{2}<0$ (subluminal tachyons). 
Since the poles do not lie on the real axis, no Feynman prescription is needed in this case and thus $\gamma_{tot}=0$ for stable tachyons\footnote{Utilizing the $i\epsilon$-prescription and following the standard procedure above would not change the final result.}. Therefore, the argument function in Eq.~\eqref{eq:Fpoles} would simply be $\pi$ in the case $E_{\vec{q}}^{2}<0$ and thus the first line of Eq.~\eqref{eq:Fpoles} yields the poles $q_{0,pole}^\pm=\pm i|E_{\vec{q}}|$. The corresponding $\chi(t)$ reads
\begin{equation}\label{eq:tachyonsmodes}
	\begin{aligned}
		\chi(t)=\frac{1}{2|E_{\vec{q}}|}\left(e^{+|E_{\vec{q}}|t}\theta(-t)+e^{-|E_{\vec{q}}|t}\theta(t)\right)\,\,,
	\end{aligned}
\end{equation}
Thus, accounting for all expressions of $\chi(t)$ above and their validity ranges, we conclude that the behavior of tachyons propagating with $E_{\vec{q}}^{2}>0$ is oscillatory as in the case of standard particles (as expected), while in regimes where tachyons propagate with $E_{\vec{q}}^{2}<0$, the Fourier modes of the propagator are exponentially decaying (cf. Fig.~\ref{fig:Fourier-modes-tachyons}). Note that using the plane-wave expansion~\eqref{eq:freefield} to construct the propagator as a time-ordered product would yield, in the case of stable tachyons, a quite different result: the signs of the exponentials in Eq.~\eqref{eq:tachyonsmodes} would have been swapped and the propagator would have been an exponentially growing function of time. Once again, this apparent inconsistency comes from the ambiguity in the quantization procedure of tachyons (cf. Sect.~\ref{sect:ambitachy}) and the fact that the free field expansion~\eqref{eq:freefield} cannot be applied in this case. Regardless of the resolution of the ambiguities in the (canonical) quantization procedure of tachyons, which to our best knowledge is an open problem, our result~\eqref{eq:tachyonsmodes} only relies on the assumption that free tachyons propagate with a standard $(q^2+c_{th}m^2)^{-1}$ propagator, which seems to be plausible. Note that here the $i\epsilon$-prescription does not play a role, since in the case of stable tachyons with $E_{\vec{q}}^2<0$ the poles are located at $q_{0,pole}^{\pm}\simeq i|E_{\vec{q}}|\mp \epsilon/2|E_{\vec{q}}|$, with $|E_{\vec{q}}|>0$. Thus, $\epsilon$ can  only shift the poles ``horizontally'', and its sign cannot change which pole contributes to Eq.~\eqref{chifunction} when closing the contour for $t>0$ or $t<0$. The absence of exponentially growing modes in the propagator of a free tachyon with $E_{\vec{q}}^2<0$ should not be surprising: even if exponentially growing modes appear in the solutions to the quantum field equations, due to exponentials like $e^{at}$ and $e^{-at}$ with $a>0$, the time ordering in the definition of the Feynman propagator enforces the appearance of theta functions such that only the combinations $\theta(\pm t)e^{\mp a t}$ can arise in its Fourier modes. This does not imply that the instabilities are removed from the theory.

Summarizing, no tachyonic instability (exponentially growing modes) arises at the level of the propagators (cf. Fig.~\ref{fig:Fourier-modes-particles}), no matter what the signs of $c_{gh}$, $c_{th}$ and~$\gamma_{tot}$ are. Such instabilities may only arise at the level of the solutions to the quantum field equations stemming from the corresponding effective action.
Some final remarks and observations are in order:
\begin{itemize}
	\item \textit{Causality and Wick rotation}: Since violations of causality arise any time complex poles appear in the first and/or third quadrants of the $q_0$-complex plane, causality could also be related to the possibility of performing an analytic Wick rotation connecting the Euclidean and Lorentzian theories. However, while a violation of causality implies the impossibility of defining an analytic continuation, the absence of complex poles is not enough to guarantee an analytic Wick rotation. In order to perform an analytic continuation, no essential singularities should occur at infinity. 
	\item \textit{Field redefinitions}: Since complex-conjugate poles can only appear in loops (in principle they do not contribute to the spectral density, even though a modified K\"{a}llen-Lehmann representation accounting for complex-conjugate poles could change this conclusion~\cite{Hayashi:2018giz,Kondo:2019ywt}) one could think of performing a field redefinition to remove them, indicating that the theory be consistent and causal. However, to ensure invariance of scattering amplitudes according to the equivalence theorem, field redefinitions should not change the number of complex and real poles of the propagator~\cite{Veltman:1963th,tHooft:1973wag,Vilkovisky:1984st}. Moreover, as we mentioned already, violations of microcausality due to the exchange of virtual particles could potentially leave detectable signatures, indicating that these poles are physical, even if they do not correspond to asymptotic states.
	\item \textit{Tachyonic modes with $\gamma_{tot}<0$}: Tachyons with negative width can potentially violate all physical principles on the market. There could be a violation of unitarity, since $c_{gh}\gamma_{tot}<0$. There would be a (microscopic) violation of causality and vacuum instabilities might occur, because $\gamma_{tot}<0$. And there would be a tachyonic instability, as $c_{th}<0$. However, the latter instability might not be a serious problem, as we comment on in the next point. 
	\item \textit{Tachyonic modes with $\gamma_{tot}\geq0$, causality and tachyonic instability}:
	Tachyonic excitations can propagate subluminally, even if the group velocity can be superluminal~\cite{Feinberg:1967zza,Csonka:1970az}. Thus, tachyons can be compatible with Einsten's locality and thereby with microcausality~\cite{Aharonov:1969vu}. In particular, the standard commutation relations between space-like-separated events are not affected by the sign of $c_{th}m^{2}$. As we have seen in this section, as long as $\gamma_{tot}\geq0$, tachyons are also compatible with causality, in the sense of forward propagation at the level of the propagator, and with (vacuum) stability. The only problem related to the existence of tachyons is the (tachyonic) instability occurring for $E_{\vec{q}}^2<0$: even if no instability arises at the level of the propagator (which is actually decaying), exponentially growing modes could potentially arise in the solutions to the quantum field equations (cf. Sect.~\ref{subsect:instabsdefs}). However, the negative mass square, $c_{th}m^{2}<0$ that generates the instability could indicate that the theory is quantized on a wrong, unstable vacuum (a maximum instead of a minimum) and that this configuration is unstable. Interactions and non-perturbative effects could trigger a tachyonic condensation, e.g., a spontaneous symmetry breaking~\cite{Sen:2002an}, as it happens in the case of the Higgs field. Therefore, the existence of tachyonic instabilities could be cured by a suitable  condensation mechanism occurring fast enough to counterbalance the instability, provided that a stable vacuum exists in the theory. In turn, the existence of such a stable vacuum is tied to the presence of interaction terms in the effective action,  cf.~Sect.~\ref{sect:interactionssavetachyons}.
	\item \textit{Causality violation and vacuum instability in the case $\gamma_{tot}<0$}: While the propagation does not display any tachyonic instability and the microscopic violation of causality might still be compatible with observations, the fact that the forward-propagating degrees of freedom with $\gamma_{tot}<0$ carry negative energy, implies that the vacuum is unstable~\cite{Sbisa:2014pzo}: the decay of the vacuum in modes carrying opposite energies (and opposite signs of~$\gamma_{tot}$) is entropically favored, unless a microscopic breaking of locality or Lorentz invariance takes place~\cite{Cline:2003gs}. However in this latter case a careful fine-tuning of the Lorentz breaking at high energies would be required to avoid incompatibility with observations~\cite{Burgess:2002tb,Eichhorn:2019ybe}. It is worth mentioning that unstable particles with negative width and complex-conjugate poles are not part of the spectrum of asymptotic states, and would decay after a certain lifetime, Eq.~\eqref{eq:tauviolation}. Nonetheless, conservation of energy would require the decay products (the type of allowed decay products depend on the interaction vertices of the full theory, and can be either stable or unstable) to have negative energies. This could potentially lead to a cascade of decays involving degrees of freedom with both positive and negative energies, and it is not clear whether a stabilization mechanism exists. In the absence of this mechanism, we will exclude poles with a negative width and complex-conjugate poles from the set of acceptable poles of a physical propagator.
\end{itemize}
Thus, summarizing: preserving unitarity requires $c_{gh}\gamma_{tot}>0$, avoiding tachyonic instabilities (no runaway solutions) requires $c_{th}>0$, preserving causality (in the sense of forward propagation) and vacuum stability requires $\gamma_{tot}\geq0$. Table~\ref{tab:polesconditions} summarizes what type of poles satisfy these conditions.\\
\begin{table}[t!]
	\begin{center}
\begin{tabular}{|c|c|p{1.3cm}|c|c|c|c|}
	\hline 
	\multicolumn{2}{|c|}{Pole} & Width & Causality & Unitarity & \parbox{1.5cm}{\centering Vacuum \\ Stability} & \parbox{1.5cm}{\centering Tachyonic \\ Stability}\tabularnewline
	\hline 
	\hline 
	\multirow{3}{*}{
			\parbox{3.6cm}{\centering Particle \\ ($c_{gh}=1$, $c_{th}=1$)}} & Stab. & $\gamma_{tot}=0$ & $\checkmark$ & $\checkmark$ & $\checkmark$ & $\checkmark$\tabularnewline
	\cline{2-7} \cline{3-7} \cline{4-7} \cline{5-7} \cline{6-7} \cline{7-7} 
	& \multirow{2}{*}{Unst.} & $\gamma_{tot}>0$ & $\checkmark$ & $\checkmark$ & $\checkmark$ & $\checkmark$\tabularnewline
	\cline{3-7} \cline{4-7} \cline{5-7} \cline{6-7} \cline{7-7} 
	&  & $\gamma_{tot}<0$ & $\times$ & $\times$ & $\times$ & $\checkmark$\tabularnewline
	\hline 
	\hline 
	\multirow{3}{*}{\parbox{3.6cm}{\centering Ghost \\ ($c_{gh}=-1$, $c_{th}=1$)}} & Stab. & $\gamma_{tot}=0$ & $\checkmark$ & $\times$ & $\checkmark$ & $\checkmark$\tabularnewline
	\cline{2-7} \cline{3-7} \cline{4-7} \cline{5-7} \cline{6-7} \cline{7-7} 
	& \multirow{2}{*}{Unst.} & $\gamma_{tot}>0$ & $\checkmark$ & $\times$ & $\checkmark$ & $\checkmark$\tabularnewline
	\cline{3-7} \cline{4-7} \cline{5-7} \cline{6-7} \cline{7-7} 
	&  & $\gamma_{tot}<0$ & $\times$ & $\checkmark$ & $\times$ & $\checkmark$\tabularnewline
	\hline 
	\hline 
	\multirow{3}{*}{\parbox{3.6cm}{\centering Tachyon \\ ($c_{gh}=1$, $c_{th}=-1$)}} & Stab. & $\gamma_{tot}=0$ & $\checkmark$ & $\checkmark$ & $\checkmark$ & $\times$\tabularnewline
	\cline{2-7} \cline{3-7} \cline{4-7} \cline{5-7} \cline{6-7} \cline{7-7} 
	& \multirow{2}{*}{Unst.} & $\gamma_{tot}>0$ & $\checkmark$ & $\checkmark$ & $\checkmark$ & $\times$\tabularnewline
	\cline{3-7} \cline{4-7} \cline{5-7} \cline{6-7} \cline{7-7} 
	&  & $\gamma_{tot}<0$ & $\times$ & $\times$ & $\times$ & $\times$\tabularnewline
	\hline 
	\hline 
	\multirow{3}{*}{\parbox{3.6cm}{\centering Tachyonic Ghost \\ ($c_{gh}=-1$, $c_{th}=-1$)}} & Stab. & $\gamma_{tot}=0$ & $\checkmark$ & $\times$ & $\checkmark$ & $\times$\tabularnewline
	\cline{2-7} \cline{3-7} \cline{4-7} \cline{5-7} \cline{6-7} \cline{7-7} 
	& \multirow{2}{*}{Unst.} & $\gamma_{tot}>0$ & $\checkmark$ & $\times$ & $\checkmark$ & $\times$\tabularnewline
	\cline{3-7} \cline{4-7} \cline{5-7} \cline{6-7} \cline{7-7} 
	&  & $\gamma_{tot}<0$ & $\times$ & $\checkmark$ & $\times$ & $\times$\tabularnewline
	\hline 
	\hline 
	\multicolumn{2}{|c|}{\parbox{4.5cm}{\centering Complex-conjugate poles \\($c_{gh}=1$, $c_{th}=1$)}} & \parbox{1.5cm}{ $\gamma_{tot}>0$ \\ $\gamma_{tot}<0$} & $\times$ & $\checkmark$ & $\times$ & $\checkmark$\tabularnewline
	\hline 
\end{tabular}\caption{Summary of poles of the dressed propagator satisfying ($\checkmark$) or violating ($\times$) the conditions of causality (no backward propagation of modes with positive energies), unitarity (no negative-norm states), vacuum stability (no modes with negative energy coupled with modes with positive energies), and tachyonic stability (no runaways in the free-field expansion).  \label{tab:polesconditions}}
\end{center}
\end{table}

\noindent\fbox{\parbox{\linewidth}{Based on the considerations of this section, \textit{to preserve unitarity, causality along all scales and avoid vacuum instabilities, the dressed (graviton) propagator should not have complex poles with negative width}. Additionally, avoiding runaways at the level of the free-field expansion requires the absence of tachyonic poles.}}

\section{Unitarity, causality, and stability: logarithmic interactions}\label{sect:complexpoles}

In this section we shall investigate the pole structure of  propagators constructed with logarithmic interactions. This class of propagators plays an important role in QFT, as logarithmic form factors generally arise as one-loop corrections to the tree-level action. Examples of such propagators are those derived from Lee-Wick QED~\cite{Lee:1971ix} and one-loop QG~\cite{Donoghue:2015nba,Donoghue:2015xla}. Their scalar part read
\begin{equation}
D_{LW-QED}(q^{2})=q^{-2}\left(1-\frac{\alpha}{3\pi}\log\left(1-\frac{q^{2}}{m_{th}^{2}}\right)-\frac{q^{2}}{M^{2}}\right)^{-1}\,,\label{eq:propQED}
\end{equation}
\begin{equation}
D_{QG}(q^{2})=q^{-2}\left(1-\frac{q^{2}G}{\pi}\log\left(1-\frac{q^{2}}{m_{th}^{2}}\right)-\frac{q^{2}}{M^{2}}\right)^{-1}\,.\label{eq:propQG}
\end{equation}
In both cases there is a massless pole, $q^{2}_{pole}=0$, describing  the photon and the graviton, respectively. Note that in the argument of the logarithm the constant term is zero in many physically-relevant models; nonetheless, we will keep it non-zero to render the analysis more general. One can easily check that our conclusions remain unchanged. 

In what follows we shall study the existence of additional real or complex poles for this class of propagators. This boils down to studying the zeros of the $P$-functions
\begin{equation}\label{Pqed}
P_{LW-QED}(z)=1-a\log\left(1-z\right)-bz \,,
\end{equation}
\begin{equation}\label{Pqg}
P_{QG}(z)=1-a\,z\log\left(1-z\right)-bz \,,
\end{equation}
with $z=q^{2}/m_{th}^{2}\equiv x+iy$,
in terms of the parameters $a$ and $b$. The physical cases of Lee-Wick QED and one-loop QG are obtained by restricting to
\begin{equation}
	b=\frac{m_{th}^{2}}{M^{2}}\leq1\,,
\end{equation}
and fixing 
\begin{equation}
a_{LW-QED}=\frac{\alpha}{3\pi}\ll1\,,
\end{equation}
\begin{equation}
a_{QG}=\frac{G_N m_{th}^{2}}{\pi}=\frac{m_{th}^{2}}{\pi M_{Pl}^{2}}\ll1\,,
\end{equation}
in accordance with the corresponding perturbative computations~\cite{Lee:1971ix,Donoghue:2015nba,Donoghue:2015xla}.
We will show that, beyond the massless pole at $q^2=0$, propagators of the type above (with the full, analytically-continued, complex logarithm) have either additional real poles (describing stable, possibly tachyonic, degrees of freedom) or a pair of complex-conjugate poles (leading at least to a violation of causality). Moreover, an unstable ghost appears due to the presence of a branch cut singularity.

\subsection{Pole structure: the case of Lee-Wick QED}

To study the zeros of the $P$-functions under consideration, we first note that the principal branch of the complex logarithm can be conveniently expanded as
\begin{equation}
\log(1-x-iy)=\log\left[\left((1-x)^{2}+y^{2}\right)^{1/2}\right]+i\,\text{atan2}(1-x-iy)\,\,.
\end{equation}
This expression comes from the use of the representation $w=|w|e^{i\text{Arg}(w)}$, with $w=1-x-iy$. The single-valued function $\text{Arg}(w):\mathbb{C}\to(-\pi,\pi]$ is the principal argument of the the complex number $w$ and can be written as
\begin{align}\label{eq:argdec}
\text{Arg}(w) & =-i\log\left(\frac{x+iy}{|x+iy|}\right)\equiv\text{atan2}(x+iy)=\begin{cases}
\arctan\left(\frac{y}{x}\right)\,, & \text{if }\ensuremath{x>0}\,,\\
\arctan\left(\frac{y}{x}\right)+\pi\,, & \text{if }\ensuremath{x<0}\text{ and }\ensuremath{y\geq0}\,,\\
\arctan\left(\frac{y}{x}\right)-\pi\,, & \text{if }\ensuremath{x<0}\text{ and }\ensuremath{y<0}\,,\\
+\pi/2\,, & \text{if }\ensuremath{x=0}\text{ and }\ensuremath{y>0}\,,\\
-\pi/2\,, & \text{if }\ensuremath{x=0}\text{ and }\ensuremath{y<0\,,}\\
\text{undefined}\,, & \text{if \ensuremath{x=y=0}\,.}
\end{cases}
\end{align}
It is important to notice that this function is odd with respect to
$y$, while for any $x<0$ it is discontinuous in $y=0$. The function $\text{Arg}(w)$ thus has a branch cut along the negative real axis, i.e., at $y=0$ and $x<0$\footnote{The location of the branch cut is strictly related to the convention that $\text{Arg}(w):\mathbb{C}\to(-\pi,\pi]$. This can however be changed, as we show in Sect.~\ref{sect:goodpropy}.}. Crossing the branch cut, its value jumps from $-\pi$ (excluded, as it belongs to another branch) to $\pi$. Note that the existences of a zero on one Riemann sheet  of a multi-valued function also implies the presence of ``shadow zeros'' on all other sheets below the one where the zero occurs for the first time.

Using Eq.~\eqref{eq:argdec}, the real and imaginary parts of the function $P_{QED}(z)$ in Eq.~\eqref{Pqed} can be written as
\begin{align}
&\mathrm{Re}\left[P_{LW-QED}\right]=1-bx-\frac{a}{2}\log\left((1-x)^{2}+y^{2}\right) \,\,, \nonumber \\
&\mathrm{Im}\left[P_{LW-QED}\right]=-by-a\,\text{atan2}(1-x-iy) \,\,.
\end{align}
Since the function $\mathrm{atan2}$ is odd with respect to $y$, it follows that
\begin{itemize}
\item $A(x,y)=\mathrm{Re}\left[P_{LW-QED}\right]$ is even with respect to $y$, and
\item $B(x,y)=\mathrm{Im}\left[P_{LW-QED}\right]$ is odd with respect to $y$,
\end{itemize}
Therefore, if $z=(x_{0},y_{0})$ is a solution to $P_{QED}=0$ (making $A=B=0$), also its complex conjugate $\bar{z}=(x_{0},-y_{0})$ is a solution. This observation is at the core of the complex conjugate root theorem. The theorem states that if $f(x)$ is a polynomial function with real number coefficients and $z$ is a complex zero of $f$, then so is its complex conjugate $\bar{z}$. Roughly, the theorem comes from the fact that replacing $z=x+iy$ into $f(z)$, {its real and imaginary part are either even or odd in $y$}.

\begin{figure}
\begin{center}
\includegraphics[scale=0.65]{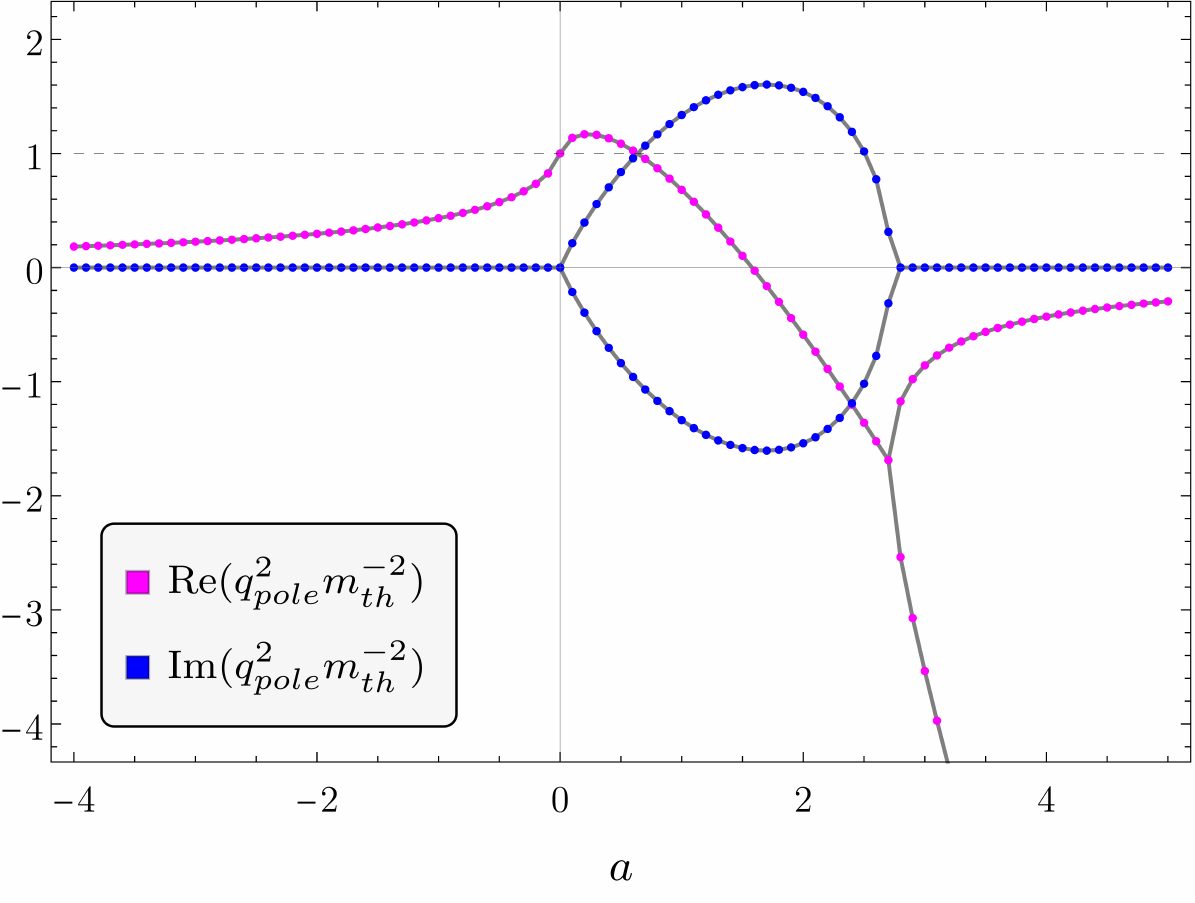}
\caption{Pole structure of the Lee-Wick $P$-function in Eq.~\eqref{Pqed} as a function of $a$, and for $b=1$. The figure shows the real (magenta dots) and imaginary (blue dots) part of the zeros of  $P_{LW-QED}$. Beyond the massless pole, the propagator has a stable ghost for $a<0$, while for $a>e$ there are two tachyonic ghosts. For $0<a<e$, the massive real pole splits into a pair of complex-conjugate poles. Since $a_{LW-QED}\ll1$, in Lee-Wick QED there is a pair of complex-conjugate poles. \label{polestructureLWQED}}
\end{center}
\end{figure}

Let us now determine explicitly the zeros of $P_{QED}$, since this gives us information on the additional degrees of freedom of the theory beyond the massless one. The real part of $P_{QED}$ is zero for
\begin{equation}
y_{0}=\pm\sqrt{\exp\left(-\frac{2(b\,x_{0}-1)}{a}\right)-1+2x_{0}-x_{0}^{2}}\,.
\end{equation}
We first look for real solutions, setting $y_{0}=0$.
In the case $b=1$, i.e., when $M^{2}=m_{th}^{2}$, these solutions can be found analytically. The condition $\mathrm{Im}\left[P_{QED}\right]=0$ is compatible with the expression of $y_{0}$ if $a=0$ or $x_{0}<1$. Real zeros of $P_{LW-QED}$ are thus found solving
\begin{equation}
\sqrt{\exp\left(-\frac{2(x_{0}-1)}{a}\right)-1+2x_{0}-x_{0}^{2}}=0\,\qquad x_{0}<1\,\,.
\end{equation}
This gives the solutions:
\begin{equation}
x_{0}=\begin{cases}
1+b\,\text{ProductLog}[-1/a] & \text{if }b<0\,||\,a\geq e\,,\\
1+b\,\text{ProductLog}[-1,-1/a] & \text{if }a\geq e \,,
\end{cases}
\end{equation}
where ``$e$'' is Euler's number.
In addition, one can easily verify that for $0\leq b<e$ there is a pair of complex-conjugate zeros.
Therefore, for $b=1$ the LW-QED propagator has the following additional poles
\begin{equation}
\begin{cases}
\text{one stable ghost} & \text{for}\quad  a<0\,,\\
\text{one pair of complex-conjugate poles} & \text{for}\quad  0\leq a<e\,,\\
\text{two stable tachyonic degrees of freedom} & \text{for}\quad a\geq e\,.
\end{cases}
\end{equation}
The pole structure of the propagator with the $P$-function given by Eq.~\eqref{Pqed} is shown in Fig.~\ref{polestructureLWQED} for the case $b=1$. The case $b\neq1$ can be studied numerically, and it can be seen that the pole structure is qualitatively the same. The value of $b$ only changes the upper bound of the region $[0,e)$ where $P_{LW-QED}$ has a pair of complex-conjugate zeros.

\subsection{Pole structure: the case of one-loop QG}

\begin{figure}
\begin{center}
\includegraphics[scale=0.65]{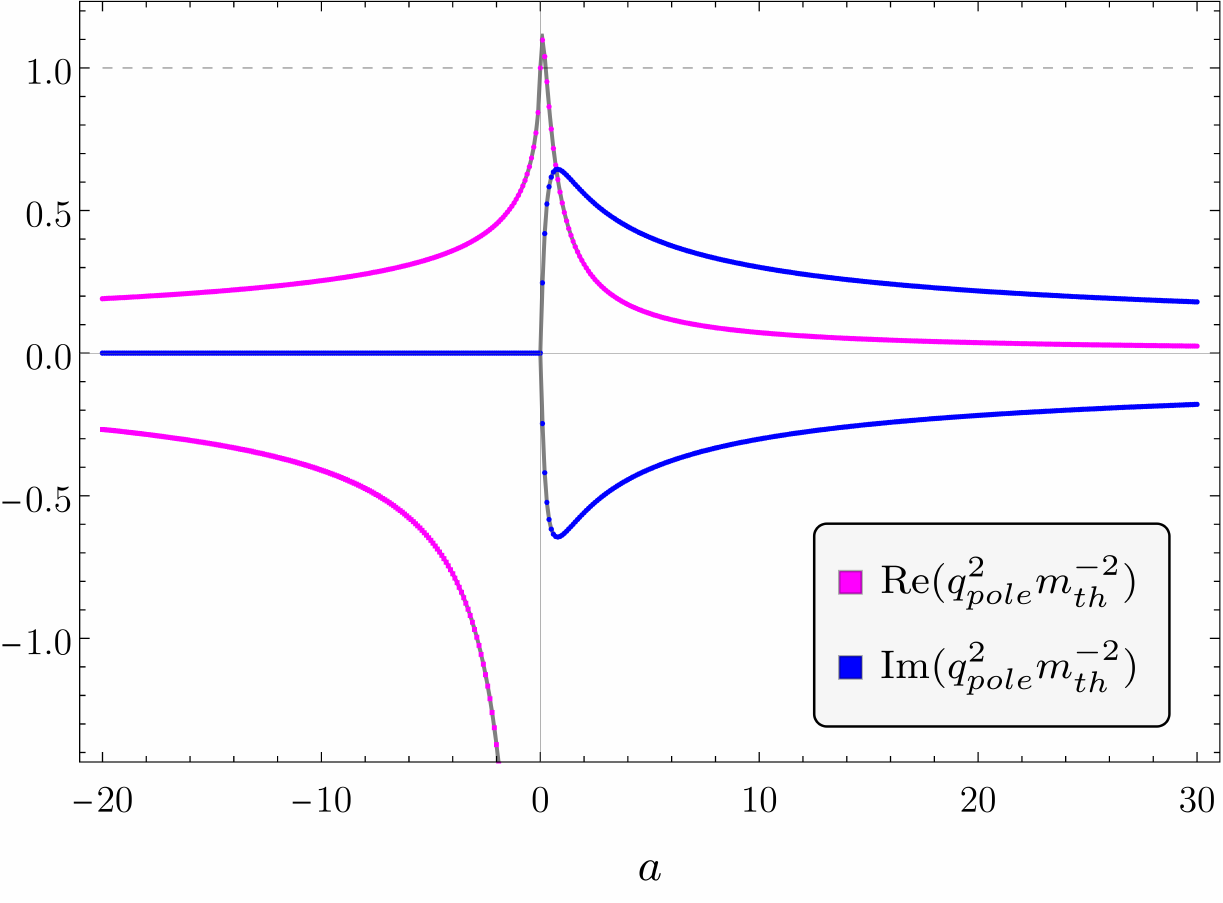}\caption{Pole structure of the one-loop graviton propagator as a function of $a$, and  for $b=1$. When the coupling $a$ is negative, the propagator displays one stable ghost pole and one stable tachyonic ghost pole. The latter approaches infinity when $a\to0$. The former splits into a pair of complex-conjugate poles when $a$ becomes positive. \label{PolestructureQG}}
\end{center}
\end{figure}

We now analyse the pole structure of the flat one-loop graviton propagator. Restricting again to the principal branch of the complex logarithm, one can isolate the real and imaginary parts of~$P_{QG}(z)$, which read
\begin{align}
&\mathrm{Re}\left[P_{QG}\right]=1-bx+ay\,\text{atan2}(1-x-iy)-\frac{ax}{2}\log\left((1-x)^{2}+y^{2}\right) \,\,, \nonumber \\
& \mathrm{Im}\left[P_{QG}\right]=-by-ax\,\text{atan2}(1-x-iy)-\frac{ay}{2}\log\left((1-x)^{2}+y^{2}\right) \,\,.
\end{align}
The real and imaginary parts of $P_{QG}(z)$ have the same key properties as those of $P_{LW-QED}(z)$: $A(x,y)=\mathrm{Re}\left[P_{QG}\right]$ is even with respect to $y$ and $B(x,y)=\mathrm{Im}\left[P_{QG}\right]$ is odd with respect to~$y$. Therefore, as can be explicitly checked, also in the case of one-loop QG the propagator can have either real or complex-conjugate poles. In the case of QG, the additional $q^{2}$-term in front of the logarithm makes the propagator more involved and it is not possible to 
look for its poles analytically. However, it is still possible to search 
the zeros of the $P_{QG}$-function numerically, by looking for points
$(x_{0},y_{0})\in\mathbb{C}$ such that $A(x_{0},y_{0})=B(x_{0},y_{0})=0$.
In the case $b=1$, as shown in Fig.~\ref{PolestructureQG}, one-loop QG is characterized by
\begin{equation}
\begin{cases}
\text{two stable degrees of freedom (one of them tachyonic)} & \text{for}\quad  a<0\,,\\
\text{one pair of complex-conjugate poles} & \text{for}\quad  a\geq0\,.
\end{cases}
\end{equation}
As in QG the parameter $a$ is $a=a_{QG}\ll1$, the one-loop dressed graviton propagator has a pair of complex-conjugate poles on the first Riemann sheet (for the specific case $a=a_{QG}$ this was already noted in~\cite{Tomboulis:1977jk,Hasslacher:1980hd,Modesto:2015ozb}). 
Finally we remark that in Euclidean signature the logarithm takes the form
\begin{equation}
\begin{aligned}
\log(1-q_{L}^{2}/m_{th}^{2})=&\log(1-(q_{0,L}^{2}-\vec{q}^{2})/m_{th}^{2})\to\\
\to &\log(1-(-q_{0,E}^{2}-\vec{q}^{2})/m_{th}^{2})=\log(1+q_{E}^{2}/m_{th}^{2})\,,
\end{aligned}
\end{equation}
with $q_{E}^{2}\in[0,+\infty)$. Thus, provided that the mass $m_{th}$ results from the integration of non-tachyonic fields, the Euclidean version of the propagator can only have real poles~\cite{Tomboulis:1977jk,Antoniadis:1986tu}.

\subsection{Complex poles, branch cut region and ghost resonance}

We have seen that form factors involving sums of polynomial and logarithmic interactions lead to dressed propagators having real or complex-conjugate poles. This result is expected to hold more generally, based on the complex conjugate root theorem and the observation that the Taylor expansion of combinations of real polynomial and logarithmic terms is still a polynomial with real coefficients. 

\begin{figure}
\begin{center}
\includegraphics[scale=0.48]{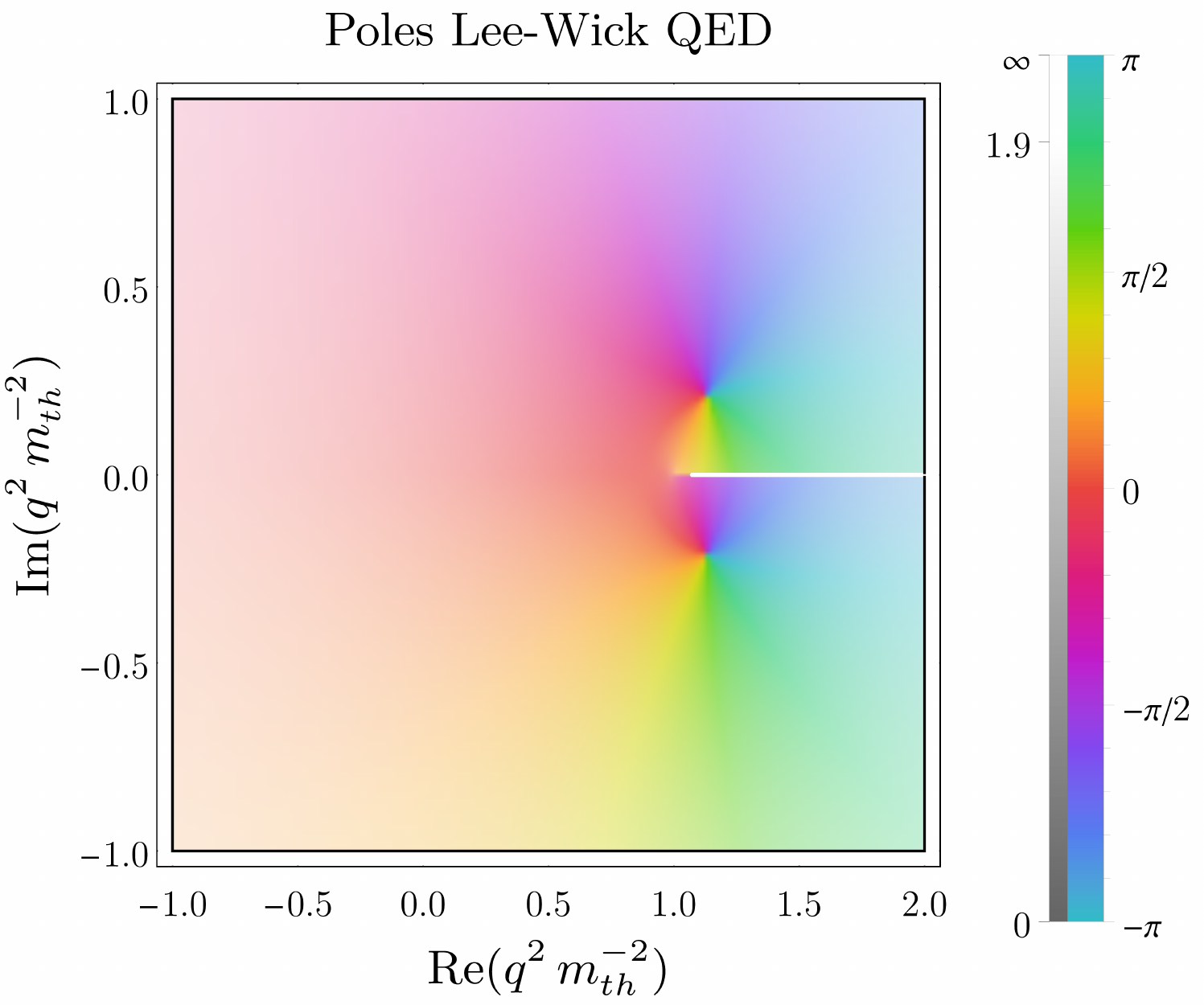}\includegraphics[scale=0.48]{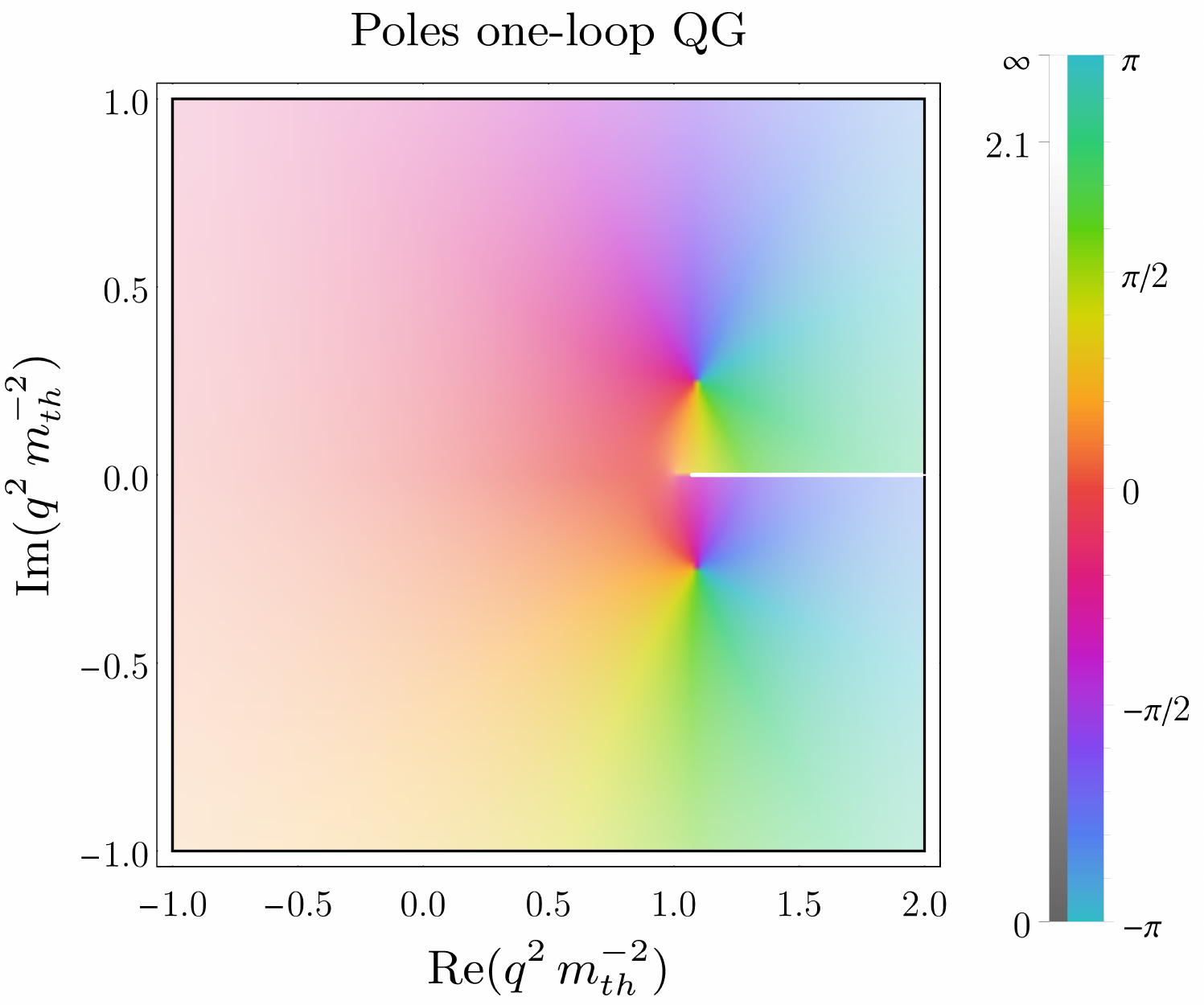}
\end{center}
\caption{Pole structure of the propagator in the complex $q^2$-plane, in the case of Lee-Wick QED (left panel) and one-loop QG (right panel), for $\alpha=\beta=1$. As is evident from the plots, the two theories share very similar features: they both develop complex-conjugate poles (entailing a violation of causality) and a single branch cut, associated with an unstable ghost degree of freedom. \label{fig:polesleewickqedandqg}}
\end{figure}

Whether there are additional real or complex-conjugate poles depends on the specific values of the couplings.
In the case of one-loop QG and ``standard'' Lee-Wick QED, the coupling $a$ is positive, with $a\ll1$, and $b\leq1$. Therefore, in both cases there is a pair of complex-conjugate poles. Moreover, the logarithmic propagators typically have at least one branch cut singularity. The location of the branch cut and the complex-conjugate poles are displayed in Fig.~\ref{fig:polesleewickqedandqg} for the case $\alpha=\beta=1$.

Using the Cauchy integral formula~\eqref{eq:fullydressed}, the structure of both propagators is
\begin{equation}\label{eq:exprropy}
D(q^{2})=\frac{R_{s}}{q^{2}+i\epsilon}+\frac{R_{c}}{q^{2}-m_{c}^{2}+i\epsilon}+\frac{R_{c}^{*}}{q^{2}-(m_{c}^{2})^{*}+i\epsilon}+\int_{m_{th}^{2}}^{\infty}\frac{\sigma(q^{2})}{q^{2}-s+i\epsilon}ds\,,
\end{equation}
where $\sigma(q^2)$ is the continuum part of the spectral density and is to be positive (for asymptotic states, cf.~Sect.~\ref{subsect:assumpunitarity}) in order to preserve unitarity. Its specific form determines the physical interpretation of the branch cut, i.e., whether it describes a resonance or a multi-particle state. 

The physical sheet of the complex $q^2$ plane is defined by applying the Feynman prescription $q^{2}\to q^{2}+i\epsilon$. The latter plays
a key role in determining the  sign of the imaginary part of the self-energy. Only after computations one can safely take the limit~$\epsilon\to0^{+}$. Restricting to $x\in\mathbb{R}$ and applying
the Feynman prescription to the propagator yields
\begin{equation}\label{eq:logdec}
\log(1-x)=\lim_{\epsilon\to0^+}\{\log(|1-x-i\epsilon|)+i \text{atan2}(1-x-i \epsilon)\}
=\log(|1-x-i\epsilon|)-i\pi\,\theta(x-1)\,.
\end{equation}
Note that if one does not use the Feynman prescription, the imaginary part of the logarithm on the real axis comes with an opposite sign. The above equation, which holds for $y\equiv\epsilon\ll1$, justifies the  expression~\eqref{eq:logwiththeta}, where the imaginary part of the logarithm is approximated by a Heaviside step function. Let us remark however that this approximation is only reliable close to the real axis, while in general the complex logarithm should be treated as in the previous two subsections.

\begin{figure}
\hspace{-0.5cm}\includegraphics[scale=0.43]{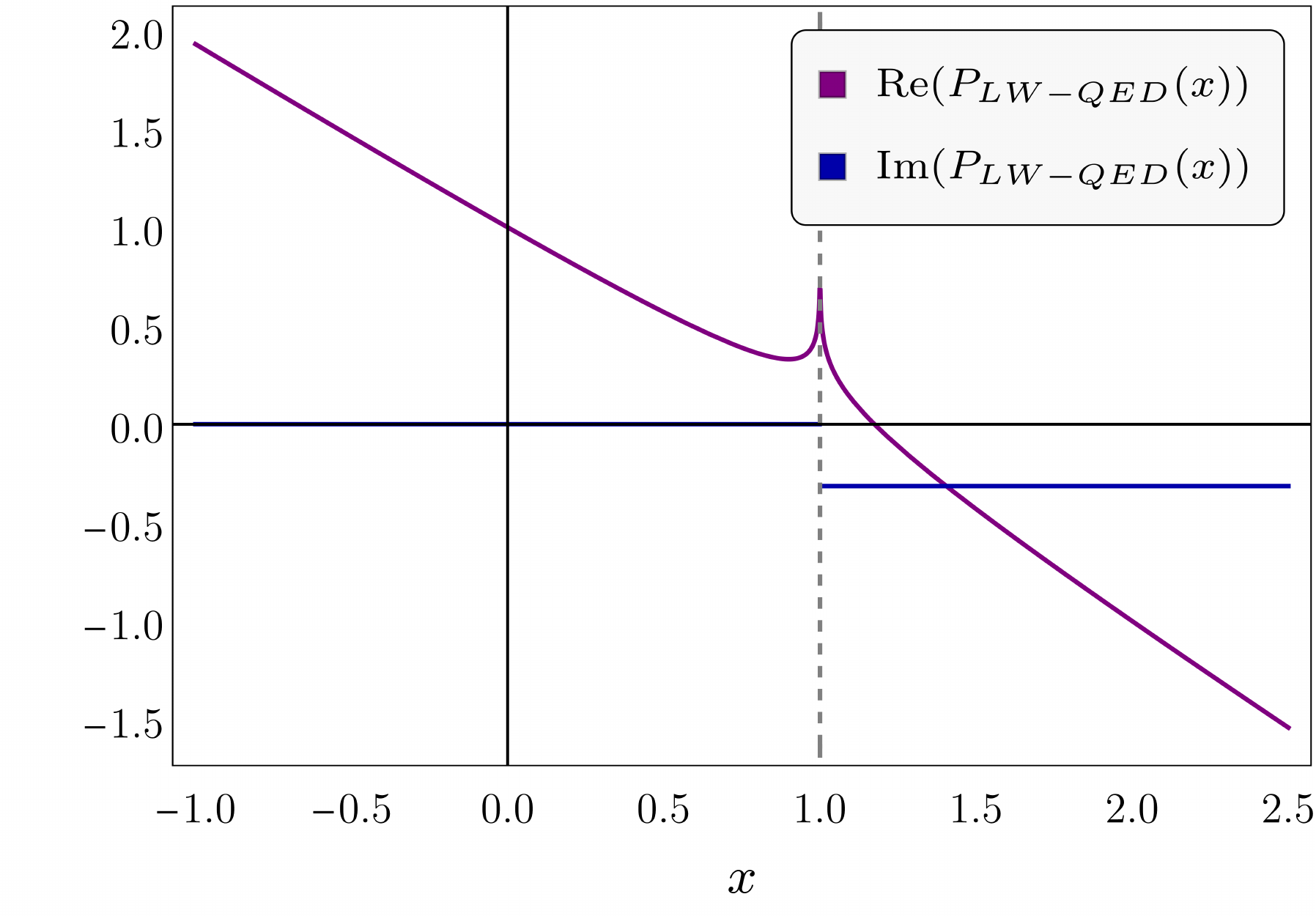}\;\includegraphics[scale=0.43]{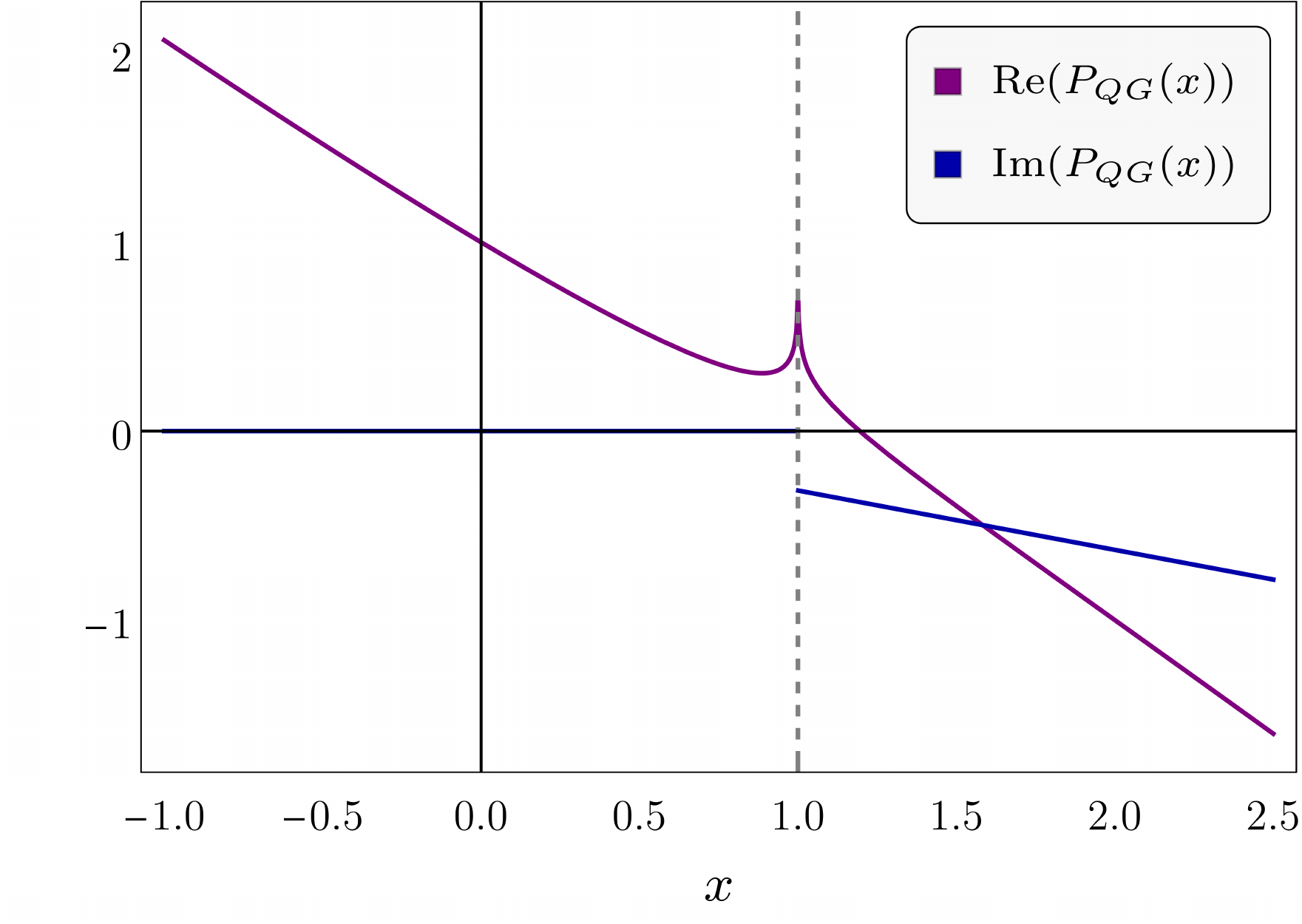}\caption{Real and imaginary parts of the $P$-functions in Eq.~\eqref{Pqed} and Eq.~\eqref{Pqg} as functions of $x\equiv \mathrm{Re}(q^2)$. The plots are obtained for $a=0.1$ and $b=1$. Both real parts have a zero in the region where the imaginary part is non zero. These zeros describe (unstable) ghost-like particles. \label{resonancepoles}}
\end{figure}

All in all, when evaluated on the real axis, the $P$-functions read
\begin{align}
&P_{LW-QED}(x)=1-b\,x-a\left[\log(|1-x|)+i\pi\,\theta(x-1)\right]\,,\nonumber\\
&P_{QG}(x)=1-b\,x-a x\left[\log(|1-x|)+i\pi\,\theta(x-1)\right]\,.
\end{align}
These are the expressions to be used to determine the contribution of the branch cut region to the scalar part of the propagator, i.e., to evaluate the integral in Eq.~\eqref{eq:exprropy}.
The functions $P_{LW-QED}$ and $P_{QG}$ have no real zeros. On the other hand, the existence of zeros at $q^2\geq m_{th}^2$ for the real part of $P(x)$ (cf. Fig.~\ref{resonancepoles}) tells us that in both cases there is an unstable degree of freedom, specifically, a ghost. Unitarity thus crucially depends on the positivity of $\sigma$. In turn, this is determined by the propagator $D(q^2)$ along the cut
\begin{equation}
D(q^{2})|_{cut}=-\frac{1}{\pi}\int_{m_{th}^{2}}^{\infty}\frac{\text{Im}\,D(s+i\epsilon)}{q^{2}-s+i\epsilon}ds\,.
\end{equation}
Here $D(q^2)$ can be written as 
\begin{equation}
D(q^{2})=\frac{\mathcal{R}}{p^{2}-m_0^{2}-\Sigma(p^{2})}\,\,,
\end{equation}
where $\mathcal{R}$ is a constant, $\Sigma(q^2)$ is the self-energy contribution to the propagator and the (real) solution of $m^{2}=m_0^{2}+\text{Re}\Sigma(m^{2})$ defines the mass of the resonance (if there is no real solution, the cut describes a multi-particle state). For instance, in the case of Lee-Wick QED, the real part of the inverse propagator has a real zero, and 
\begin{equation}
\Sigma(q^{2})|_{cut}=(b q^4+a\,\log(1-q^2/m_{th}^2))|_{cut}=b q^4+a\,\log(|1-q^2/m_{th}^2|)-ia\pi\,\,,
\end{equation}
so that $\mathrm{Im}\Sigma(q^{2})|_{cut}=-a\pi$. Therefore, in this case the imaginary part of the self-energy is small and negative (the same holds true in one-loop QG) and, additionally, it is constant along the cut.

If the cut describes a resonance of mass $m$ and if the interaction coupling that makes this degree of freedom unstable is small, $\mathrm{Im}\Sigma(s)\ll m^2$ (as in the case of one-loop QG and Lee-Wick QED, where $a\ll1$), one can approximate
\begin{equation}
D(q^2)|_{cut}\simeq D(q^{2}\simeq m^{2})\simeq\frac{Z \,\mathcal{R}}{p^{2}-m^{2}-i\,Z\,\text{Im}\Sigma(m^{2})}\simeq\frac{Z\,\mathcal{R}}{p^{2}-(m^{2}-i\,\Gamma/2)^{2}}\,\,,
\end{equation}
where the decay width is given by $\Gamma\simeq\frac{Z}{m}\text{Im}\Sigma(m^{2})$ and $Z$ is defined as
\begin{equation}
Z=(1-\partial_{m^2}\mathrm{Re}\Sigma(m^2)-i \partial_{m^2}\mathrm{Im}\Sigma(m^2))^{-1}\,\,.
\end{equation}
The corresponding spectral density is thus approximated by a Breit-Wigner distribution
\begin{equation}
\sigma(q^{2}\simeq m^{2})\simeq-\frac{1}{\pi}\frac{\mathcal{R}\,Z^2\,\mathrm{Im}\Sigma(m^2)}{(q^{2}-m^{2})^{2}+(Z\,\mathrm{Im}\Sigma(m^2))^2}\,.
\end{equation}
For an unstable ghost the constant $\mathcal{R}$ is negative and thus unitarity requires the decay width~$\Gamma$ to be negative. As we already mentioned, this is the case in Lee-Wick QED and also in one-loop QG~\cite{Donoghue:2019ecz,Donoghue:2019fcb}. However, due to the negative decay width, the pole corresponding to the unstable ghost lies on the first (physical) Riemann sheet~\cite{Donoghue:2019ecz,Donoghue:2019fcb}\footnote{This is to be contrasted with the case of standard resonances, which come with positive width and appear in the second Riemann sheet only.}. As detailed in Sect.~\ref{sect:causality}, preserving unitarity in this case comes at the expense of vacuum instabilities and a violation of causality on time scales~$\tau\simeq 1/\Gamma$. In addition, the presence of the complex-conjugate poles in the  $q_0^2$-complex plane, beyond bringing additional acausalities, makes it impossible to Wick-rotate. In order to perform an analytical continuation, one would need to resort to modified integration contours, such as those determined by the Lee-Wick (LW)~\cite{Lee:1971ix} or Cutkosky-Landshoff-Olive-Polkinghorne (CLOP) prescriptions~\cite{Cutkosky:1969fq}, to translate the Lorentzian description into the Euclidean one (and vice versa).

Summarizing, due to the complex conjugate root theorem,  logarithmic quantum corrections to bare higher-derivative  theories typically yield dressed propagators displaying additional real (possibly tachyonic and ghost-like) poles or complex-conjugate poles. In the former case, one would loose unitarity (possibly gaining tachyonic instabilities), in the latter the theory would be prone to acausalities (and eventually to vacuum instabilities) and it would not be possible to perform an analytic Wick rotation. In both cases, the unstable ghost associated with the branch cut would lead to acausalities. 

Nonetheless, one-loop corrections are not the end of the story. Resumming all quantum fluctuations is expected to result in non-local effective actions with more elaborate form factors. In the next section we will discuss some  examples of propagators that are free from additional poles and for which an analytic continuation can be performed. \\

\noindent\fbox{\parbox{\linewidth}{In accordance with the literature, logarithmic quantum corrections to higher-derivative theories can generally add complex-conjugate poles, tachyonic degrees of freedom and unstable ghosts. Thus, \textit{preserving causality and unitarity while avoiding instabilities requires more sophisticated non-local quantum corrections}, beyond the one-loop approximation.}}

\section{Consistent graviton propagators: some examples}\label{sect:goodpropy}

The integration of quantum gravitational fluctuations (or, more generally, massless fields) at the level of the path integral is expected to yield a non-local  effective action. In the one-loop approximation, the form factors are logarithmic. As we discussed in the previous section, logarithmic form factors are enough to make the theory unitary but typically introduce acausalities and instabilities. This suggest that the coexistence of unitarity, causality and stability in theories whose bare action has higher derivatives requires resumming all quantum effects into a more sophisticated non-local effective action.

In this section we discuss and compare two examples of propagators--that proposed in~\cite{Platania:2020knd} and the one derived  in~\cite{Wetterich:2020cxq,Wetterich:2021ywr,Wetterich:2021hru}--compatible with all requirements of causality, unitarity and stability discussed throughout the manuscript, including the possibility to perform an analytic Wick rotation.
We shall see that a common feature of these propagators is the presence of two symmetric branch cuts. However, for the propagator in~\cite{Wetterich:2020cxq,Wetterich:2021ywr,Wetterich:2021hru} performing an analytic Wick rotation may require a specific prescription for the choice of location of branch cuts. Finally, we will discuss and compare three classes of propagators which might play a role in some approaches to QG (see also~\cite{Knorr:2021iwv}), and which could implement a certain form of dimensional reduction~\cite{Carlip:2017eud}.

\subsection{Wetterich propagators}

In QFT, unitarity requires that no ghosts (at least, no stable ghosts) appear in the physical spectrum of the theory. In the case of non-gauge theories, this is equivalent to saying that the dressed propagator should not display additional real poles beyond the massless one (cf. Sect.~\ref{sect:Non-perturbative-unitarity}). Causality and (vacuum) stability further restrict the pole structure of the propagator: no complex poles should be located in the first or third quadrant of the physical Riemann sheet of the $q_0$-complex energy plane\footnote{Standard (i.e., non-ghost) unstable particles are typically described by complex poles lying in the first quadrant of the complex energy plane. However, since their decay width is positive, the corresponding poles are located on the second Riemann sheet. Wick rotation is thereby not forbidden by the presence of this type of complex poles.} (cf. Sect.~\ref{sect:causality}). Analytic continuation additionally requires the absence of essential singularities.

\begin{figure}
\centering\includegraphics[scale=0.5]{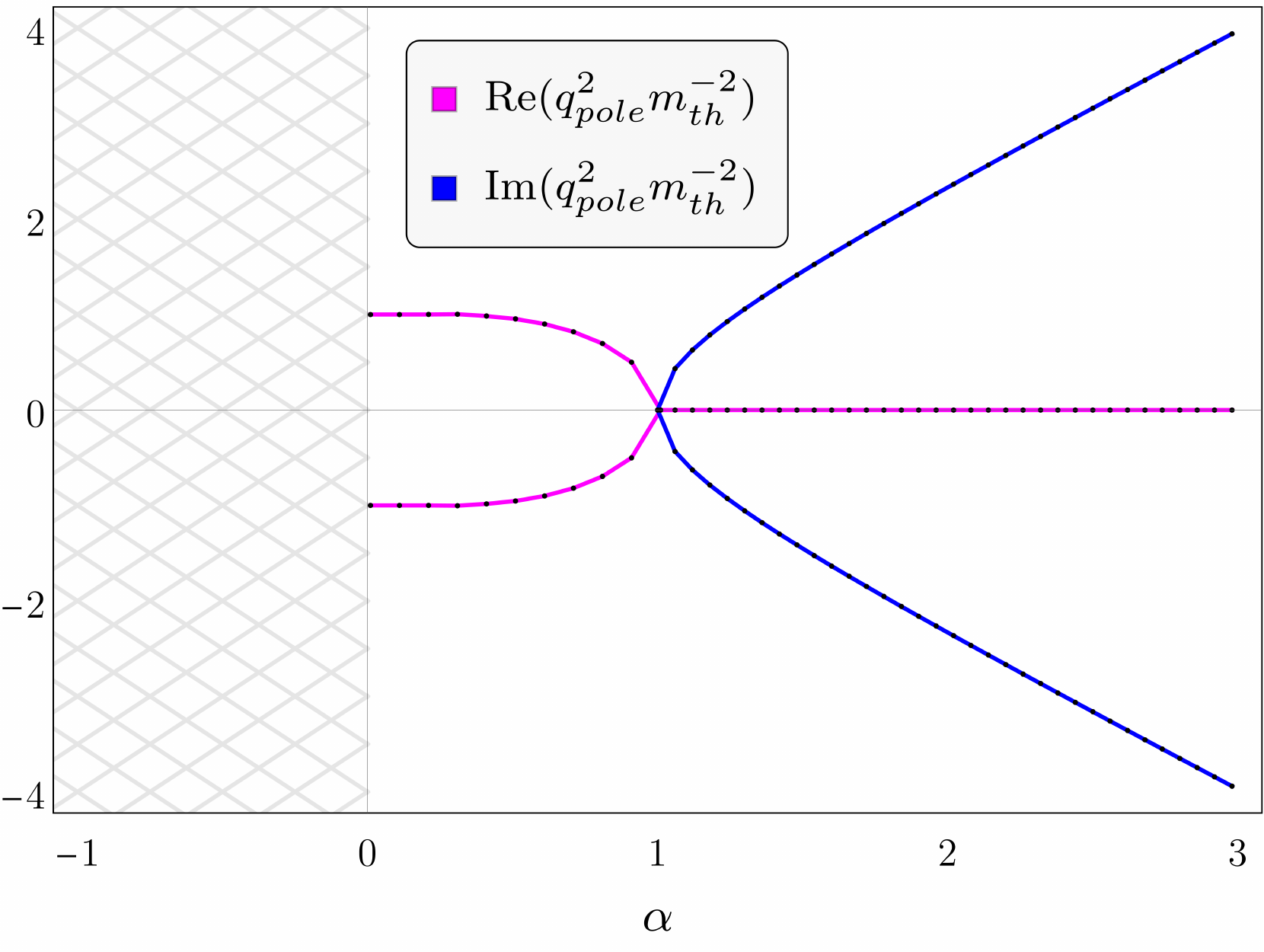}
\caption{Poles of the propagator $\eqref{eq:goodpropy}$ as a function of $\alpha$.
For $\alpha<0$ there are no additional poles, beyond the massless
one. For $0<\alpha<1$ there are two stable ghost poles, and one of
them is a tachyonic ghost. Finally, for all $\alpha>1$ there is a
pair of massless complex-conjugates poles. \label{fig:Polestructure}}
\end{figure}

\begin{figure}
\hspace{-0.5cm}\includegraphics[scale=0.52]{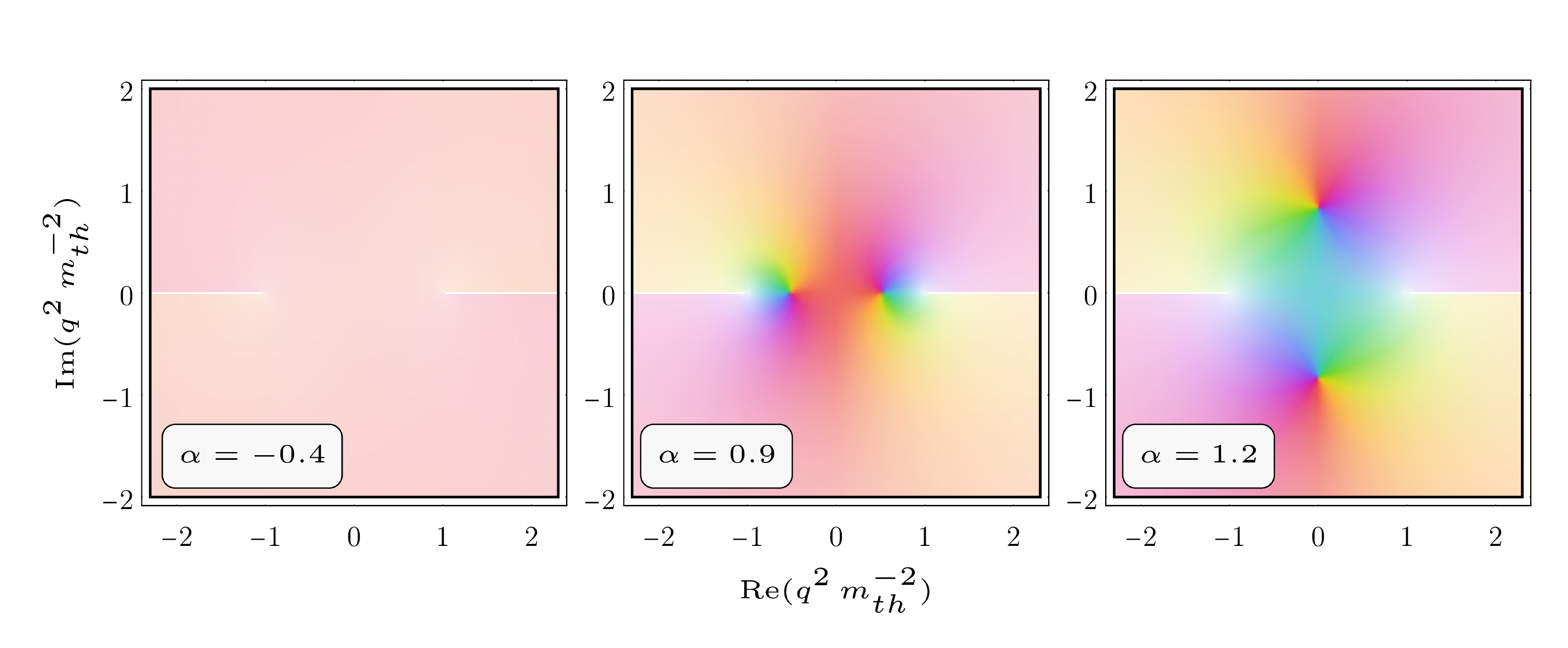}
\caption{Pole structure of the function $P(z)$  in Eq.~\eqref{eq:goodpropy} in the complex $q^{2}$-plane for three different values of $\alpha$, corresponding to the cases of complex-conjugate poles ($\alpha>1$), two real poles ($0<\alpha<1$) and no additional poles ($\alpha<0$). These are the only cases realized within the propagator of Eq.~\eqref{eq:goodpropy}, as also shown in Fig.~\ref{fig:Polestructure}. \label{fig:Polestructure2}}
\end{figure}

A propagator satisfying all these requirements has been proposed in~\cite{Platania:2020knd}. Its scalar part in Lorentzian signature reads
\begin{equation}
iD(q^{2})=\frac{i}{q^{2}\left(1+\alpha\,\frac{m^{2}}{q^{2}}\,\,\mathrm{arctanh}\left[-\frac{q^{2}}{m^{2}}\right])\right)}\,\,,\qquad\alpha<0\,,\label{eq:goodpropy}
\end{equation}
where $m^{2}$ is a mass scale. As detailed in~\cite{Platania:2020knd}, this propagator is endowed with two branch cut singularities along the real $q^2$ axis. If $\alpha<0$ no additional poles exist beyond the massless one (cf. Fig.~\ref{fig:Polestructure} and Fig.~\ref{fig:Polestructure2}). Since the real part of the inverse propagator is always positive, the branch cut singularities do not describe an unstable (ghost) degree of freedom, but rather two symmetric multi-particle states. The spectral density is positive-definite for $\alpha<0$, thus no violation of unitarity is expected in this case. Moreover, as shown in Fig.~\ref{fig:Polestructure2}, for $\alpha<0$ there are no complex poles and thus, based on the arguments we discussed in Sect.~\ref{sect:causality}, no violation of causality ought to occur. Finally, the absence of complex poles in the first and third quadrants of the $q_0$-complex plane, together with the absence of singularities at infinity, make it possible to connect the Euclidean and  Lorentzian versions of the theory via an analytic Wick rotation.

\begin{figure}
\includegraphics[scale=0.425]{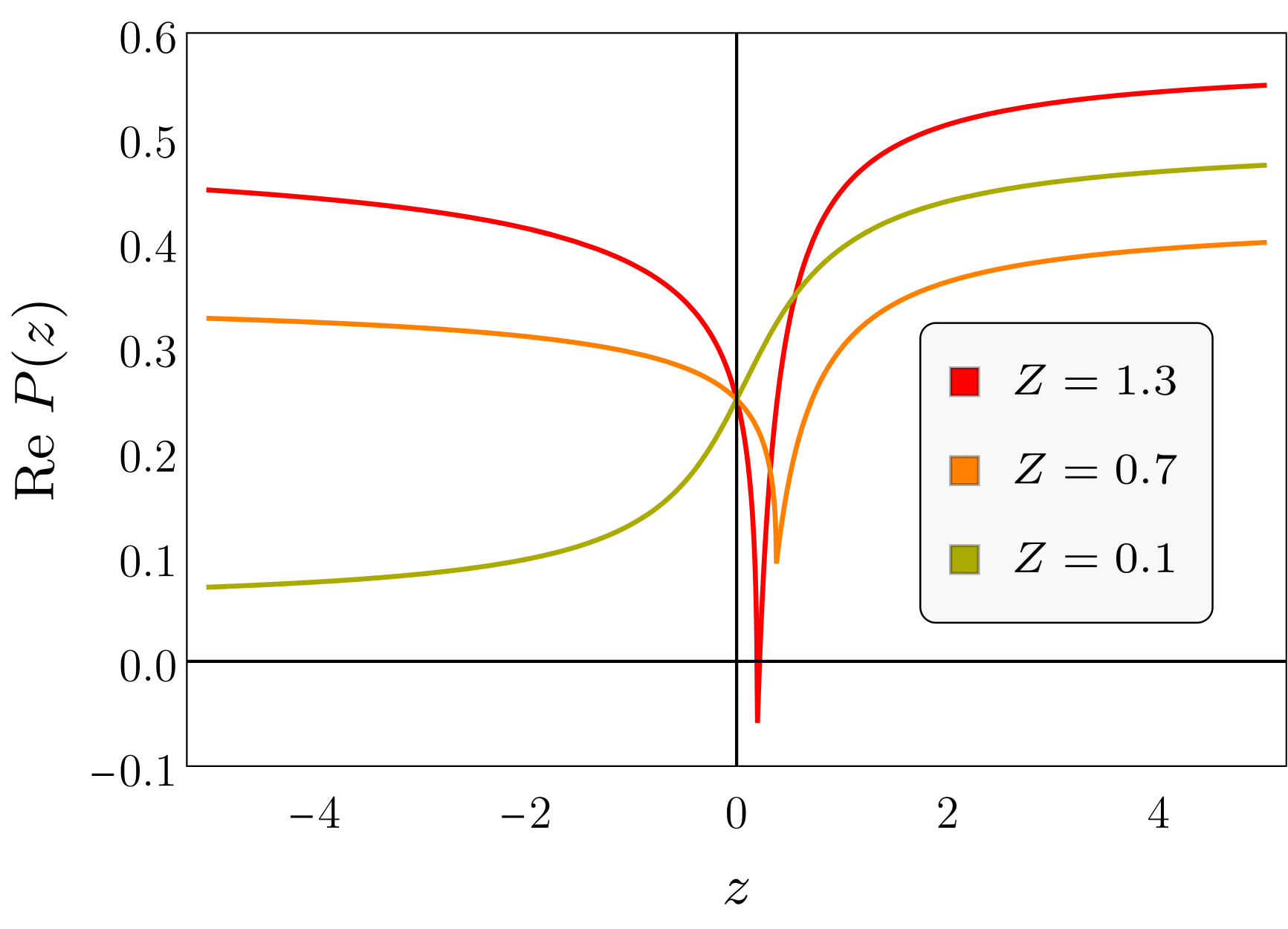}
\includegraphics[scale=0.425]{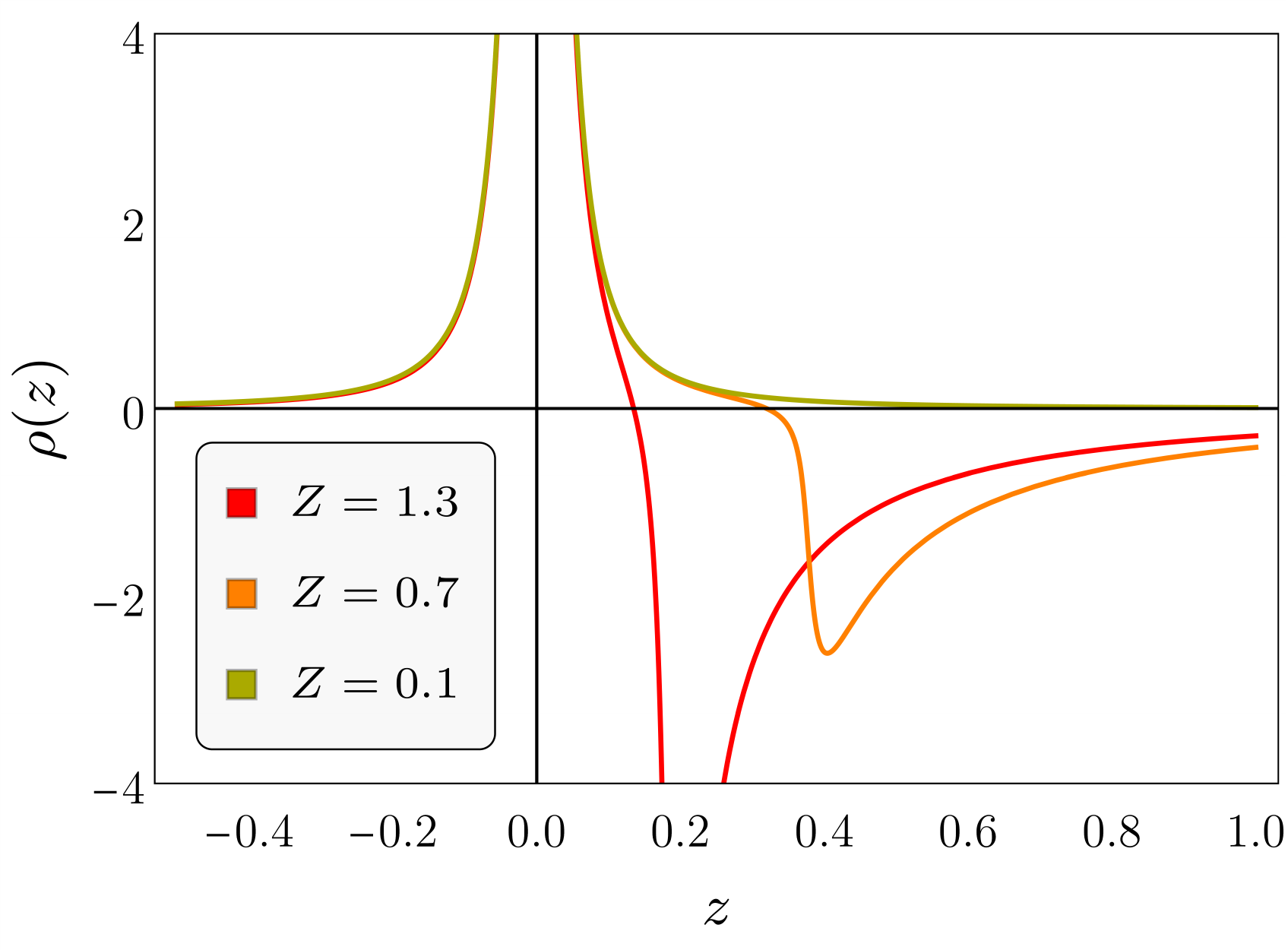}
\caption{Real part of the $P$ function (left panel) and spectral density $\rho(z)$ (right panel), with $z=q^{2}m^{-2}$, associated
with the graviton propagator~$\eqref{eq:gravprop}$ for three different values of $Z$, and for $m^2=2$ and $M^2=1$ (in Planck units). These values are chosen to illustrate the three qualitatively-different cases listed after Eq.~\eqref{eq:brpoints}. \label{fig:Spectral-density}}
\end{figure}

\begin{figure}
\hspace{-0.4cm}\includegraphics[scale=0.42]{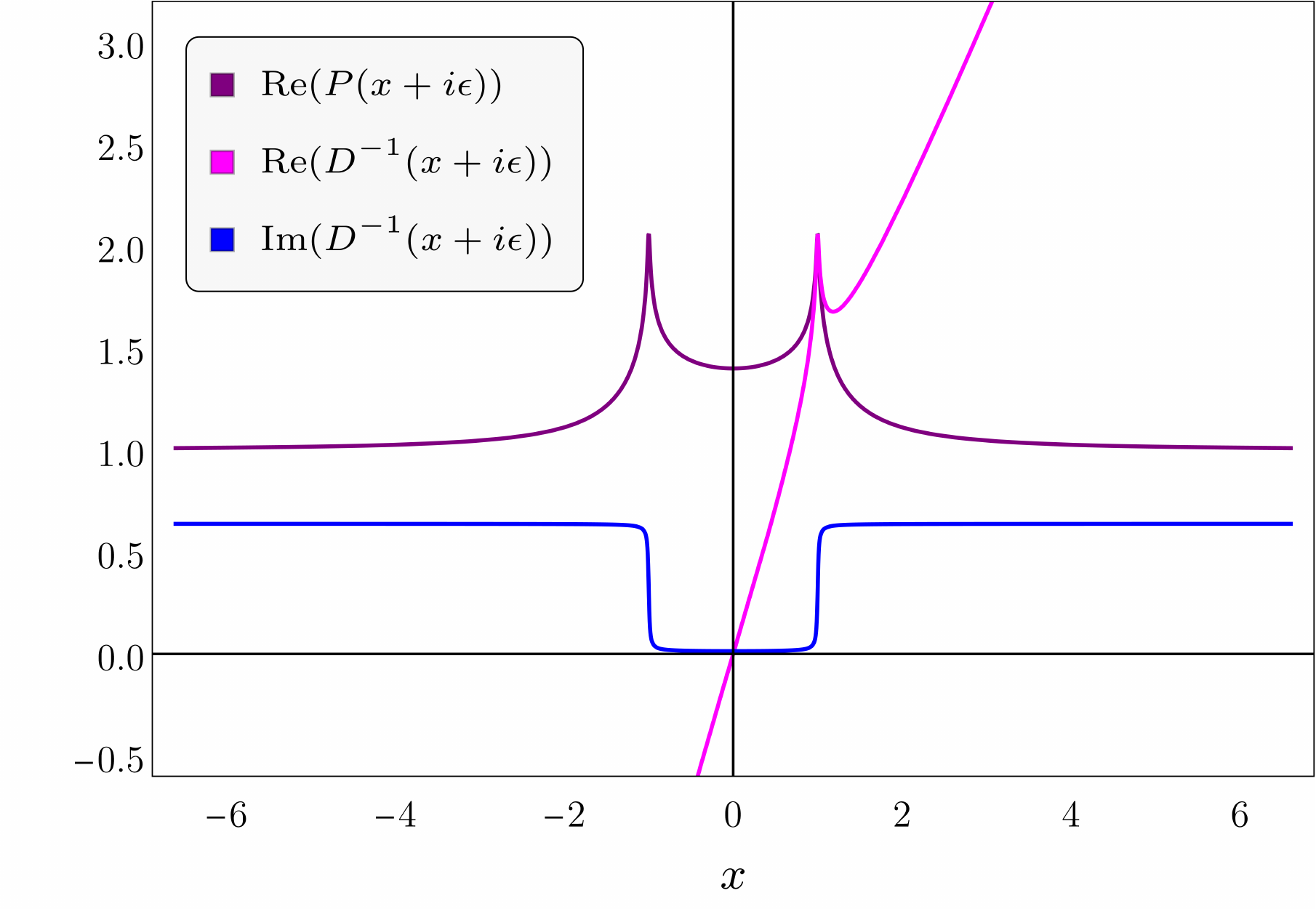}
\includegraphics[scale=0.42]{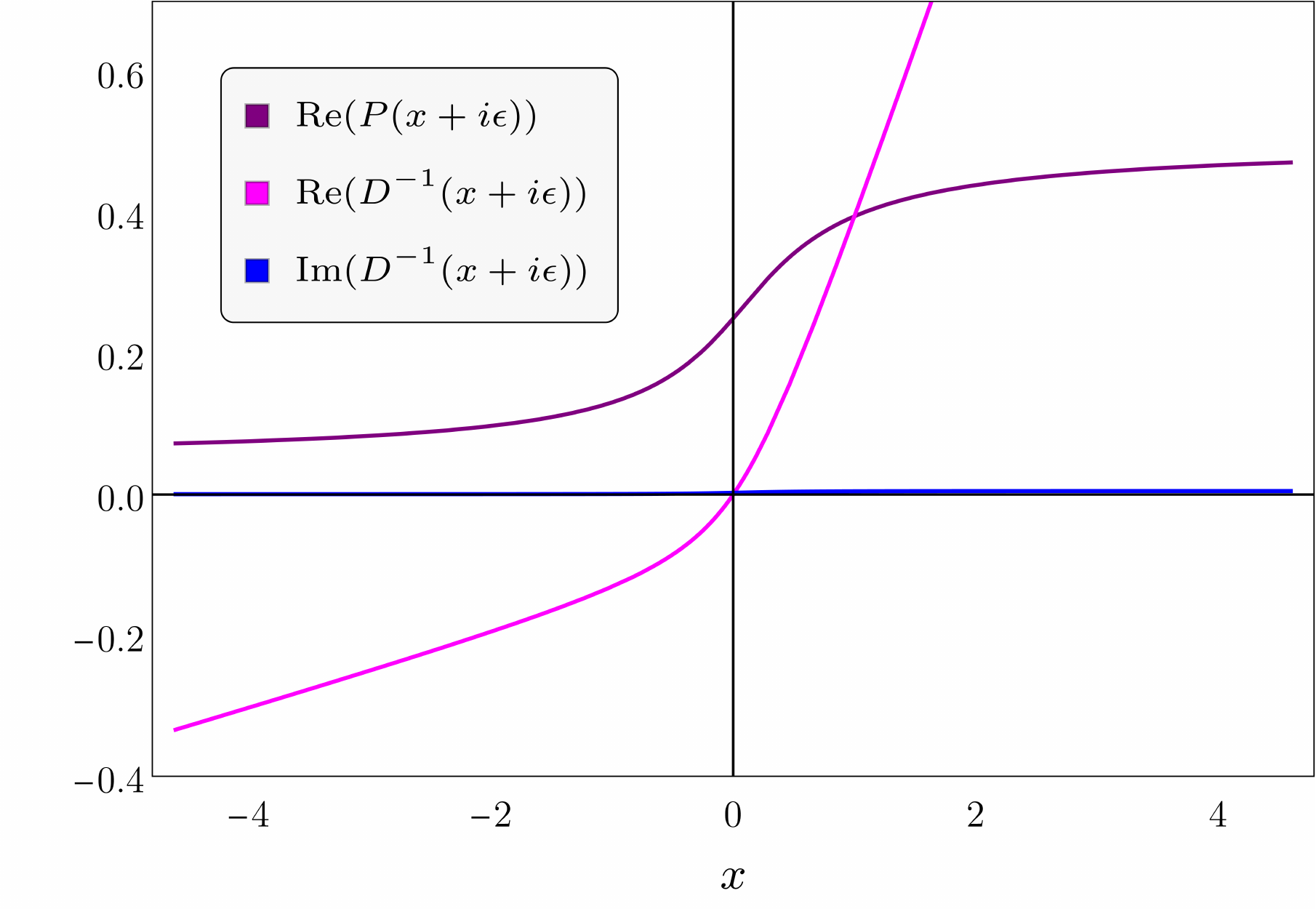}
\caption{Comparison of the real and imaginary parts of the propagators in~Eq~\eqref{eq:goodpropy} (case $\alpha<0$, with $\alpha=-0.2$ to produce the plot in~\cite{Platania:2020knd}, here in the left panel) and~\eqref{eq:gravprop} (case $Z<M^2/m^2$, with $Z=0.1$ to produce the plot in the right panel). In both cases, the real part of~$P(x)=x^{-2}D^{-1}(x)$
(purple line) is positive, while the imaginary part of $D^{-1}(x)$ (blue line) is non-zero only along the branch cuts. Since for the case $Z<M^2/m^2$ the branch cuts are not along the real $q^2$-axis, the imaginary part of $D^{-1}(z)$ is zero for $z\equiv x=\mathrm{Re}z$. The real part
of the inverse propagator $D^{-1}(z)$ (magenta line) has thereby one single massless pole.\label{fig:ReImPropy}}
\end{figure}

Let us now discuss the graviton propagator derived in~\cite{Wetterich:2020cxq,Wetterich:2021ywr,Wetterich:2021hru} and compare it with the one proposed in~\cite{Platania:2020knd} and reviewed above. The scalar part of the propagator derived in~\cite{Wetterich:2020cxq,Wetterich:2021ywr,Wetterich:2021hru} 
reads
\begin{equation}\label{eq:gravprop}
D^{-1}=\frac{m^2}{8\varepsilon}\left[m^2-M^2- (Z+1)\varepsilon q^2-\sqrt{(m^2-M^2-(Z-1)\varepsilon q^2)^2-4 \varepsilon\frac{q^2}{m^2}(m^2-M^2)^2}\,\right],
\end{equation}
where $\varepsilon=-1$ in the Lorentzian and $\varepsilon=1$ in the Euclidean signature, $m^2,M^2>0$, and $Z$ is a parameter. In the following we will use the dimensionless momentum square $z=q^2/m_{th}^2$ with $m_{th}=m$.
In order to have a massless pole, one has to require $m^2\geq M^2$. One additional ghost pole at $z = (-m^2 M^2 + M^4)/(m^4 Z \varepsilon)$ appears under the signature-independent condition
\begin{equation}
Z \geq Z_c= \frac{M^2}{m^2-M^2}\,.
\end{equation}
In what follows, we shall restrict ourselves to the case $Z<Z_c$ and $\varepsilon=-1$.
Due to the presence of the square root, there are branch cuts in the complex energy plane. The branch points are 
\begin{equation}\label{eq:brpoints}
z_\pm=-\frac{1}{m^4(Z-1)^2}\left[m^2 M^2 (3 + Z)-2 M^4 - m^4 (1 + Z) \pm2\sqrt{(m^2-M^2)^3(Z m^2-M^2)}\right]\,,
\end{equation}
and they are located along the positive real axis if $Z>M^2/m^2$. In contrast, in Euclidean signature they would be located along the negative real axis. Finally, independent of the signature, for $Z<M^2/m^2$ the branch points $z_\pm$ are complex conjugates and two  branch cuts appear parallel to the imaginary axis. It is also important to notice that $Z_c\geq M^2/m^2$ if $M^2\geq m^2$, so that one can distinguish three cases:
\begin{itemize}
\item $0<Z<M^2/m^2$: The branch points $z_\pm$ are complex conjugates and two  branch cuts appear parallel to the imaginary axis. There are no additional ghost poles and the spectral density is positive.
\item $M^2/m^2<Z<Z_c$: Branch cuts appear along the real $q^2$-axis, and the inverse propagator develops an imaginary part (the spectral density thus has a continuum part, corresponding to a multi-particle state). In this case there are no additional ghost poles and Wick rotation is straighforward, as in~\cite{Platania:2020knd}; however, for any $Z$ in this range, the $P$-function develops an imaginary part along $\mathrm{Re}(q^2)$ where the spectral density becomes negative.
\item $Z>Z_c$: In this case there are multi-particle states with negative spectral density and additional ghost poles. Unitarity is thus violated.
\end{itemize}
The spectral density associated with these three cases is depicted  in Fig.~\ref{fig:Spectral-density} for the case $m^2=2$ and $M^2=1$ (in Planck units), while Fig.~\ref{fig:ReImPropy} compares the propagators in Eqs~\eqref{eq:goodpropy} and~\eqref{eq:gravprop}.
Ghosts and a negative spectral density are avoided for~$Z<M^2/m^2$. In the next subsection we will explain how to perform an analytic continuation in this case.

\subsubsection{Digression: Wick rotation via Sommerfeld branch cuts}

\begin{figure}
\begin{flushleft}
\hspace{-0.35cm}\includegraphics[scale=0.315]{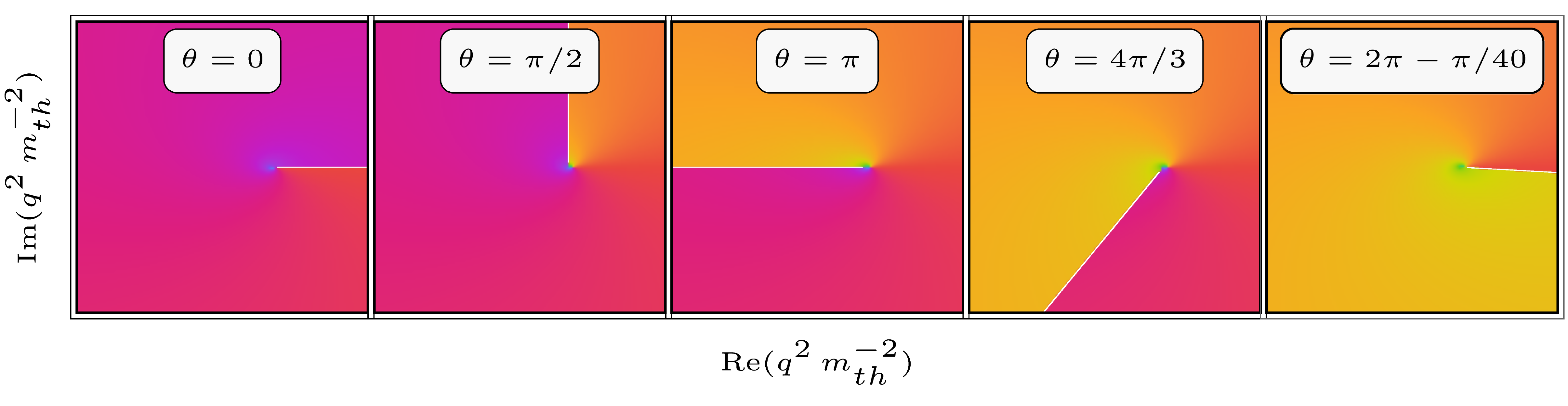}\\
\hspace{-0.35cm}\includegraphics[scale=0.315]{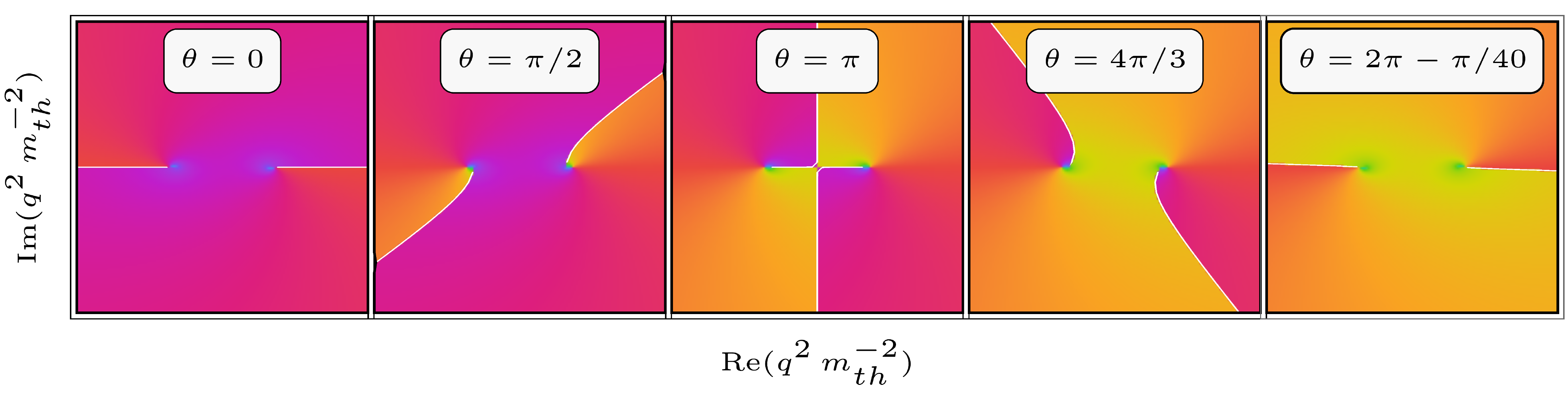}
\end{flushleft}
\caption{Rotation of branch cuts of a simple test function, $f=1+\log[(1 + 3 q^2/m_{th}^2 + \xi ( q^2/m_{th}^2)^2)]$, for various angles $\theta\in[0,2\pi)$, and in the case of one (top panels, case $\xi=0$) and two (bottom panel, case $\xi=1$) branch cuts. The variation of the branch cut is implemented according to Eq.~\eqref{BCrule}.\label{fig:BranchRotation}}
\end{figure}

\begin{figure}
\includegraphics[scale=0.5]{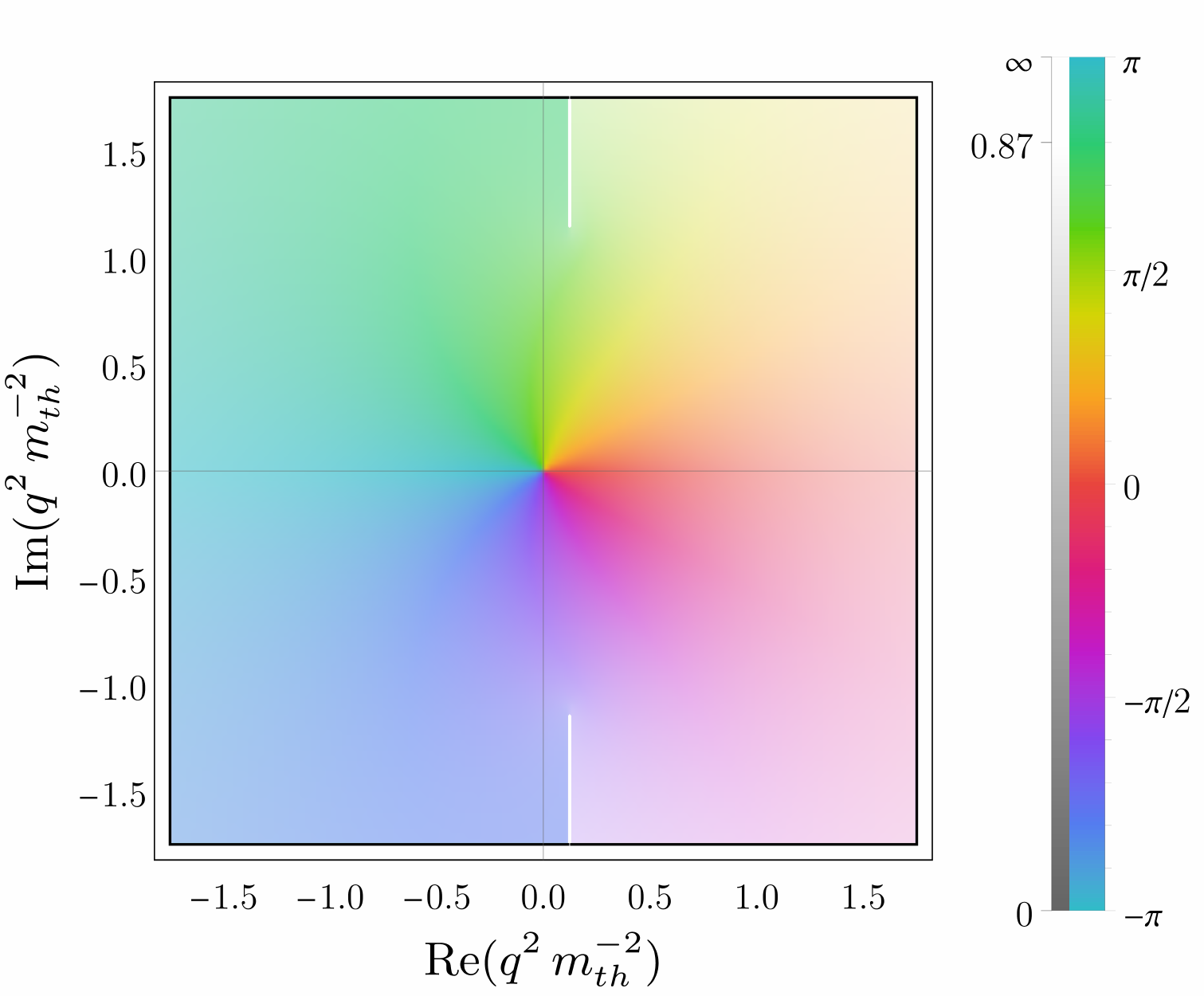}\includegraphics[scale=0.5]{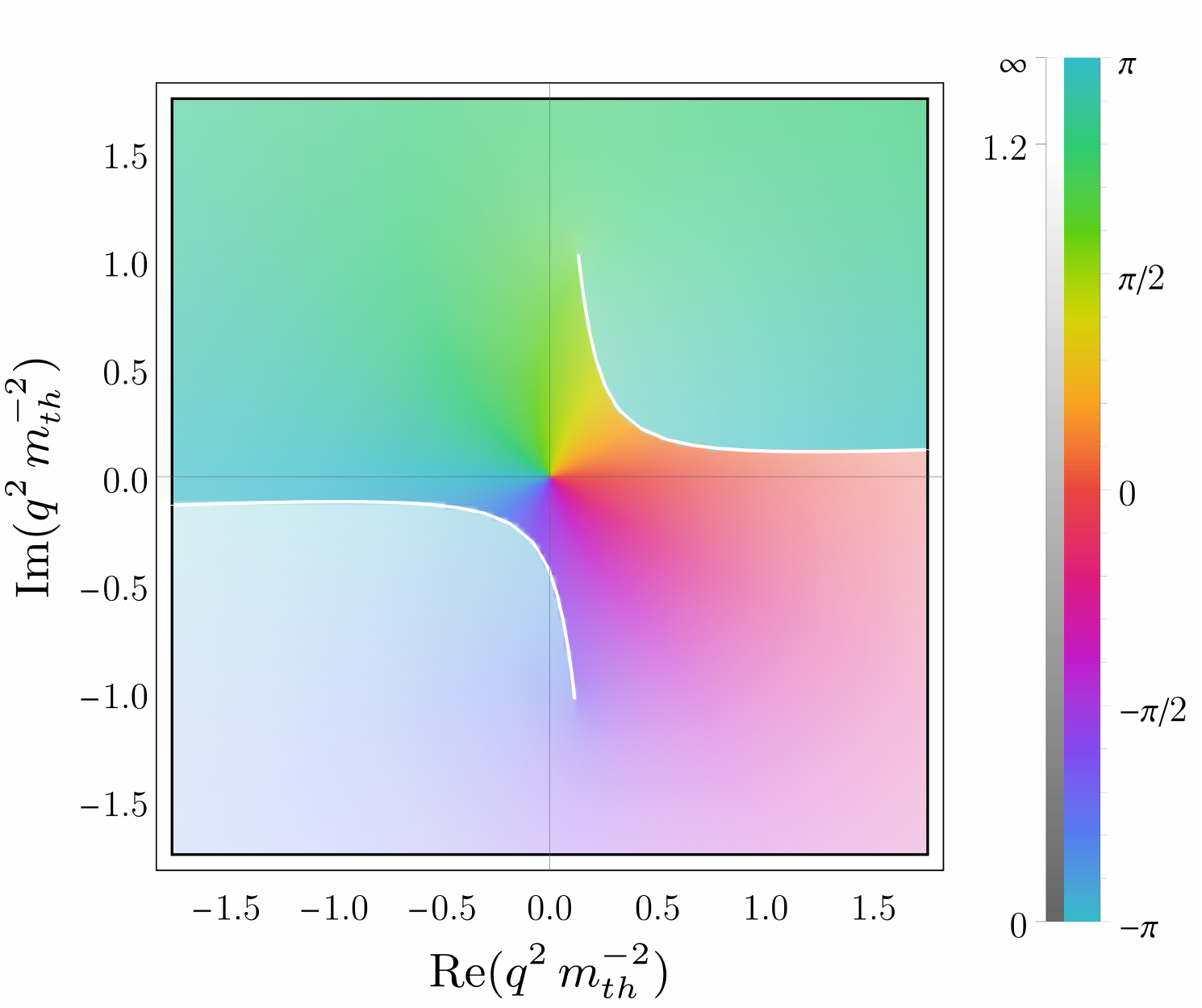}\\
\includegraphics[scale=0.5]{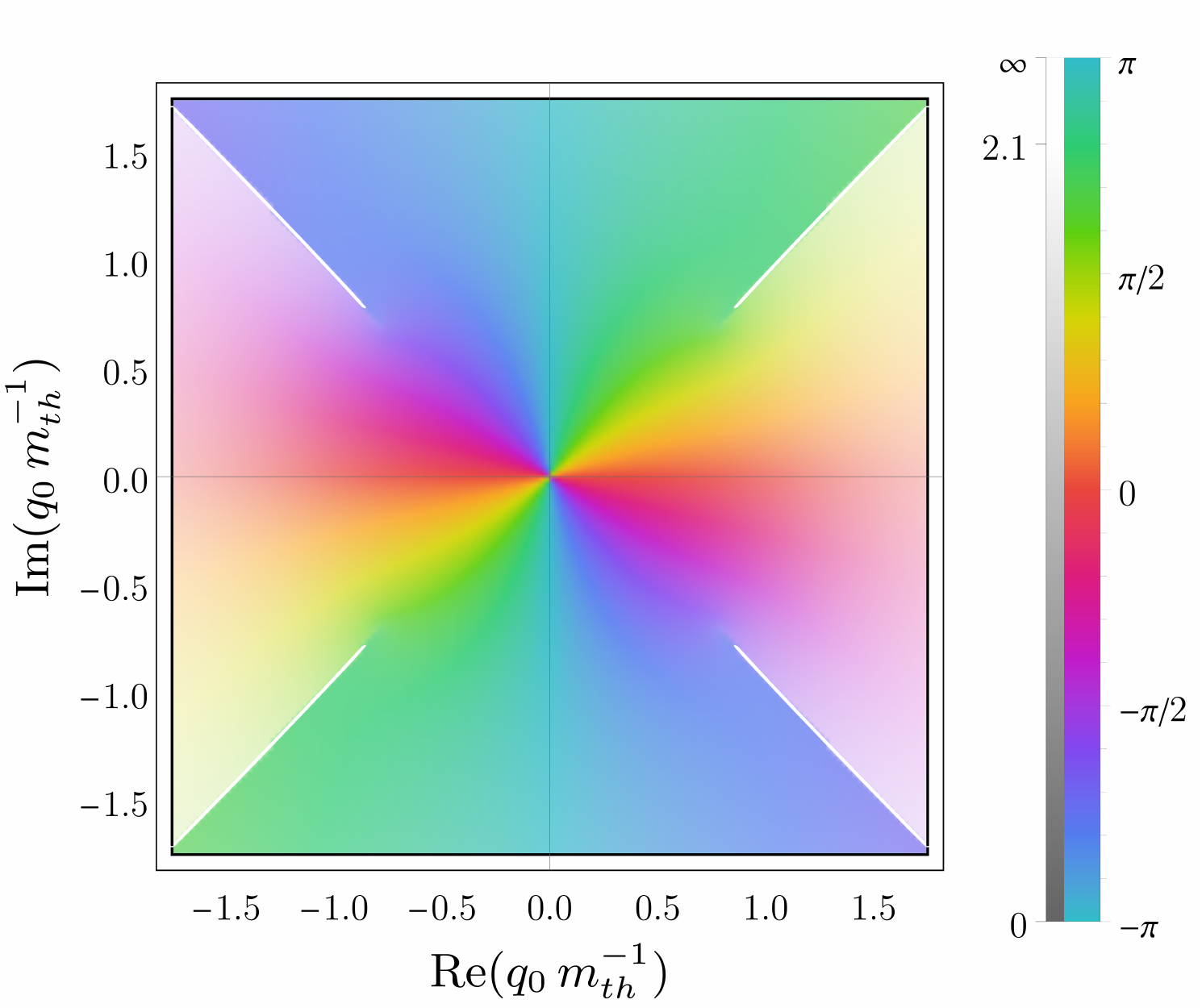}\includegraphics[scale=0.5]{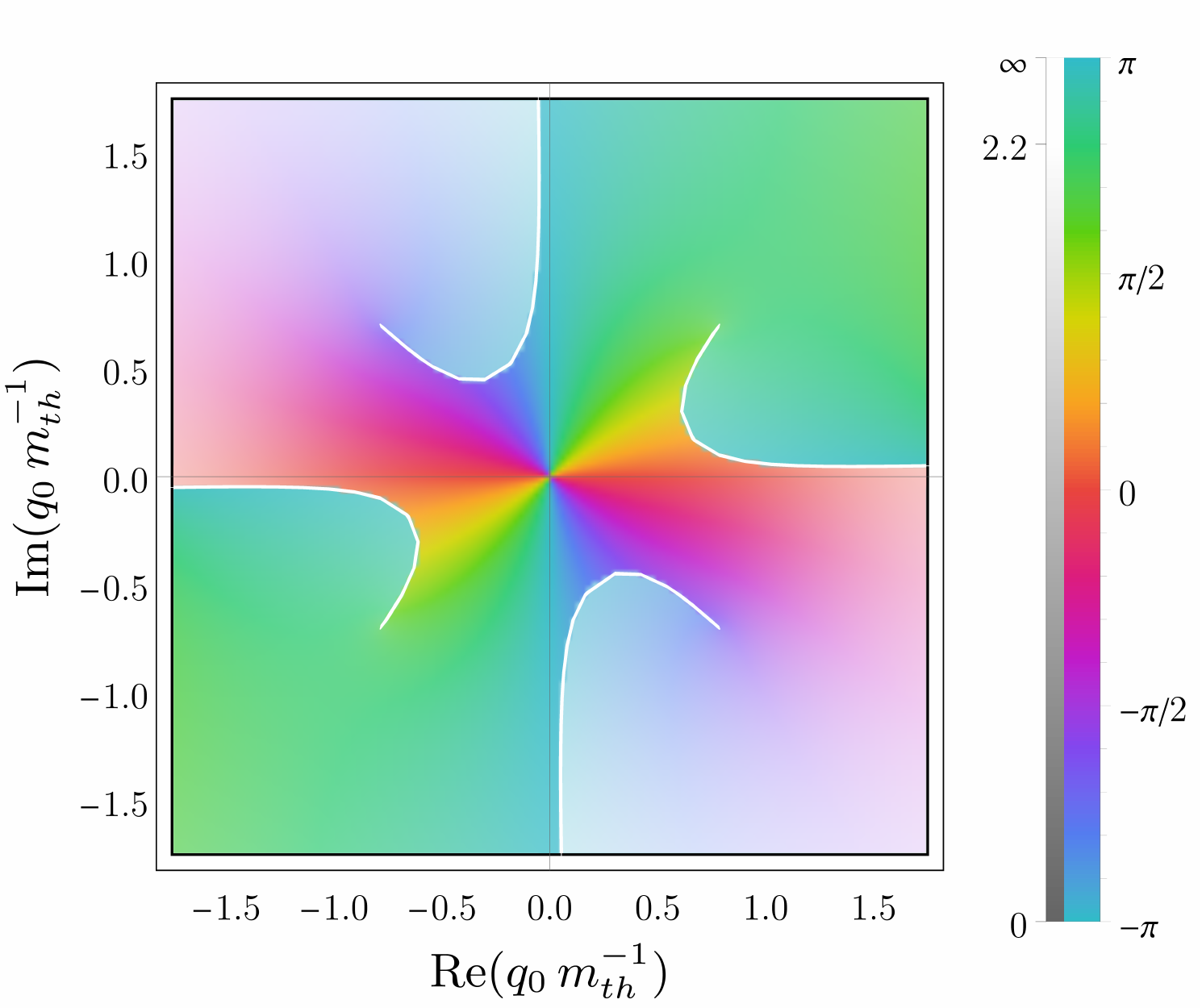}\\
\includegraphics[scale=0.5]{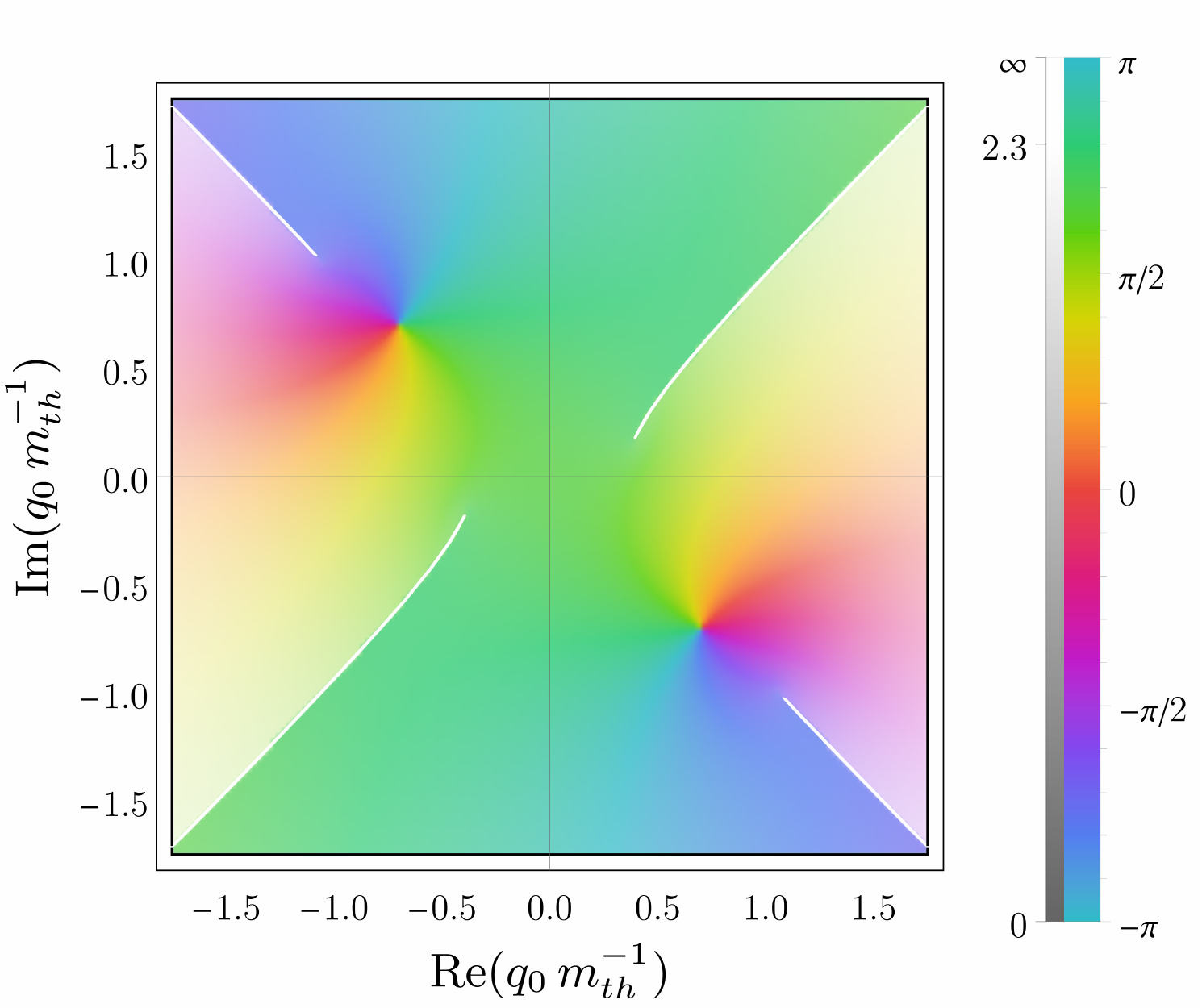}\includegraphics[scale=0.5]{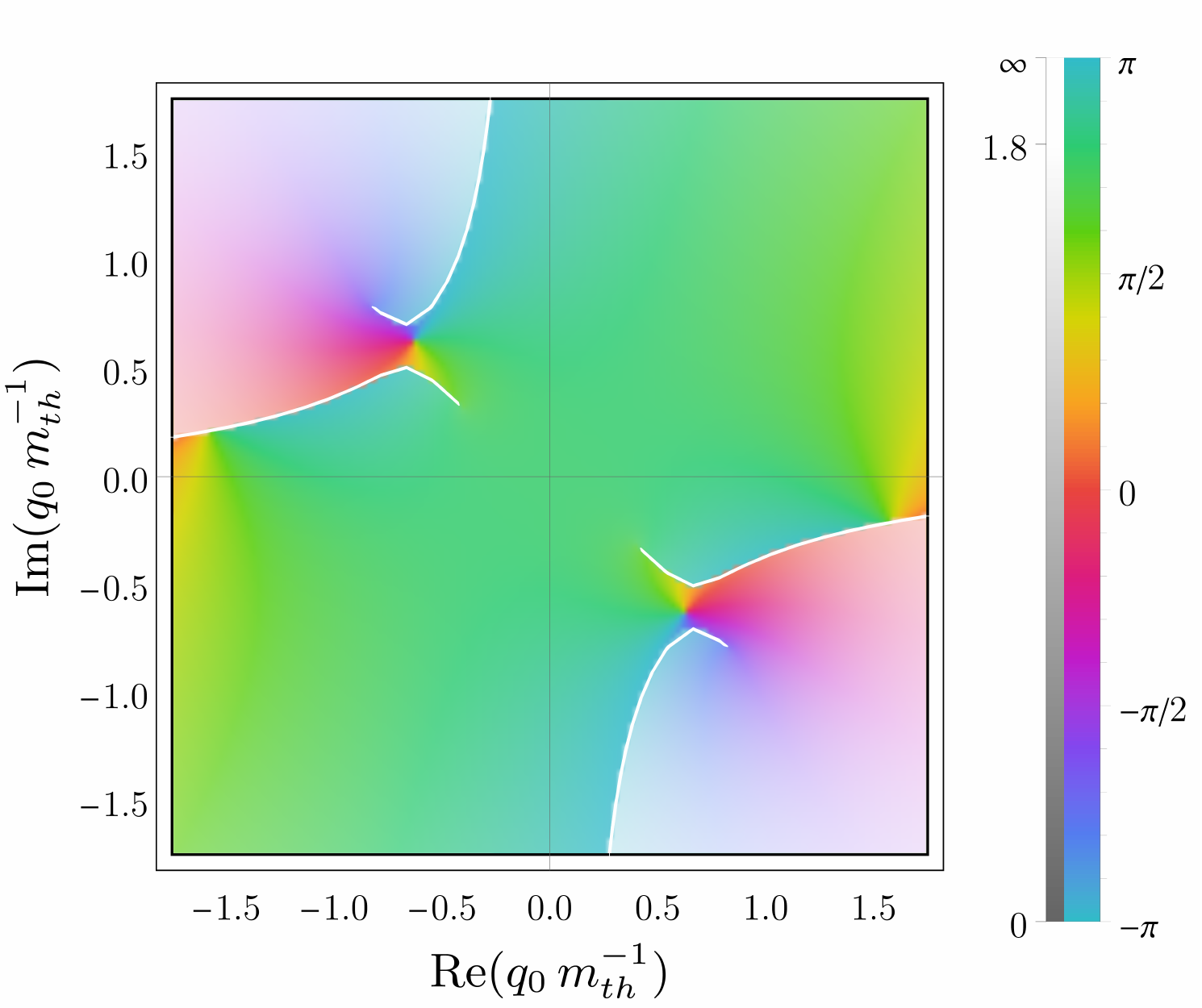}
\caption{Pole structure of the graviton propagator for $Z<M^2/m^2$ in the $q^2$-plane (top panels) $q_0$-plane (center panels) and $q_0$-plane with the $i\epsilon$-prescription implemented (bottom panels). The figures in the left column are obtained with the ``standard'' prescription for the position of the branch cuts. The figures on the right instead implement the rotation of branch cuts according to Eq.~\eqref{BCrule} with $\theta=\pi/30$. This prescription, in combination with the  $i\epsilon$-prescription, is used here to move branch cuts and branch points suitably to allow for an analytic Wick rotation from the Euclidean to the Lorentzian and vice versa. Indeed, when both $i\epsilon$-prescription and branch cut rotations are applied, the first and third quadrants of the $q_0$-plane are free from obstructions (figure on the bottom-right panel) and an analytic continuation can be performed.\label{fig:Poles-structure-New}}
\end{figure}

We now focus on the possibility of performing an analytical Wick rotation in the presence of branch cuts that are parallel to the imaginary axis. To this end, we note that in the $q_0$-plane, if $Z>M^2/m^2$ there are four symmetric branch points along the real axis in Lorentzian, and along the imaginary axis in Euclidean signature. If instead $Z < M^2/m^2$, the two branch points on the $q^2$-plane translate in two pairs of complex-conjugate branch points in the $q_0$-complex plane, independently of the signature. Thus, in all these cases, the branch cuts of the propagator~\eqref{eq:gravprop} cross all quadrants of the complex energy plane. Wick rotation thus cannot be performed, at least not straightforwardly. In particular, since the branch cuts are a continuum of singular points, one cannot use a modified contour à la Lee-Wick to circumvent the problem. On the other hand, since the location of the branch cuts depends on the convention on the domain of the principal argument, one can change the location of the branch cuts while keeping the location of the branch points fixed. In this procedure one has to make sure not to alter the values of the $P$-function along the real axis, since the reality of the $P$-function along the real axis is related to the stability properties of the given degrees of freedom. We thus use the replacement rule
\begin{equation}\label{BCrule}
\mathrm{Arg}(z)\longrightarrow \mathrm{Arg}(z\,e^{i (\theta-\pi)})-(\theta-\pi)\,,
\end{equation}
with $\theta\in[0,2\pi)$. Specifically, the choice $\theta=\pi$ corresponds to the conventional case where $\mathrm{Arg}(z):\mathbb{C}\to[-\pi,\pi)$, while the choice $\theta=0$ reproduces in this case the Sommerfeld (hyperbolic) branch cut.
Examples of this implementation are shown in Fig.~\ref{fig:BranchRotation} for the case of one and two branch points and cuts.
In our case, the branch points are located in all quadrants in the $q_0$-complex plane and, as depicted in Fig.~\ref{fig:Poles-structure-New}, after rotating the branch cuts, one ends up with four ``hooks'' connecting the four branch points with $\pm\infty$ and $\pm i \infty$. Next, by employing the  $i\epsilon$-prescription, the branch points can be moved towards the second and fourth quadrant of the complex energy plane. Therefore, combining an appropriate rotation of the branch cuts with the Feynman $i\epsilon$-prescription one can remove them from the first and third quadrants (cf. Fig.~\ref{fig:Poles-structure-New}), thereby allowing for an analytic Wick rotation. 

\subsection{Propagators mimicking dimensional reduction versus exponential form factors}

In the last subsection we examined two propagators satisfying all properties required by unitarity, stability, and causality that we discussed throughout the manuscript. Despite their nice features in this regard, these propagators cannot be compatible with a well-defined UV completion in terms of an UV fixed point of the RG flow. The latter would indeed require a certain form of effective dimensional reduction~\cite{Carlip:2017eud}.

In this final subsection we shall compare the features of the propagators stemming from three different effective actions or derivations. We will work here with dimensionless momenta $q$ for shortness. Chronologically, the first is 
\begin{equation}\label{eq:propyD1}
	D_1^{-1}(q^2)=q^2 e^{q^2}\,,
\end{equation}
and it comes from the simplest non-local gravity model~\cite{Tomboulis:1997gg,Modesto:2011kw,Biswas:2011ar,Tomboulis:2015esa}. As a caveat, within non-local gravity the exponential form factor is assumed to arise at the level of the bare action\footnote{Note that despite its interpretation as a bare action, the corresponding action is not derived from first-principle computations, e.g., as an RG fixed point of some theory. Moreover, since the bare action is non-local, these theories violate  locality at a fundamental level.}, and it is not clear whether its form is preserved along the RG flow, since there is no known symmetry or mechanism preventing other non-exponential operators from being generated by quantum fluctuations. This also means that some properties that seem to be satisfied at a perturbative level w.r.t. the bare action could be spoiled by quantum fluctuations (see, e.g.,~\cite{Shapiro:2015uxa} for an example of this mechanism). To allow for a consistent comparison with the other propagators we will introduce, we shall assume that quantum fluctuations indeed preserve the structure of this exponential form factor at quadratic order, and we shall thus re-interpret $D_1(q^2)$ as (the scalar part of) a dressed propagator.
The second propagator we will consider is
\begin{equation}\label{eq:propyD2}
	D_2^{-1}(q^2)=q^2 (1+q^2 \tanh{q^2})\,,
\end{equation}
and comes from the effective action proposed in~\cite{Draper:2020bop}. The motivation behind the proposal was to provide a proof of principle that effective actions can exist, that are compatible with positivity and causality bounds (at the level of scattering amplitudes), as well as with Weinberg's asymptotic safety condition. 
Finally, we will analyze the properties of the dressed graviton propagator numerically-derived in~\cite{Fehre:2021eob} using a spectral version of the FRG~\cite{Braun:2022mgx} which also accounts for Lorentzian signature. The spectral density in this case is analytically approximated by the interpolating function 
\begin{equation}\label{eq:spectralD3}
	\rho_3(\lambda^2)=\delta(\lambda^2)+\frac{1}{\pi}\left({\frac{8.4984 \,\lambda^{1.04/2}}{2.5593+0.6668\sqrt{\lambda^2}+\lambda^2}+\frac{3.5593}{1+0.8170 \sqrt{\lambda^2}+0.9151\, \lambda^2}}\right)\,,
\end{equation}
and the corresponding scalar part of the propagator for the transverse-traceless is given by
\begin{equation}\label{eq:propyD3}
	D_3(q^2)=\int_0^{\infty} d\lambda^2 \frac{\rho_3(\lambda^2)}{q^2-\lambda^2+i\epsilon} \,.
\end{equation}
Note that in the expression above the $i\epsilon$-prescription is crucial to avoid unphysical singularities at timelike momenta.

In what follows we shall compare these three propagators and some of their properties. See also~\cite{Knorr:2021iwv} for a complementary comparison of the first two form factors. For a given propagator $D_i$, a momentum-dependent version of the anomalous dimension $\eta_i(q^2)$ may be implicitly defined by 
\begin{equation}
	D_i(q^2)=q^{-2(1-\eta_i(q^2)/2)}\,.
\end{equation}
Nonetheless, this map between $D_i$ and $\eta_i$ is not bijective in $q^2=1$ and therefore a na\"ive inversion of this formula can result in unphysical divergences of $\eta_i$ at finite momentum. A definition avoiding this issue is
\begin{equation}
	\eta_i(q^2)\equiv-q \partial_{q}\log \left( q^{-2}D_i^{-1}(q^2) \right)\,,
\end{equation}
and we will employ the latter to determine the anomalous dimension associated with each~$D_i$.

\begin{figure}
	\includegraphics[scale=0.64]{"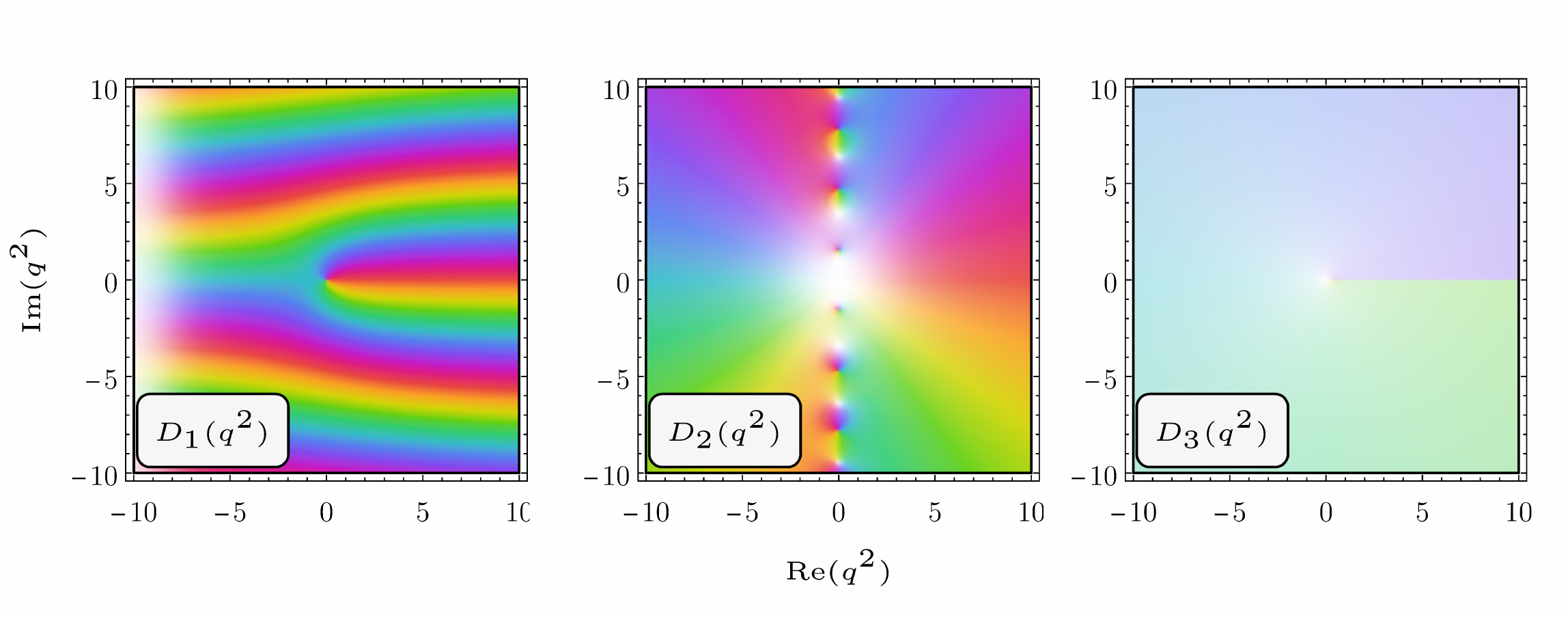"}\\
	\includegraphics[scale=0.6]{"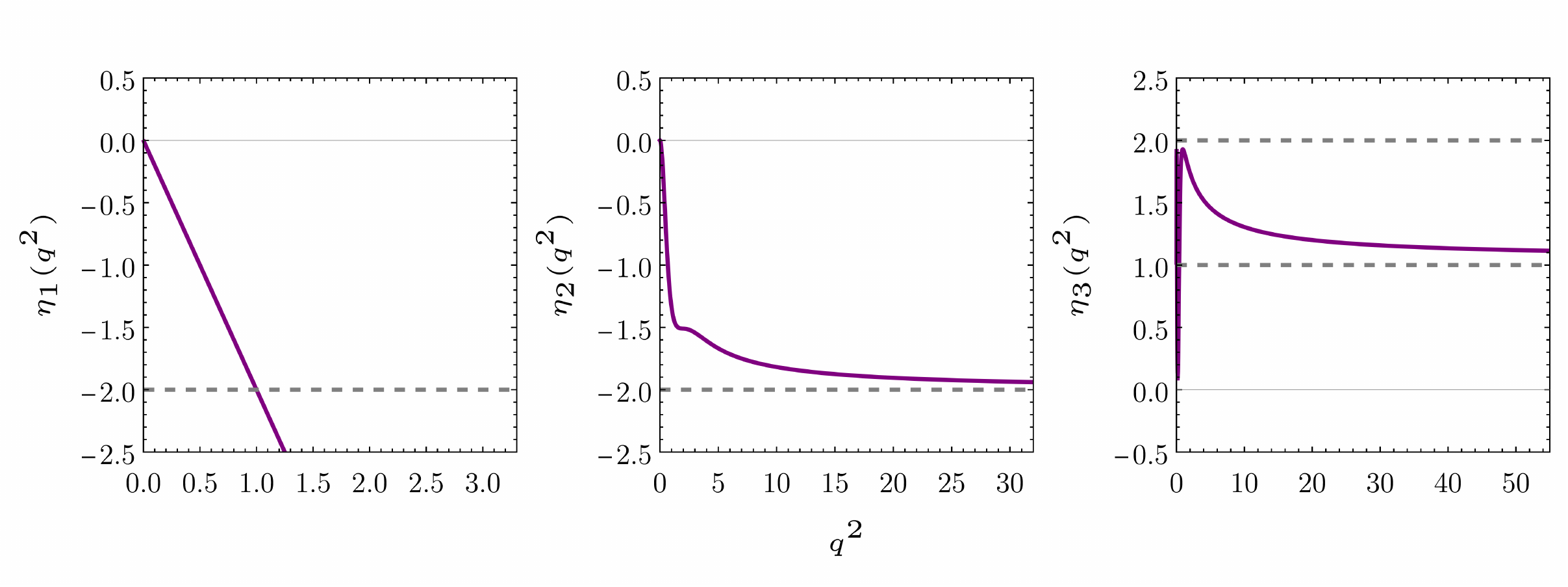"}
	\caption{Pole structure in the $q^2$-complex plane (top panel) and anomalous dimension $\eta_i$ for the propagators $D_1$ (figures on the left), $D_2$ (figures in the center), and $D_3$ (figures on the right). Aside from the massless pole at $q^2=0$, the propagator $D_1$ does not have additional poles but has an essential singularity at infinity. Its anomalous dimension varies from zero to negative infinity. In contrast to the case of $D_1$, the anomalous dimensions associated with the propagators $D_2$ and $D_3$ are bounded from above and from below. Yet, $D_2$ and $D_3$ display a very different pole structure: the first has no branch cut singularities and is characterized by a tower of massless complex-conjugated poles, while the second does not have poles beyond the massless one and has instead a branch cut for~$q^2>0$.  \label{fig:propydimred}}
\end{figure}

The properties of the three propagators are depicted in Fig.~\ref{fig:propydimred}, which shows their pole structure (top panel) and the momentum-dependent anomalous dimension (bottom panel).
Focusing on the propagator $D_1$, the absence of additional poles beyond the massless one indicates that causality and perturbative unitarity are preserved at the level of the two-point function. Nonetheless, the presence of essential singularities at infinity forbids performing an analytic Wick rotation (this problem could be alleviated employing a different prescription for the analytic continuation~\cite{Buoninfante:2022krn}) and makes it difficult to generalize some results in QFT whose derivation makes use of the Cauchy integral formula\footnote{See~\cite{Pius:2016jsl} for first steps towards overcoming this issue.}. A final important observation is that the momentum-dependent anomalous dimension $\eta_1(q^2)$ diverges to minus infinity in the limit $q^2\to\infty$, potentially indicating that this type of exponential form factors might not be compatible with a standard UV completion defined by an RG fixed-point action. This is compatible with the findings in~\cite{Fraaije:2022uhg}: UV non-locality requires the presence of fundamental scales in the fixed-point effective action, but such mass or length scales would explicitly break the scale invariance required for the theory to be UV-completed by a standard RG fixed point. 

The behavior of the anomalous dimensions $\eta_2$ and $\eta_3$ is instead qualitatively similar. The function $\eta_2$ varies continuously in the range $[0,-2)$, $\eta_*=-2$ being the limiting value for large momenta, and thus entails an effective dimensional reduction~\cite{Carlip:2017eud} from four (small momenta) to two (large momenta) spacetime dimensions. The anomalous dimension $\eta_3$ instead varies in the range $(0,2)$, and approaches the value $\eta_*\simeq 1.04$ as $q^2\to\infty$. Both behaviors are compatible with the existence of a non-trivial fixed point in the UV~\cite{Christiansen:2014raa,Draper:2020bop,Fehre:2021eob}. Despite these similarities, the propagators $D_2$ and $D_3$ have a very different pole structure: aside from the massless pole located at the origin of the $q^2$-complex plane, the former is characterized by an infinite number of massless complex-conjugate poles, while the latter does not have any additional poles and only displays a branch cut singularity for $q^2>0$. The tower of massless complex-conjugate poles characterizing $D_2$ is key to satisfy both causality bounds and Weinberg's asymptotic safety condition in~\cite{Draper:2020bop}. Yet, according to the discussion in Sect.~\ref{sect:causality}, such complex-conjugate pairs entail a violation of causality in terms of backward propagation of modes with positive energy (similarly to Lee-Wick theories) and vacuum instabilities as defined in the same section (unless a stabilizing mechanism exists\footnote{The question of whether a stabilizing mechanism exists in general is insofar unexplored. It is worth mentioning that the case of~\cite{Draper:2020bop} should be explored separately, since the interplay of infinitely many complex-conjugate poles might give rise to non-trivial effects, and could differ from the case of a single pair of complex-conjugate poles.}). The branch cut in $D_3$ is instead expected, since at low energies one should recover the one-loop logarithmic corrections to the classical action (cf. Sect.~\ref{sect:complexpoles}). Together with the absence of complex poles, this makes $D_3$ fully consistent w.r.t. all properties discussed throughout this manuscript and with standard EFT results.
At the same time, it is an open question whether scattering amplitudes determined using the techniques in~\cite{Fehre:2021eob} will display the same desirable properties as those stemming from the form factors studied in~\cite{Draper:2020bop}. 

\section{Discussion and conclusions} \label{sect:conclusions}

The perturbative quantization of General Relativity yields a theory that is perturbatively non-renormalizable but unitary. Adding a finite number of higher derivatives improves the ultraviolet behavior of gravity, but introduces ghosts.

In an attempt to reconcile renormalizability and unitarity, many theories of quantum gravity have been advanced.
Yet, independent of the specific ultraviolet completion of gravity, it is a key question what the properties of the dressed graviton propagator should be, at least ideally, and whether there exist propagators satisfying all these properties. 
Motivated by these questions, in this work we have discussed various aspects of causality, unitarity and stability, and their implications for the properties of the dressed (graviton) propagator.

First, we argued that unitarity of quantum field theories is best understood at the level of the effective action, and that truncations of the latter based on both entire and non-entire form factors lead to the appearance of fictitious ghosts. In particular, we provided numerical evidence (see.~Sect.~\ref{sect:polology}) for the residue decoupling mechanism of fictitious ghosts theorized in~\cite{Platania:2020knd}.  Second, we collected various inequivalent notions and definitions of causality that appear in the literature, and we discussed their relations. We also provided a detailed analysis of ghost and tachyonic instabilities, at the classical and quantum level, showing that at the quantum level tachyonic instabilities can be less severe than the vacuum instabilities provoked by unstable degrees of freedom with negative width and complex-conjugate poles. In this course, we highlighted a key difference between quantum ghost and tachyonic instabilities: the former is a ``kinetic problem'', and its resolutions in a non-perturbative setup is to be sought in the \emph{momentum dependence} of the effective action. The latter is a ``potential problem'' and its resolution lies in the \emph{field dependence} of the effective action. An explicit example of how interaction terms in the effective action can cure tachyonic instabilities was provided in Sect.~\ref{sect:interactionssavetachyons}. The relation between poles of a dressed propagator, and some forms of causality, unitarity and stability is summarized in Tab.~\ref{tab:polesconditions}. An exact formula quantifying the causality violation 
is derived in Sect.~\ref{subsect:propycauscond}, cf. Eq.~\eqref{eq:tauviolation}, and is depicted in Fig.~\ref{fig:decaylifetime}.
Third, we showed that logarithmic corrections to the classical action are not sufficient to ameliorate the behavior of the theory, as the corresponding propagator displays either complex-conjugate poles (implying a violation of causality and vacuum instabilities) or stable tachyonic ghosts (implying a violation of unitarity, as well as tachyonic instabilities). Accounting for the infinitely many derivatives at the level of the effective action is  crucial, but not sufficient. Thus, we discussed and compared some examples of consistent field theories whose dynamics---encoded in the quantum effective action---preserves causality, (vacuum and tachyonic) stability and unitarity at the level of the two-point function, and whose Euclidean theory is connected to the Lorentzian one via an analytical Wick rotation. In particular, the propagator computed in~\cite{Fehre:2021eob} seems ideal in this respect, it being compatible with all properties listed and analyzed here, as well as with Weinberg's asymptotic safety condition.
Our work thus provides further support to the possibility of formulating a consistent and stable quantum theory of gravity compatible with all fundamental principles of quantum field theory.

\section*{Acknowledgements}

The author thanks D. Anselmi, I. Basile, L. Buoninfante, J. Donoghue, A. Eichhorn, B. Knorr, R. Percacci, A. Pereira, M. Reichert and C. Wetterich for many insightful discussions. The author is also grateful to M. Reichert for providing the interpolating function in Eq.~\eqref{eq:spectralD3} and to B. Knorr for very helpful comments on the manuscript. During part of the development of this work, A.P. was supported by the Alexander von Humboldt Foundation. A.P. acknowledges support by Perimeter Institute for Theoretical Physics. Research at Perimeter Institute is supported in part by the Government of Canada through the Department of Innovation, Science and Economic Development Canada and by the Province of Ontario through the Ministry of Colleges and Universities.

\bibliographystyle{JHEP}

\addcontentsline{toc}{section}{\refname}
\bibliography{AleBib}

\end{document}